\documentclass[11pt]{article}

\usepackage[T1]{fontenc}
\usepackage[utf8]{inputenc}

\usepackage{fancyhdr}

\usepackage[inline]{enumitem}
\usepackage{multicol,multirow,booktabs}

\usepackage{times}

\newcommand{\arxiv}{1}

\usepackage{comment}

\usepackage[hang,small,labelfont=bf,up,textfont=it,up]{caption}
\usepackage{subcaption}
\usepackage{amsmath,amsfonts,amsthm,amssymb,bm,dsfont}

\setlist{nolistsep} 

\usepackage{graphicx}
\usepackage[a4paper,top=2.5cm,bottom=2.5cm,left=2.2cm,right=2.2cm]{geometry}

\usepackage{numprint}
\npthousandsep{,}\npthousandthpartsep{}\npdecimalsign{.}

\usepackage{thmtools}
\usepackage{thm-restate}

\newtheorem{theorem}{Theorem}
\newtheorem{proposition}[theorem]{Proposition}

\newtheorem{definition}{Definition}

\newtheorem{assumption}{Assumption}

\DeclareMathOperator{\rank}{rank}

\newcommand{\Pmat}{\mathbf{P}}
\newcommand{\Amat}{\mathbf{A}}

\newcommand{\Vmat}{\mathbf{V}}
\newcommand{\UmatA}{\mathbf{U}_{\mathbf{A}}}
\newcommand{\VmatA}{\mathbf{V}_{\mathbf{A}}}
\newcommand{\UmatP}{\mathbf{U}_{\mathbf{P}}}
\newcommand{\VmatP}{\mathbf{V}_{\mathbf{P}}}
\newcommand{\DmatA}{\mathbf{D}_{\mathbf{A}}}
\newcommand{\DmatP}{\mathbf{D}_{\mathbf{P}}}
\newcommand{\Wmat}{\mathbf{W}}
\newcommand{\Dmat}{\mathbf{D}}
\newcommand{\Ltilde}{\tilde{\mathbf{L}}}
\newcommand{\Rtilde}{\tilde{\mathbf{R}}}
\newcommand{\XmatP}{\mathbf{X}_{\Pmat}}
\newcommand{\YmatP}{\mathbf{Y}_{\Pmat}}
\newcommand{\Xmat}{\mathbf{X}}
\newcommand{\Ymat}{\mathbf{Y}}
\newcommand{\hXm}{\hat{\Xmat}}
\newcommand{\hYm}{\hat{\Ymat}}
\newcommand{\Qmat}{\mathbf{Q}}
\newcommand{\Rmat}{\mathbf{R}}
\newcommand{\tti}{2 \to \infty}
\newcommand{\Mmat}{\mat{M}}

\newcommand{\Kn}{K_n}
\newcommand{\Tn}{T_n}
\newcommand{\K}{K}
\newcommand{\T}{T}

\newcommand{\Op}{O_{\mathbb{P}}}
\newcommand{\OmegaP}{\Omega_{\mathbb{P}}}
\newcommand{\UmatAi}{\mathbf{U}^{(i)}_{\mathbf{A}}}

\newcommand{\maxKT}{\max(\Kn, \Tn)}

\newcommand{\abs}[1]{\vert#1\vert}
\newcommand{\norm}[1]{\|#1\|}
\newcommand{\bignorm}[1]{\left\|#1\right\|}
\newcommand{\bigabs}[1]{\left\vert\ #1\ \right\vert}
\newcommand{\bigabsnogap}[1]{\left\vert#1\right\vert}

\newcommand{\twoinf}[1]{\|#1\|_{\tti}}
\newcommand{\frob}[1]{\norm{#1}_{F}}

\newcommand{\GL}{\mathrm{GL}}
\newcommand{\Od}{\mathbb{O}}
\newcommand{\Id}{\mat{I}}

\newcommand{\bDelta}{{\bm\Delta}}
\newcommand{\tDelta}{\tilde{\bm\Delta}}

\newcommand{\WX}{{\mat W}_{X}}
\newcommand{\WY}{{\mat W}_{Y}}

\renewcommand{\vec}[1]{\boldsymbol{#1}}

\newcommand{\mat}[1]{\mathbf{#1}}

\newcommand{\titledoc}{Doubly unfolded adjacency spectral embedding of dynamic multiplex networks}
\newcommand{\titleshort}{Doubly unfolded adjacency spectral embedding of dynamic multiplex networks}

\providecommand{\keywords}[1]{\noindent{\small{\textbf{\textit{Keywords --}} #1}}}

\usepackage{natbib}
\usepackage{setspace}

\usepackage{authblk}

\usepackage{mathtools}
\mathtoolsset{showonlyrefs}

\usepackage[ruled,linesnumbered,noend]{algorithm2e}
\SetKwInput{KwInput}{Input}
\let\oldnl\nl
\newcommand{\nonl}{\renewcommand{\nl}{\let\nl\oldnl}}
\makeatother

\usepackage{graphicx}
\usepackage[colorlinks=true, allcolors=blue]{hyperref}
\usepackage[english]{babel}
\usepackage{listings}
\usepackage{placeins}
\usepackage{csquotes}

\author{Maximilian Baum}
\author{Francesco Sanna Passino}
\author{Axel Gandy}

\affil{Department of Mathematics, Imperial College London \\ 180 Queen's Gate, SW7 2AZ, London}

\date{}

\title{\Huge\textbf{\titledoc}}

\allowdisplaybreaks

\graphicspath{{Figures/}}

\usepackage{titlesec}
\titleformat{\paragraph}[runin]{\normalfont\itshape}{\theparagraph}{1em}{}
\titlespacing*{\paragraph}{0pt}{1.5ex plus 1ex minus .2ex}{0.5em}

\usepackage{multibib}
\newcites{SM}{Supplementary references}

\begin{document}

\maketitle


\begin{abstract}
    Many real-world networks evolve dynamically over time and present different types of connections between nodes, often called layers. 
In this work, we propose a latent position model for these objects, called the \textit{dynamic multiplex random dot product graph} (DMPRDPG), which uses an inner product between layer-specific and time-specific latent representations of the nodes to obtain edge probabilities.
We further introduce a computationally efficient spectral embedding method for estimation of DMPRDPG parameters, called \textit{doubly unfolded adjacency spectral embedding} (DUASE). The DUASE estimates are proved to be both consistent and asymptotically normally distributed. A key strength of our method is the encoding of time-specific node representations and layer-specific effects in separate latent spaces, which allows the model to capture complex behaviours while maintaining relatively low dimensionality. The embedding method we propose can also be efficiently used for subsequent inference tasks. In particular, we highlight the use of the ISOMAP algorithm in conjunction with DUASE as a way to efficiently capture trends and global changepoints within a network, and the use of DUASE for graph clustering. Applications on real-world networks describing geopolitical interactions between countries and financial news reporting demonstrate practical uses of our method. 

\end{abstract}


\section{Introduction}

A theme of growing interest in modern statistics is the analysis of graphs or networks, representing mathematical objects which can capture a notion of connectedness between entities. These structures may be either dynamic or static in nature, and may involve distinct types of connections, which are usually referred to as \textit{layers} of the graph \citep[see, for example,][]{DeDomenico13}. Such objects arise in many different fields, such as transportation systems, where different modes of transport represent the layers, or computer networks, where connection between hosts could occur on different ports. 
When a network evolves over time, it is usually called a \textit{dynamic graph}. Similarly, if a network is characterised by different edge types, it is typically called a \textit{multilayer graph}. When temporal evolution and multiple edge types occur at the same time, then the network is called a 
\textit{dynamic multilayer graph}.
These structures are actively studied in the machine learning and artificial intelligence literature, where they are often known as \textit{temporal knowledge graphs} \citep[see, for example,][]{Fensel2020}.
Within the mathematical statistics literature, extensive work has been carried out to study dynamic networks and multilayer networks \textit{separately}, such as \cite{sewell2015latent, athreya2023euclidean, Billio24} for dynamic networks and \cite{Sosa22, Huang23, Lei23} for multilayer graphs, among others. Alternatively, techniques that can be applied interchangeably to dynamic or multilayer graphs have been proposed \citep[for example,][]{Han15, jones2021multilayer}.
On the other hand, only a limited number of studies analyse dynamic and multilayer graphs \textit{simultaneously}. Notable examples are
\cite{Oselio14, Hoff15, Durante17, Lopez22, Loyal23, Wang23}. 

In this work, we propose a spectral embedding method for dynamic graphs with multiple layers. 
We call our method doubly unfolded adjacency spectral embedding (DUASE), where the 
two levels of unfolding arise from the temporal and layer dimensions of the graph. 
We also propose a \textbf{d}ynamic \textbf{m}ulti\textbf{p}lex \textbf{r}andom \textbf{d}ot \textbf{p}roduct \textbf{g}raph model (DMPRDPG),  and prove desirable asymptotic properties of the DUASE estimator for estimation of the DMPRDPG parameters. In particular, we establish consistency and central limit theorems when the number of nodes in the graph grows to infinity. 
We use these results for subsequent inference tasks, such as community detection under a dynamic multilayer stochastic blockmodel \citep[see, for example,][]{Lopez22}, or global graph
changepoint detection \citep{athreya2023euclidean}.


\subsection{Background on graphs}

A graph is mathematically represented as 
$\mathcal G=(\mathcal V,\mathcal E)$, where $\mathcal V=[n]$ is a set of $n\in\mathbb N$ nodes, with $[n]=\{1,\dots,n\}$, and $\mathcal E\subseteq\mathcal V\times\mathcal V$ is an edge set, where $(i,j)\in\mathcal E$ only if nodes $i\in\mathcal V$ and $j\in\mathcal V$ are connected. Graphs are usually represented mathematically via an adjacency matrix $\Amat=\{\Amat_{i,j}\}_{i,j\in\mathcal V}\in[0,1]^{n\times n}$ with $\Amat_{i,j}=\mathds 1_{\mathcal E}\{(i,j)\}$, where $\mathds 1_{\cdot}\{\cdot\}$ denotes the indicator function. 
It is usually assumed that the graph has no self loops, implying that $(i,i)\notin\mathcal E$ for all $i\in\mathcal V$, corresponding to $\Amat_{i,i}=0$. 
Additionally, the graph is 
undirected if and only if $(i,j)\in\mathcal E$ implies $(j,i)\in\mathcal E,\ i,j\in\mathcal V$. Otherwise, the graph is directed. 

One of the foundational statistical models used to describe a single network adjacency matrix is the \emph{independent edge model}, where entries in the adjacency matrix are assumed to be independent, with probability $\mathbb P(\Amat_{i,j}=1)=p_{i,j}\in[0,1]$ for all $i,j\in[n]$. Therefore, in expectation, $\mathbb E(\Amat)=\Pmat\in[0,1]^{n\times n}$, where $\Pmat_{i,j}=p_{i,j}$. Many existing statistical models for networks then attempt to ``borrow stregth'' between nodes and the edges they form, by assuming that the matrix $\Pmat$ can be encoded via a \emph{low dimensional structure} \citep[see, for example,][]{RubinDelanchy20}.
In particular, in the latent position model \citep{hoff2002latent}, also known as the latent space model (LSM), adapted to directed graphs, each node is equipped with latent features, represented by vectors $\Xmat_i\in\mathcal X\subseteq\mathbb R^d,\ \Ymat_i\in\mathcal Y\subseteq\mathbb R^d,\ i=1,\dots,n,$ in some low-dimensional spaces $\mathcal X$ and $\mathcal Y$, where usually $d\ll n$. These latent vectors probabilistically characterise the connectivity via a kernel function $\kappa:\mathcal X\times\mathcal X\to[0,1]$, which gives the probability of a link between two nodes $i\in\mathcal V$ and $j\in\mathcal V$ as follows:
\begin{equation}
\mathbb{P}(\Amat_{i,j} = 1) = \kappa(\Xmat_i,\Ymat_j),\ i,j\in[n]. 
\end{equation}
If the link function is known, the inferential objective becomes to estimate the 
latent position coordinates $\Xmat_1,\dots,\Xmat_n$ and $\Ymat_1,\dots,\Ymat_n$, which represent the behaviour of each node as a source and destination of a connection respectively.
In the setting where $\kappa(\Xmat _i, \Ymat_j) = \Xmat_i^\intercal\Ymat_j$ we refer to the model as a directed random dot product graph \citep[RDPG; see, for example,][]{athreya2018statistical}. The RDPG will the be at the core of the model presented in this work, building upon the existing rich literature and theoretical results including consistent estimation of latent postions \citep{sussman2013consistent}  and distributional results around spectral estimators \citep{athreya2016limit}. In particular, our work focuses on extending an RDPG-based model for multiplex graphs, called the multilayer RDPG \citep[MRDPG,][]{jones2021multilayer} to a time-varying setting. In multiplex graphs, nodes in $\mathcal V$ can exhibit connectivity through $K\in\mathbb N$ different connection types 
encoded via type-specific edge sets $\mathcal E_k,\ k=1,\dots,K$.
In this way, the graph can describe a richer class of behaviors, such as different modes of commuting in transportation networks, or connections via different ports in cyber-security networks.
We remark that, in many disciplines, the terminology \textit{multiplex networks} indicates a subset of the wider class of \textit{multilayer networks} \citep[see, for example,][]{DeDomenico13}, where layer-specific node sets are used and connections can also occur between nodes in different layers. In this article, we assume that $\mathcal V$ is shared across layers and we do not observe connections between layers, but only \textit{within} layers, implying that we work under the multiplex graph framework.

In our work, we propose an embedding method for time-varying multiplex graphs and evaluate its theoretical properties. 
More formally, we consider the case of a dynamic multiplex network consisting of $\K\in\mathbb N$ layers observed at $\T\in\mathbb N$ time points, written $\mathcal G=(\mathcal V,\{\mathcal E^{k,t}\}_{k\in[\K],\ t\in[\T]})$. 
The set of $n\in\mathbb N$ nodes $\mathcal V$ is \textit{shared} across layers and time points, and $\mathcal E^{k,t}\subseteq\mathcal V\times\mathcal V$ is an edge set of connections between nodes, where $(i,j)\in\mathcal E^{k,t}$ if and only if there is a connection of type $k$ between nodes $i\in\mathcal V$ and $j\in\mathcal V$ at the $t$-th time point. 
For any pair of edge sets $\mathcal E^{k,t}$ and $\mathcal E^{k^\prime,t^\prime}$, $k,k^\prime\in[K],\ t,t^\prime\in[T]$, we \textit{do not} assume $\mathcal E^{k,t}\cap\mathcal E^{k^\prime,t^\prime}=\varnothing$, implying that connections between the same pair of nodes $i\in\mathcal V$ and $j\in\mathcal V$ can occur across multiple layers and time points. 
Similarly to the case of the static monoplex network introduced before, network $\mathcal G$ can be encoded via a collection of adjacency matrices $\{ \Amat^{k,t}\}_{k\in[\K], t\in[T]}$, where $\Amat^{k,t}\in\{0,1\}^{n\times n}$, with entries $\Amat^{k,t}_{i,j}=\mathds{1}_{\mathcal{E}^{k,t}}\{(i,j)\}$. 
In this work, we consider the graph to be directed and make comments where necessary to accommodate the case of undirected graphs. 
Following 
an approach that share similarities to RDPGs for static networks and MRDPGs \citep{jones2021multilayer}, we propose a model which postulates a low-rank structure for the matrices $\mathbb E(\Amat^{k,t})=\Pmat^{k,t}, k\in[K], t\in[T]$.

\subsection{Notation}

Before describing the proposed modelling framework and its properties, we establish the notation used in this work. We follow the standard notation utilised in the literature about random dot product graphs \citep[see][]{athreya2018statistical, RubinDelanchy22}.

\paragraph{Matrices.}
In this work, matrices are generically denoted via uppercase bold letters, such as $\mat M$. Its transpose is written $\mat M^\intercal$, and the identity matrix is usually written $\mat I$. The matrix elements are denoted $\mat M_{i,j}$, 
whereas $\mat M_i$ represents the $i$-th row of $\mat M$. 
For a matrix $\mat M\in\mathbb R^{m\times m}$, $\lambda_\ell(\mat M)$ is the $\ell$-th largest eigenvalue in magnitude of the matrix, for $ \ell\in[m]$. Similarly, $\sigma_\ell(\mat M)$ is used to denote the $\ell$-th largest singular value of a matrix $\mat M\in\mathbb R^{m_1\times m_2}$. 
Generally, $\frob{\mat M}$ is the Frobenius norm of $\mat M$, whereas $\norm{\mat M}$ is the spectral norm, corresponding to $\sqrt{\lambda_1(\mat M^\intercal \mat M)}$ or equivalently $\sigma_1(\mat M)$. 
The two-to-infinity norm of a matrix is written as $\twoinf{\cdot}$, corresponding to the maximum of the Euclidean norms of the rows: $\twoinf{\mat M} = \max_{i\in[m_1]}\norm{\mat M_i}_2$ \citep{cape19two}, where $\norm{\cdot}_2$ is the Euclidean vector norm. 
Matrices in a sequence are usually denoted via superscripts. For example, a sequence of $r\in\mathbb N$ matrices is written $\mat M^{1},\dots,\mat M^{r}\in\mathbb R^{m_1\times m_2}$. Also, we write $[\mat M^{1}\mid\cdots\mid \mat M^{r}]\in\mathbb R^{rm_1\times m_2}$ and $[\mat M^{1},\dots,\mat M^{r}]\in\mathbb R^{m_1\times rm_2}$ denote the vertical and horizontal stacking of the matrices, respectively. Additionally, we let $\GL(d)$ be the general linear group of dimension $d$, containing all $d\times d$ \textit{invertible} matrices. 
Similarly, we let $\Od(d)$ be the orthogonal group of dimension $d$, containing all matrices $\mat M\in\mathbb R^{d\times d}$ such that $\mat M^\intercal\mat M=\Id$. 

\paragraph{Asymptotic notation.}
The theoretical results in this work will be shown to hold probabilistically as the number of nodes $n$ tends to infinity. Hence, we introduce notation to characterise the asymptotic behaviour of functions and random variables.
For two real-valued functions $f$ and $g$, we write $g(n) = \Omega\{f(n)\}$ and $f(n) = O\{g(n)\}$ as $n\to\infty$ if there exist $n^\ast\in\mathbb R$ and a constant $C>0$ such that $\abs{f(n)} \leq Cg(n)$ for all
$n>n^\ast$. 
Also, we write $f(n) = o\{g(n)\}$ and
$g(n) = \omega\{f(n)\}$  as $n\to\infty$ if $\lim_{n\to\infty} f(n)/g(n) = 0$.
To more precisely define the probabilistic nature of our bounds, we also adopt the notion of \textit{overwhelming probability} \citep[see, for example,][]{tao2010random}. 

\begin{definition}[Overwhelming probability] \label{definition:op}
    An event $E_n$ depending on $n$ holds with ``overwhelming probability'' if for every constant $\gamma>0$, there exists a finite $C_\gamma>0$ (depending on $\gamma$ but independent of $n$) such that $\mathbb{P}(E_n) \geq 1 - C_\gamma n^{-\gamma}$.
\end{definition}

Note that if $E_n$ is a sequence of events of cardinality
$O\{n^{O(1)}\}$, each holding with uniformly overwhelming probability, then the intersection $\bigcap_n E_n$ holds with overwhelming probability \citep{tao2010random}.
The definition of overwhelming probability is used to establish notation around the limiting behaviour related to sequences of random variables. In particular, for a real-valued random variable $Z$ and a real-valued function $f(n)$, we write $\abs{Z} = \Op\{f(n)\}$ if, for any $\gamma >0$ there exists an $n^\ast\in\mathbb N$ and a constant $C_\gamma>0$ such that the event $\abs{Z} \leq C_\gamma f(n)$ holds with probability greater than $1-n^{-\gamma}$ for all $n \geq n^\ast$. Similarly, $\abs{Z} = \OmegaP\{f(n)\}$ if, for any $\gamma>0$ there exists an $n^\ast\in\mathbb N$ and a constant $C_\gamma>0$ such that $\abs{Z} \geq C_\gamma f(n)$ with probability at least $1-n^{-\gamma}$ for all $n \geq n^\ast$.

We explicitly clarify here that the results in our work hold as the number of nodes $n$ tends to infinity, as opposed to the case when $n$ is fixed and we observe the network over an increasing number of points in time $\T$ or number of layers $K$. 
In some situations, $\K$ and $\T$ may also grow as $n$ tends to infinity. In the settings where we let $\K$ and $\T$ grow with $n$, we will denote these as $\Kn$ and $\Tn$ to emphasise their relationship with the number of nodes. 

\subsection{Summary}

The remainder of this work is organised as follows: Section~\ref{sec:related_literature} summarises related background literature on statistical modelling for dynamic multilayer graphs. In Section~\ref{sec:DMPRDPG}, we introduce our proposed dynamic multiplex random dot product graph (DMPRDPG) and the related doubly unfolded adjacency spectral embedding (DUASE) estimator. The main results related to the asymptotic properties and stability guarantees of DUASE are established in Section~\ref{sec:theoretical_results}, followed by examples on subsequent inference tasks in Section~\ref{sec:subsequent_inference}. 

\section{Background}
\label{sec:related_literature}

The DMPRDPG proposed in this work fits into a rich literature around latent space models for both the dynamic and multiplex regimes. Usually, the dynamic and multiplex cases are treated separately, and a small number of techniques exist to cover both cases simultaneously. A particularly relevant example within this framework is \cite{Oselio14}, which introduces a general hierarchical model for dynamic multiplex graphs in which a set of latent positions is shared between layers and each layer is independent after conditioning on this set. 
\citet{Oselio14} mainly focus on the task of recovering the true underlying network adjacency matrix via a Bayesian model, when only a noisy version is observed.
Popular proposals for modeling the evolution of the time-varying latent variables include smoothness penalization and auto-regressive processes \citep{sewell2015latent}, as well as Gaussian processes \citep{Durante17} or latent position processes \citep{athreya2023euclidean}. 
In particular, \cite{Durante17} proposes to model the edge probabilities as functions of shared and layer-specific latent positions evolving as a Gaussian process. 
\cite{Loyal23} propose an eigenmodel where the latent positions are time-varying and multiplied with time-invariant layer-specific matrices of homophily coefficients in the log-odds space. \cite{Wang23} builds upon the MRDPG \citep{jones2021multilayer} to formulate an online change point detection framework for dynamic multilayer graphs.
Beyond latent space models for dynamic multilayer graphs, 
\cite{Lopez22} propose a dynamic multilayer stochastic blockmodel which utilises layer-specific Markov-chains to model the formation of node communities.

More generally, our work is linked to the literature around \textit{graph embedding methods}. 
The concept of a graph embedding 
refers to a low-dimensional representation of each node, that can be used to explain much of the structure of the original graph. In a latent space model, this generally translates to the problem of estimating the node latent positions based on an observed network adjacency matrix. To this end, a number of streams of research have emerged, most notably including spectral methods \citep{sussman2012consistent}, Bayesian methods \citep{sewell2015latent} and deep-learning methods \citep{grover2016node2vec}. Within the setting of spectral methods one of the foundational techniques used is adjacency spectral embedding \citep[ASE; see, for example,][]{athreya2018statistical}, where the eigenvalues and eigenvectors of the graph adjacency matrix $\Amat$ are used to obtain a latent position estimate. In the setting of multilayer graphs, there are several prominent methodologies within the spectral embedding framework, including omnibus spectral emdedding \citep[OMNI;][]{levin2017central}, mean adjacency spectral embedding \citep[MASE;][]{Arroyo21} and unfolded adjacency spectral embedding \citep[UASE;][]{jones2021multilayer,gallagher2021spectral}.
In particular, for a sequence of $\K$ adjacency matrices $\Amat^1,\dots,\Amat^{\K}\in\{0,1\}^{n\times n}$, OMNI considers an embedding based on a singular value decomposition (SVD) of the matrix
\begin{equation}
    \tilde{\Amat} = \begin{bmatrix} 
    \Amat^1 & (\Amat^1 + \Amat^2)/2 & \cdots & (\Amat^1 + \Amat^{\K})/2 \\
    (\Amat^2 + \Amat^1)/2 & \Amat^2 & \cdots & (\Amat^2 + \Amat^{\K})/2 \\
    \vdots & \vdots & \ddots & \vdots \\
    (\Amat^{\K} + \Amat^1)/2 & (\Amat^{\K} + \Amat^2)/2 & \cdots & \Amat^{\K}
    \end{bmatrix}\in\{0,1\}^{n\K\times n\K}.
\end{equation}
On the other hand, UASE constructs the embedding via the SVD of $\tilde{\Amat} = [\Amat^1, \dots,\Amat^{\K}]\in\{0,1\}^{n\times n\K}$. 
The DUASE method proposed in this work provides layer-specific and time-specific embeddings simultaneously for each node, by stacking the adjacency matrices horizontally and vertically, combining the ideas underlying UASE and OMNI. 

The UASE method for graph embedding is a natural spectral estimator for the MRDPG \citep{jones2021multilayer} for adjacency matrices $\{\Amat^k\}_{k\in[K]}$, which assumes a shared set of latent positions $\Xmat_i\in\mathbb R^d,\ i\in[n]$, across all layers, and layer-specific positions $\Ymat_i^k\in\mathbb R^d,\ i\in[n],\ k\in[K]$, such that, in its simplest version, $\mathbb P(\Amat_{i,j}^k=1)=\Xmat_i^\intercal \Ymat_j^k$ independently for all $i,j\in[n], i\neq j$.  
The model for dynamic multiplex graphs proposed in this work, in its undirected version, is also closely related to the common subspace independent edge (COSIE) 
model of \cite{Arroyo21}. In COSIE, originally developed for modelling a sequence of adjacency matrices $\{\Amat^k\}_{k\in[K]}$, it is assumed that each node has a latent position $\Xmat_i\in\mathbb R^d$ shared across layers, and $\mathbb P(\Amat^k_{i,j}=1)=\Xmat_i^\intercal \mat{S}_k\Xmat_j$ independently for all $i,j\in[n],\ i\neq j$, where $\mat{S}_k\in\mathbb R^{d\times d},\ k\in[K]$, are called \emph{score matrices}, and $\Xmat=[\Xmat_1\mid\cdots\mid\Xmat_n]\in\mathbb R^{n\times d}$ is the \emph{matrix factor}. 
\cite{Wang23} extend COSIE and MRDPG to a dynamic MRDPG for matrices $\{\Amat^{k,t}\}_{k\in[K], t\in[T]}$, by assuming that $\mathbb P(\Amat^{k,t}_{i,j}=1)=\Xmat_i^\intercal \mat{S}_{k,t}\Xmat_j$ independently, where $\mat{S}_{k,t}\in\mathbb R^{d\times d},\ k\in[K],\ t\in[T]$ is the \emph{weight matrix sequence}.  

A further natural method for embedding dynamic and multiplex networks is the use of tensor-based methods, via adjacency tensors. These techniques represent a graph in the form of a higher-dimensional tensor object and perform inference on that object directly. 
For example, \cite{ke2019community} employs a higher-dimensional generalization of singular value decomposition to address  the problem of community detection in hypergraphs. 
\cite{zhen2023community} address the same problem but make use of a likelihood-based estimator which consistently recovers true underlying node communities. While these methods have been used primarily in the context of hypergraph analysis, it is possible to use similar mathematical tools to analyse dynamic or multiplex graphs. Some early steps in this direction have been taken as \cite{malik2021dynamic} use deep learning and graph neural network methods to analyse dynamic networks as a stacked adjacency tensor. 







\section{Dynamic multiplex random dot product graphs}
\label{sec:DMPRDPG}

In this section, we present two of the main contributions of our work: the dynamic multiplex random dot product graph, and an estimator for the model parameters, called doubly unfolded adjacency embedding. 

\subsection{Model}

For each layer $k\in[K]$ and $t\in[T]$, we propose to use a random dot product graph to model each adjacency matrix $\Amat^{k,t},\ k\in[\K],\ t\in[\T]$, with latent positions shared across different layers and time points, \textit{borrowing strength} 
from multiple adjacency matrices. 
In particular, we assume that each adjacency matrix has the following low-rank decomposition: 
\begin{equation}
\mathbb E(\Amat^{k,t}) = \Pmat^{k,t} = \Xmat^k \Ymat^{t\intercal},\quad k\in[K],\ t\in[T], 
\end{equation}
where $\Xmat^{k} = [\Xmat^{k}_1\mid\dots\mid\Xmat^{k}_n]\in\mathbb R^{n\times d}$ and $\Ymat^{t} = [\Ymat^{t}_1\mid\dots\mid\Ymat^{t}_n] \in\mathbb R^{n\times d}$, with $\Xmat_i^k,\ k\in[\K]$ and $\Ymat_i^t,\ t\in[\T]$ denoting node-specific latent positions in embedding spaces $\mathcal{X}^{k} \subseteq \mathbb{R}^d,\ k\in[\K]$, and $\mathcal{Y}^{t} \subseteq \mathbb{R}^d,\ t\in[\T]$. The latent position $\Xmat_i^k\in\mathcal{X}^k$ is a time-invariant representation of node $i$, specific to layer $k$, which combines information across all time points, 
whereas $\Ymat_j^t\in\mathcal Y^t$ is a time-varying global representation of node $j$, which combines information across all layers, resulting in a unique representation at time $t$. 
The connection probability between nodes $i$ and $j$ at time $t$ in layer $k$ is given by the inner product of these positions: $\mathbb P(\Amat_{i,j}^{k,t}=1) = \Xmat_i^{k\intercal}\Ymat_j^t$. 
Adopting a low-rank factorization of the probability matrices via $\K+\T$ $d$-dimensional latent positions for each node significantly reduces the model complexity compared to independent RDPG models applied to each $\Amat^{k,t},\ k\in[\K],\ t\in[\T]$, which would have resulted in $\K\times\T$ $d$-dimensional latent features for each node. 
This is a useful paradigm in applications such as cyber-security, where nodes are \textit{hosts} within an enterprise computer network, with edges representing connections between machines on different \textit{ports} over time \citep[see, for example,][Section 5]{jones2021multilayer}.
Each server behaves differently depending on the port being used. For example, a particular web server would form connections over port 443 (HTTPS, secure hypertext transfer protocol), but it would be less active over port 22 (SSH, secure shell). On the other hand, the activity of client machines changes over time, depending on factors like security vulnerabilities, time of day, or network activity.
The proposed modelling framework is  formalised as a dynamic multiplex random dot product graph, defined below.


\begin{definition}[DMPRDPG -- Dynamic multiplex random dot product graph]\label{def:DMPRDPG}
For integers $n, d,\T,\K\in\mathbb N$, let $\mathcal{X}^{1}, \dots, \mathcal{X}^{\K}, \mathcal{Y}^{1}, \dots, \mathcal{Y}^{\T} \subset \mathbb{R}^d$ be defined such that $x^\intercal y \in [0,1]$ for any $x \in \mathcal{X}^k$ and $y \in \mathcal{Y}^t$,\ $k\in[\K],\ t\in[\T]$. 
We let $F$ be a distribution on the product space $[\bigotimes_{k=1}^K \bigotimes_{i=1}^n \mathcal{X}^{k}] \bigotimes~[\bigotimes_{t=1}^T \bigotimes_{j=1}^n \mathcal{Y}^{t}]$, such that 
$
    \Xmat^{1}_1, \dots, \Xmat^{1}_n, \dots, \Xmat^{\K}_n, \Ymat^{1}_1, \dots, \Ymat^{1}_n, \dots, \Ymat^{\T}_n \sim F.
$
The components $\Xmat_i^k,\ i\in[n],\ k\in[\K]$ and $\Ymat_j^t,\ j\in[n],\ t\in[\T]$ are called latent positions. 
Organise the latent positions 
as $\Xmat^{k} = [\Xmat^{k}_1\mid\dots\mid\Xmat^{k}_n]\in\mathbb R^{n\times d}$ and $\Ymat^{t} = [\Ymat^{t}_1\mid\dots\mid\Ymat^{t}_n] \in\mathbb R^{n\times d}$ via vertical stacking, and 
define $\Xmat = [\Xmat^{1}\mid\dots\mid \Xmat^{\K}]\in \mathbb{R}^{n\K\times d}$ and $\Ymat = [\Ymat^{1}\mid\dots\mid\Ymat^{\T}]\in \mathbb{R}^{n\T \times d}$. Additionally, we define the $n\times n$ connection probability matrices for each time point and layer as $\Pmat^{k,t} = \Xmat^{k}\Ymat^{t\intercal}$ and we refer to the doubly unfolded probability matrix as
\begin{equation}
    \Pmat = \begin{bmatrix}
        \Pmat^{1,1} & \dots & \Pmat^{1,\T}\\
        \vdots & \ddots & \vdots \\
        \Pmat^{\K,1} & \dots & \Pmat^{\K,\T}
        \end{bmatrix} = \Xmat\Ymat^\intercal \in \mathbb{R}^{n\K \times n\T}.
\end{equation}
Given a sequence of adjacency matrices $\Amat^{k,t} \in \{0,1\}^{n \times n},\ k\in[\K],\ t\in[\T]$, we define the doubly unfolded adjacency matrix $\Amat \in \{0,1\}^{n\K \times n\T}$ as
\begin{equation}
\Amat = \begin{bmatrix}
\Amat^{1,1} & \dots & \Amat^{1,\T}\\
\vdots & \ddots & \vdots \\
\Amat^{\K,1} & \dots & \Amat^{\K,\T}
\end{bmatrix}.
\label{eq:double_unfolding}
\end{equation}
We can then say that $(\Amat,\Xmat,\Ymat) \sim 
\mathrm{DMPRDPG}(F)$ if, conditional on $\Xmat^{k}$ and $\Ymat^{t}$, the matrix $\Amat^{k,t}$ has independent entries with distribution 
\begin{equation}
    \Amat^{k,t}_{i,j} \sim \mathrm{Bernoulli}(\Pmat^{k,t}_{i,j}),
\end{equation}
for all $i,j\in\{1,\dots,n\}$, $i \neq j,\ k \in [\K],\ t \in [\T]$. 
\end{definition}

In the setting where the individual graphs $\Amat^{k,t}$ are undirected, we introduce the requirement that for each $k$ and $t$, with probability one, there exists a symmetric matrix $\mat{G}^{k,t} \in \mathrm{GL}(d)$ such that $\Xmat^k = \Ymat^t [\mat{G}^{k,t}]^{-1}$. This assumption ensures that the resulting matrices $\Pmat^{k,t}$ are symmetric by constraining the distribution of the right latent positions in a manner similar to the MRDPG \citep{jones2021multilayer} and the COSIE model \citep{Arroyo21}. Additionally, in the undirected case, the proposed approach reduces to a version of the dynamic MRDPG \citep{Wang23}. In particular, a layer-specific COSIE structure  can be recovered by writing $\Pmat^{k,t} = \Xmat^k {\Ymat^t} ^{\intercal} = \Xmat^k \left( {\Xmat^k \mat{G}^{k,t}} \right)^{\intercal} = \Xmat^k {\mat{G}^{k,t}}^{\intercal} {\Xmat^k}^{\intercal}$. Under this parametrization, the matrix $\mat{G}^{k,t}$ takes the role of the score matrix 
in the standard COSIE framework, whereas the layer-specific invariant subspaces defined by $\Xmat^k$ are the matrix factors. Similarly, our DMPRDPG can be given a COSIE structure similar to the dynamic MRDPG, by taking a reference set of latent positions $\Xmat^{k^\ast},\ k^\ast\in[K]$, as the common subspace, and setting the weight matrix sequence to $\mat{S}_{k,t}=\mat{G}^{k^\ast,t}[\mat{G}^{k,t}]^{-1}[\mat{G}^{k^\ast ,t}]^\intercal,\ k\in[K],\ t\in[T]$. 

The proposed framework for modeling dynamic multilayer graphs has several advantages. First, it simultaneously provides separate \textit{comparable} latent positions for each node at each time point, and for each node on each layer, which can be used to evaluate how the connectivity behaviour of nodes changes across time and layers. Additionally, latent positions can be efficiently estimated via a spectral decomposition of 
the adjacency matrices, with convenient asymptotic properties.
It is important to remark that the main modelling assumption postulated under the DMPRDPG is primarily a low-rank representation of the doubly unfolded matrix $\Pmat$. When this assumption is satisfied, the implication is that it is possible to factorise the layer-specific effects and time-specific effects into separate embedding spaces via the decomposition $\Pmat^{k,t}_{i,j} = {\Xmat^k_i}^\intercal{\Ymat^t_j}$, $i,j \in [n], k\in[K],\ t\in[T]$. 
 
\subsection{Doubly unfolded adjacency spectral embedding}

In practice, the latent positions $\Xmat$ and $\Ymat$ are unknown, and must be estimated from the observed matrix $\Amat$. To this end, we propose a doubly unfolded adjacency spectral embedding (DUASE) estimator for dynamic multiplex graphs. 
Given the realized adjacency matrices $\Amat^{k,t},\ k\in[\K],\ t\in[\T]$, we make use of a truncated SVD of rank $d$ to obtain a low-rank approximation of the doubly unfolded matrix $\Amat$. 

\begin{definition}[DUASE -- Doubly unfolded adjacency spectral embedding] \label{definition:duase}
Given a set of adjacency matrices $\{\Amat^{k,t}\}_{k\in [\K],t\in [\T]}$, where $\Amat^{k,t}\in \{0,1\}^{n \times n}$ for all $k\in[\K]$ and $t\in[\T]$, consider the doubly unfolded adjacency matrix 
\begin{equation}
\Amat = \begin{bmatrix}
\Amat^{1,1} & \dots & \Amat^{1,\T}\\
\vdots & \ddots & \vdots \\
\Amat^{\K,1} & \dots & \Amat^{\K,\T}
\end{bmatrix}\in \{0,1\}^{n\K \times n\T}.
\end{equation}
Consider the singular value decomposition 
\begin{equation}
    \mat A = \mat U\mat D\mat V^\intercal + \mat U_\perp\mat D_\perp \mat V_\perp^\intercal,
\end{equation} 
where $\mat D\in\mathbb R^{d\times d}$ is a diagonal matrix containing the $d$ largest singular values of $\mat A$, $\mat U\in\mathbb R^{n\K\times d}$ and $\mat V\in\mathbb R^{n\T\times d}$ contain the corresponding left and right singular vectors respectively, and $\mat D_\perp$, $\mat U_\perp$ and $\mat V_\perp$ contain the remaining singular values, left singular vectors, and right singular vectors respectively.
Then, the doubly unfolded adjacency spectral embedding of $\{\Amat^{k,t}\}_{k\in [\K],\ t\in [\T]}$ into $\mathbb R^d$ is 
\begin{align}
\hat{\Xmat} = \mat{U}\mat{D}^{1/2}\in\mathbb R^{n\K\times d}, & & 
\hat{\Ymat} = \mat{V}\mat{D}^{1/2}\in\mathbb R^{n\T\times d}.
\end{align}


\end{definition}

Conventionally, we will refer to $\hXm$ as the \textit{left embedding} or \textit{left DUASE}, and to $\hYm$ as the \textit{right embedding} or \textit{right DUASE}.
Based on the DUASE in Definition~\ref{definition:duase}, we can also retrieve layer-specific and time-specific estimates $\hXm^k$ and $\hYm^t$ by unstacking the $n \times d$ chunks of $\hXm$ and $\hYm$ in a manner that is analogous to the stacking procedure in the DMPRDPG in Definition~\ref{def:DMPRDPG}: 
$\hat{\Xmat} = [\hXm^1\mid\cdots\mid\hXm^{\K}]$
and $\hat{\Ymat} = [\hYm^1\mid\cdots\mid\hYm^{\T}]$.

We remark that DUASE is related to a UASE procedure for a bipartite MRDPG \citep{jones2021multilayer} with $n\T$ source nodes and $n$ destination nodes, observed across $\K$ layers. Therefore, the main asymptotic properties of UASE are retained by DUASE, with the additional property of admitting the number of layers and time points to grow simultaneously with $n$, and non-random diagonal entries in the adjacency matrices.  
Additionally, DUASE inherits the 
stability properties of the MRDPG and UASE \citep{jones2021multilayer, gallagher2021spectral}, as it assigns identical positions up to noise to nodes with similar behavior within each layer across all time points (cross-sectional stability), and maintains the same position up to noise for a single node exhibiting similar behavior across layers over different times (longitudinal stability). 

\subsection{Key assumptions and sparsity} \label{sec:sparsity_considerations} 

In this section, we present additional conditions on the DMPRDPG presented in Definition~\ref{def:DMPRDPG}, 
which are not fundamental to our model, but will be used to derive theoretical properties of the DUASE estimator. As a first step, we make the 
assumption that the support of the latent positions for each layer and time point are bounded. 
\begin{assumption}\label{ass:bounded-latent-space}
     $\mathcal{X}^k$ and $\mathcal{Y}^t$ are bounded subsets of $\mathbb{R}^d$ for all $k \in [\Kn]$, $t\in[\Tn]$.
\end{assumption}
In order to better control the overall connection density of the network, we additionally introduce a global sparsity parameter. In general, if the number of edges scales sub-quadratically with the number of nodes, then the graph is said to be \textit{sparse}, otherwise it is \textit{dense} \citep[see, for example,][]{Bollobas09}.
To explicitly introduce this notion within our DMPRDPG model, we adopt a global sparsity factor $\rho_n \in (0,1]$ to control the asymptotic connection density of the network as the number of nodes in the network $n$ tends to infinity \citep[see, for example,][]{RubinDelanchy22}. We assume that the sequence $\rho_n$ either converges to $0$ as $n\to\infty$, or is equal to $1$, corresponding to the sparse and dense regime respectively. Also, to ensure that the network is sufficiently dense, we introduce Assumption \ref{ass:rho-growth} to ensure that $\rho_n$ does not converge to $0$ too quickly. 
\begin{assumption}\label{ass:rho-growth}
    The sparsity parameter satisfies $\rho_n = \omega\{\log(n) n^{-1/4}\}$.
\end{assumption}
We adopt the notation $F_X$ and $F_Y$ to denote the marginal distributions of $F$ on $\bigotimes_{k=1}^K \bigotimes_{i=1}^n \mathcal{X}^{k}$ and $\bigotimes_{t=1}^{\T} \bigotimes_{j=1}^n \mathcal{Y}^{t}$ respectively. The desired sparsity regime is achieved by setting $\boldsymbol{\xi} \sim F_X$ and $\boldsymbol{\nu} \sim F_Y$ and defining the full matrix of final latent positions to be the scaled versions of these variables: $\Xmat = \rho_n^{1/2} \boldsymbol{\xi}$ and $\Ymat = \rho_n^{1/2} \boldsymbol{\nu}$. We adopt the notation $F_\rho$ to refer to this scaled distribution. To simplify the notation, we do not add additional subscripts or superscripts related to $n$ to $F$, $F_\rho$, $F_X$, $F_Y$, $\Xmat$, $\Ymat$, $\boldsymbol{\xi}$ or $\boldsymbol{\nu}$, but we generally assume that these quantities have dimensionality dependent on the number of nodes. 
Additionally, we assume that for each $k\in[\Kn]$ and $t\in[\Tn]$ the collections $\Xmat^k_1, \dots, \Xmat^k_n$ and $\Ymat^t_1, \dots, \Ymat^t_n$ are sampled independently from a shared marginal distribution of $F$ on each $\mathcal{X}^k$ and $\mathcal{Y}^t$. This condition is formalised below as Assumption \ref{ass:independence}. 

\begin{assumption}\label{ass:independence}
    Conditional on the layer $k \in [\Kn]$ or time point $t \in [\Tn]$, the latent positions for each node are sampled independently from a shared distribution such that $\bm\xi_1^k, \dots, \bm\xi_n^k \overset{i.i.d}{\sim} F_{X,k}$ or $\bm\nu_1^t, \dots, \bm\nu_n^t \overset{i.i.d}{\sim} F_{Y,t}$. Additionally, for $\bm\xi^k \sim F_{X,k}$ and $\bm\nu^t \sim F_{Y,t}$ the second moment matrices $\bDelta_{X,k} = \mathbb{E} [\bm\xi^k \bm\xi^{k\intercal}]$ and $\bDelta_{Y,t} = \mathbb{E} [\bm\nu^t\bm\nu^{t\intercal}]$ are full rank matrices.
\end{assumption}

In order to consider the asymptotic regime where $\Kn$ and $\Tn$ grow with $n$, we introduce Assumption \ref{ass:KT-growth} to govern the rate of this growth, as well as Assumptions \ref{ass:second-moment-convergence} and \ref{ass:clt-matrices}, which are regularity conditions on the second moment matrices of the latent position distributions.  
\begin{assumption}\label{ass:KT-growth}
    $\Kn$ and $\Tn$ satisfy $\max(\Kn,\Tn) = O\{\log(n)\}$, $\frac{\Kn}{\Tn} = O(1)$ $\frac{\Tn}{\Kn} = O(1)$.
\end{assumption}
\begin{assumption}\label{ass:second-moment-convergence}
    There exist full-rank $d \times d$ matrices $\tilde \bDelta_X$ and $\tilde \bDelta_Y$ such that 
    \begin{align}
        &\lim_{n \to \infty} \bignorm{\Kn^{-1} \sum_{k=1}^{\Kn} \bDelta_{X,k} - \tilde \bDelta_X} = 0, & &
        \lim_{n \to \infty} \bignorm{\Tn^{-1} \sum_{t=1}^{\Tn} \bDelta_{Y,t} - \tilde \bDelta_Y} = 0.&
    \end{align}
\end{assumption}
\begin{assumption}\label{ass:clt-matrices}
    For any fixed latent position $\vec{x} \in \mathcal{X}^k$, $\vec{y} \in \mathcal{Y}^t$ there exist positive definite $d \times d$ matrices $\mat{V}_X (\vec{y})$ and $\mat{V}_Y (\vec{x})$ such that
    \begin{equation}
\begin{aligned}
    & \mat{V}_Y(\vec{x}) = \lim_{n \to \infty}
    \begin{cases}
      \mathbb{E} [ \Tn^{-1} \sum_{t=1}^{\Tn} \vec{x}^\intercal \boldsymbol{\nu}^t (1 - \vec{x}^\intercal \boldsymbol{\nu}^t) \cdot \boldsymbol{\nu}^t {\boldsymbol{\nu}^t}^\intercal] & \text{if}\ \rho_n = 1, \\
        \mathbb{E} [ \Tn^{-1} \sum_{t=1}^{\Tn} \vec{x}^\intercal \boldsymbol{\nu}^t \cdot \boldsymbol{\nu}^t {\boldsymbol{\nu}^t}^\intercal] & \text{if}\ \rho_n \to 0,
    \end{cases} \\ 
    & \mat{V}_X (\vec{y}) = \lim_{n \to \infty}
    \begin{cases}
       \mathbb{E} [ \Kn^{-1} \sum_{k=1}^{\Kn} \vec{y}^\intercal \boldsymbol{\xi}^k (1 - \vec{y}^\intercal \boldsymbol{\xi}^k) \cdot \boldsymbol{\xi}^k {\boldsymbol{\xi}^k}^\intercal] & \text{if}\ \rho_n = 1, \\
        \mathbb{E} [ \Kn^{-1} \sum_{k=1}^{\Kn} \vec{y}^\intercal \boldsymbol{\xi}^k \cdot \boldsymbol{\xi}^k {\boldsymbol{\xi}^k}^\intercal] & \text{if}\ \rho_n \to 0,
    \end{cases}      
\end{aligned}
\label{eq:clt_matrices}
\end{equation}
where $\boldsymbol\xi^k\sim F_{X,k}$ and $\boldsymbol\nu^t\sim F_{Y,t}$, and $F_{X,k}, F_{Y,t}$ are the marginal distributions on $\mathcal X^k$ and $\mathcal Y^t$. 
\end{assumption}

\section{Theoretical results}
\label{sec:theoretical_results}

In this section, we present key theoretical results about the DUASE estimator for the latent positions under the DMPRDPG model. These results have a natural correspondence with the theoretical properties of ASE for the RDPG \citep{athreya2018statistical} and UASE for the MRDPG \citep{jones2021multilayer}. In particular, we 
derive two key theoretical results to demonstrate the effectiveness of DUASE for the problem of latent position recovery. The first of these is a consistency result that proves that the $\twoinf{\cdot}$ norm \citep{cape19two} of the error of our estimates converges to $0$, and the second shows that for any node, conditional on either the right or left true latent position, the distribution of DUASE estimation error around the true value is asymptotically Gaussian. The $\twoinf{\cdot}$ norm is a particularly meaningful metric in the context of our model because it corresponds to the maximum Euclidean row norm, which in our case is the maximum error for the latent position estimate of any one node. Hence, $\twoinf{\cdot}$ consistency demonstrates that DUASE produces consistent estimates for the latent positions of each node individually. This result is formally stated in Theorem~\ref{result:TwotoInfNorm}.


%

\begin{restatable}[Two-to-infinity norm bound for DUASE]{theorem}{ttibound}\label{result:TwotoInfNorm}
Let $(\Amat, \Xmat, \Ymat) \sim \mathrm{DMPRDPG}(F_\rho)$ with $\Kn$ layers and $\Tn$ time points, defined as in Definition~\ref{def:DMPRDPG} and suppose that Assumptions~\ref{ass:bounded-latent-space}--\ref{ass:second-moment-convergence} are satisfied. Then, for each $k\in[\Kn]$ and $t\in[\Tn]$, there exist sequences of matrices $\WX$ and $\WY \in \GL(d)$ (dependent on $n$), where $\WX^{-1} = \WY^{\intercal}$, such that
\begin{equation}
\begin{aligned}
    \twoinf{\hXm^{k}\WX^{-1} - \Xmat^{k}} = \Op \left\{ \frac{\log^{1/2}(n)}{\rho_n^{1/2}n^{1/2} \Tn^{1/2}} \right\}, \\
    \twoinf{\hYm^{t}\WY^{-1} - \Ymat^{t}} = \Op \left\{ \frac{\log^{1/2}(n)}{\rho_n^{1/2}n^{1/2}\Kn^{1/2}} \right\}. 
\end{aligned}
\label{eq:tti_bound}
\end{equation}
\end{restatable}

The proof of Theorem~\ref{result:TwotoInfNorm} is given in Appendix~\ref{sec:proof_clt}. It is important to note that the true scale of $\Xmat$ and $\Ymat$ are fundamentally unidentifiable as any scaling of $\Xmat$ by a constant factor $c$ and a corresponding scaling of $\Ymat$ by $c^{-1}$ produces an identical matrix $\Pmat$. 
The scale of the estimates obtained via DUASE will be a function of both $\Kn$ and $\Tn$. In particular, the scaling of each $\hXm^k$ will be or order $\Tn^{1/4}\Kn^{-1/4}$ while the scaling of each $\hYm^t$ is of order $\Kn^{1/4}\Tn^{-1/4}$. As such, a corresponding rescaling can be done in practice to obtain estimates for the left and right embedding that are of approximately equal magnitude. 

The second main result proved in this work establishes a central limit theorem for the DUASE estimate of latent positions for any given node. In particular, Theorem~\ref{result:CLT} shows that, for $n\to\infty$, the DUASE estimate of the latent positions converges to their true value with Gaussian error. The theorem is formalised below.

\newcommand{\titprop}{DUASE central limit theorem}
\begin{restatable}[\titprop]{theorem}{duaseclt} \label{result:CLT}
Let $(\Amat, \Xmat, \Ymat) \sim \mathrm{DMPRDPG}(F_\rho)$ with $\Kn$ layers and $\Tn$ time points, defined as in Definition~\ref{def:DMPRDPG} and suppose that Assumptions~\ref{ass:bounded-latent-space}--\ref{ass:clt-matrices} are satisfied. Given 
latent positions $\vec{x} \in \mathcal{X}^k$ and $\vec{y} \in \mathcal{Y}^t$, then for all $\vec{z} \in \mathbb{R}^d$ and for any fixed $i\in[n]$, $k\in[\K_n]$ and $t\in[\T_n]$ there exist sequences of matrices $\WX$ and $\WY \in \GL(d)$ (dependent on $n$) such that, for $n\to\infty$:
\begin{equation}
\begin{aligned}
    & \mathbb{P} \left\{ n^{1/2}\Tn^{1/2} (\hat{\mathbf{X}}^k \WX^{-1} - \mathbf{X}^k )^\intercal_i \leq \vec{z} \mid \boldsymbol{\xi}^k_i = \vec{x} \right\} \to \Phi \left\{ \vec{z}, \tilde{\boldsymbol\Delta}_Y^{-1} \mat{V}_{Y}(\vec{x}) \tilde{\boldsymbol\Delta}_Y^{-1} \right\}, \\ 
    & \mathbb{P} \left\{ n^{1/2} \Kn^{1/2}(\hat{\mathbf{Y}}^t \WY^{-1} - \mathbf{Y}^t )^\intercal_i \leq \vec{z} \mid \boldsymbol{\nu}^t_i = \vec{y} \right\} \to \Phi \left\{ \vec{z}, \tilde{\boldsymbol\Delta}_{X}^{-1} \mat{V}_X (\vec{y}) \tilde{\boldsymbol\Delta}_X^{-1} \right\}, 
\end{aligned}
\label{eq:clt_duase}
\end{equation}
where $\Phi(\vec z, \bm\Sigma)$ is the CDF of a $d$-dimensional normal distribution centered at $\vec 0$ (the identically zero vector of dimension $d$), with covariance matrix $\bm\Sigma\in\mathbb R^{d\times d}$, evaluated at $\vec z\in\mathbb R^d$. 
The matrices $\mat V_X(\cdot)$ and $\mat V_Y(\cdot)$ are the same as Equation~\eqref{eq:clt_matrices}. 
\end{restatable}
The proof of Theorem~\ref{result:CLT} is discussed in Appendix~\ref{sec:proof_clt}.
The theorems presented in this section have significant implications for practical tasks around inference on dynamic multilayer networks, which will be discussed in Section~\ref{sec:subsequent_inference}.

\section{Subsequent inference tasks on DMPRDPG}
\label{sec:subsequent_inference}

The proposed dynamic multiplex random dot product graph and the results on the DUASE estimator have implications for several subsequent inference tasks. In this section, we focus on two cases: clustering under an extension of the stochastic blockmodel (\textit{cf.} Section~\ref{sec:sbm}), and 
network changepoint detection via the iso-mirror 
\citep[\textit{cf.} Section~\ref{sec:isomirror}]{athreya2023euclidean}.

\subsection{Clustering under the stochastic blockmodel}
\label{sec:sbm}

Community detection on graphs corresponds to the task of finding nodes exhibiting similar connectivity behaviour. A widely used model for this task is the popular stochastic blockmodel \citep[SBM;][]{holland1983stochastic}, called the stochastic co-blockmodel (co-SBM) in its directed version \citep{Rohe16}. In co-SBMs, the probability of connection between two nodes $i$ and $j\in[n]$ depends on group allocations $z_i\in[G_1]$ and $\upsilon_j\in[G_2]$, where $G_1,G_2$ are the number of groups. If $\mat B\in[0,1]^{G_1\times G_2}$ denotes a between and within group connection probability matrix, then the adjacency matrix under a stochastic co-blockmodel is $\Amat_{i,j}\sim\mathrm{Bernoulli}(\mat B_{z_i,\upsilon_j})$ for $i,j\in[n]$. 
To extend the co-SBM to a dynamic and multiplex setting, we assume that each node is characterised by two types of community memberships: (i) a \textit{global} community shared across layers, which can change over time, and (ii) \textit{layer-specific} communities which do not change over time. 

\begin{definition}[DMPSBM -- Dynamic multiplex stochastic blockmodel]
    Assume that, for a dynamic multiplex network with $\K$ layers and $\T$ time points, nodes in a graph are assigned to groups or communities, where integers $z^k_i\in[G_1],\upsilon^t_i\in[G_2],\ G_1,G_2\in\mathbb N,\ i\in[n]$, denote the group membership assigned to the $i$-th node for the $t$-th time point and $k$-th layer respectively.
    Furthermore, define matrices $\mat{B}^{k,t} \in [0,1]^{G_1 \times G_2}$ representing between-group connection probabilities for the $k$-th layer and $t$-th time point, and set $\mathcal{B} = \{\mat{B}^{k,t}\}_{k \in [K], t \in [T]}$, $\mathcal Z=\{z_i^k\}_{i\in[n],k\in[\K]}$ and $\mathcal U=\{\upsilon_i^t\}_{i\in[n],t\in[\T]}$. For a set of adjacency matrices $\{\Amat^{k,t}\}_{k \in [K], t \in [T]}$, we say that $\Amat \sim \mathrm{DMPSBM}(\mathcal{B}, \mathcal{Z}, \mathcal{U})$, where $\Amat$ is the doubly unfolded adjacency matrix, if
    \begin{equation}
        \Amat^{k,t}_{i,j} \sim \mathrm{Bernoulli}\left( \mat{B}^{k,t}_{z_i^k, \upsilon_j^t} \right)
    \end{equation}
    for each $i,j \in [n]$,  $i \neq j$, $k \in [K]$ and $t \in [T]$.
\end{definition}

   The DMPSBM can be viewed as a special case of the DMPRDPG by assuming group-specific latent positions $\bm\mu_g^k\in\mathbb R^d,\ g\in[G_1]$ and $\bm\lambda_q^t \in\mathbb R^d,\ q\in[G_2]$, such that $\bm\mu_g^{k\intercal}\bm\lambda_h^t\in[0,1]$ for all $k\in[\K],\ t\in[\T],\ g\in[G_1],\ h\in[G_2]$, and setting $\mat B^{k,t}_{g,h}=\bm\mu_g^{k\intercal}\bm\lambda_h^t$. Then, the DMPRDPG latent positions are $\Xmat_i^k=\bm\mu_{z_i^k}^k$ and $\Ymat_j^t=\bm\lambda_{\upsilon_j^t}^t$.
This corresponds to setting $F_{X,k}=\sum_{g=1}^{G_1} \pi^k_g\delta_{\bm\mu_g^k}$ and 
$F_{Y,t}=\sum_{g=1}^{G_2} \pi^t_g\delta_{\bm\lambda_g^t}$, where $\delta_\cdot$ is the Dirac delta measure, and $\bm\pi^k=(\pi^k_1,\dots,\pi^k_{G_1})$, $\bm\pi^t=(\pi^t_1,\dots,\pi^t_{G_2})$ are group allocation probabilities, such that $\pi^k_g,\pi^t_q\geq0$ for all $g\in[G_1],\ q\in[G_2]$, and $\sum_{g=1}^{G_1} \pi^k_g=1$, $\sum_{g=1}^{G_2} \pi^t_g=1$,\ $k\in[\K],\ t\in[\T]$. The values of $\bm\mu_g^k$ and $\bm\lambda_g^t$ can be recovered (up to invertible transformations) from $\mathcal B$ by applying DUASE to the 
following matrix: 
\begin{equation}
\mat B =
\begin{bmatrix}
    \mat B^{1,1} & \mat B^{1,2} & \cdots & \mat B^{1,\T} \\
    \mat B^{2,1} & \mat B^{2,2} & \cdots & \mat B^{2,\T} \\
    \vdots & \vdots & \ddots & \vdots \\
    \mat B^{\K,1} & \mat B^{\K,2} & \cdots & \mat B^{\K,\T}
\end{bmatrix}\in[0,1]^{G_1\K\times G_2\T},\label{eq:B}
\end{equation}
where the embedding dimension is $d = \rank (\mat B)$.

In stochastic blockmodels, the main inferential interest is usually to recover the latent community structure for clustering purposes \citep{holland1983stochastic}.
The DUASE CLT in Theorem~\ref{result:CLT} provides theoretical justifications to perform clustering in DMPSBMs via Gaussian mixture models on the left and right DUASE. In particular, given a fixed index $i\in\mathbb N$ and conditioning on the community allocations, Theorem~\ref{result:CLT} gives:
\begin{equation}
\begin{aligned}
    & \mathbb{P} \left\{ n^{1/2}\Tn^{1/2} \left( \hat{\Xmat}_i^k \WX^{-1} -  
    \bm\mu^k_g \right)^\intercal \leq \vec{q} \mid z_i^k = g  \right\} \to \Phi \left\{ \vec{q}, \bm\Sigma_{X,g} \right\}, && g\in[G_1],\ \vec{q}\in\mathbb R^d, \\ 
    & \mathbb{P} \left\{ n^{1/2} \Kn^{1/2}\left(\hat{\Ymat}^t_i \WY^{-1} - \bm\lambda_g^t \right)^\intercal \leq \vec{q} \mid \upsilon_i^t = g \right\} \to \Phi \left\{ \vec{q}, \bm\Sigma_{Y,g} \right\},\ &&  g\in[G_2],\ \vec{q}\in\mathbb R^d,
\end{aligned}
\label{eq:clt_sbm}
\end{equation}
where $\bm\Sigma_{X,g}$ and $\bm\Sigma_{Y,g}$ are group-specific covariance matrices obtained from \eqref{eq:clt_duase} and 
\eqref{eq:clt_matrices}. 
The result in \eqref{eq:clt_sbm} implies that nodes belonging to the same community under the DMPSBM have the same asymptotic Gaussian distribution, suggesting that Gaussian mixture modelling is an appropriate strategy for clustering on the left and right DUASE. The embedding dimension can be chosen via the scree-plot method of \cite{Zhu06} or techniques akin to \cite{SannaPassino20} and \cite{Yang20}.

As an illustration, we conduct a simulation study 
on a graph with $n=1000$ nodes, equally distributed among $G_1=G_2=4$ communities, and $\K=3,\ \T=3$. We also assume that, for all nodes $i\in[n]$, $z_i^k=\upsilon_i^t$ for all $k\in[\K],\ t\in[\T]$. The matrices 
of connection probabilities are similar to those utilised in \cite{gallagher2021spectral} for a multilayer graph: 

\newcommand{\scalejcgs}{0.85}
{
\scalebox{\scalejcgs}{%
\centering\noindent
\parbox{\textwidth}{
\centering
\begin{align}
& 
\mat B^{1,1} = \begin{bmatrix}
0.08 & 0.02 & 0.18 & 0.10\\
0.02 & 0.20 & 0.04 & 0.10\\
0.18 & 0.04 & 0.02 & 0.02\\
0.10 & 0.10 & 0.02 & 0.06\\
\end{bmatrix},
& & 
\mat B^{1,2} = \begin{bmatrix}
0.16 & 0.16 & 0.04 & 0.10\\
0.16 & 0.16 & 0.04 & 0.10\\
0.04 & 0.04 & 0.09 & 0.02\\
0.10 & 0.10 & 0.02 & 0.06\\
\end{bmatrix},
& & 
\mat B^{1,3} = \begin{bmatrix}
0.08 & 0.02 & 0.18 & 0.10\\
0.02 & 0.20 & 0.04 & 0.10\\
0.18 & 0.04 & 0.02 & 0.02\\
0.10 & 0.10 & 0.02 & 0.06\\
\end{bmatrix},
\\ & 
\mat B^{2,1} = \begin{bmatrix}
0.08 & 0.02 & 0.18 & 0.10\\
0.02 & 0.20 & 0.04 & 0.10\\
0.18 & 0.04 & 0.02 & 0.02\\
0.10 & 0.10 & 0.02 & 0.06\\
\end{bmatrix},
& &
\mat B^{2,2} = \begin{bmatrix}
0.16 & 0.16 & 0.04 & 0.10\\
0.16 & 0.16 & 0.04 & 0.10\\
0.04 & 0.04 & 0.09 & 0.02\\
0.10 & 0.10 & 0.02 & 0.06\\
\end{bmatrix},
& &
\mat B^{2,3} = \begin{bmatrix}
0.08 & 0.02 & 0.18 & 0.10\\
0.02 & 0.20 & 0.04 & 0.10\\
0.18 & 0.04 & 0.02 & 0.02\\
0.10 & 0.10 & 0.02 & 0.06\\
\end{bmatrix},
\\ &
\mat B^{3,1} = \begin{bmatrix}
0.08 & 0.08 & 0.08 & 0.08\\
0.08 & 0.08 & 0.08 & 0.08\\
0.08 & 0.08 & 0.08 & 0.08\\
0.08 & 0.08 & 0.08 & 0.08\\
\end{bmatrix},
& & 
\mat B^{3,2} = \begin{bmatrix}
0.08 & 0.08 & 0.08 & 0.08\\
0.08 & 0.08 & 0.08 & 0.08\\
0.08 & 0.08 & 0.08 & 0.08\\
0.08 & 0.08 & 0.08 & 0.08\\
\end{bmatrix},
& &
\mat B^{3,3} = \begin{bmatrix}
0.08 & 0.08 & 0.08 & 0.08\\
0.08 & 0.08 & 0.08 & 0.08\\
0.08 & 0.08 & 0.08 & 0.08\\
0.08 & 0.08 & 0.08 & 0.08\\
\end{bmatrix}.
\end{align}
}}}
\vspace*{-.75em}

\noindent
The collection of matrices $\mat{B}^{k,t}$ was designed with several distinctive features which are useful to
illustrate the theoretical properties of DUASE. In particular, the connection probabilities for each of the communities are identical for layers $k=1$ and $k=2$, corresponding to $\bm\mu^1_g=\bm\mu^2_g$ for all $g\in[G_1]$. 
Also, the same connection probability matrices were used at $t=1$ and $t=3$, by setting $\bm\lambda_g^1=\bm\lambda_g^3$ for all $g\in[G_2]$. Additionally, some of the groups exhibit identical behaviour at some time points or layers: this 
occurs for the first two communities 
at time $t=2$ ($\bm\lambda_1^2=\bm\lambda_2^2$) and for all communities in layer $k=3$ ($\bm\mu^3_1=\bm\mu^3_2=\bm\mu^3_3=\bm\mu^3_4$). 

\begin{figure}[!t]
\begin{subfigure}{\textwidth}
    \centering
    \caption{Left embedding}
    \includegraphics[width=.975\textwidth]{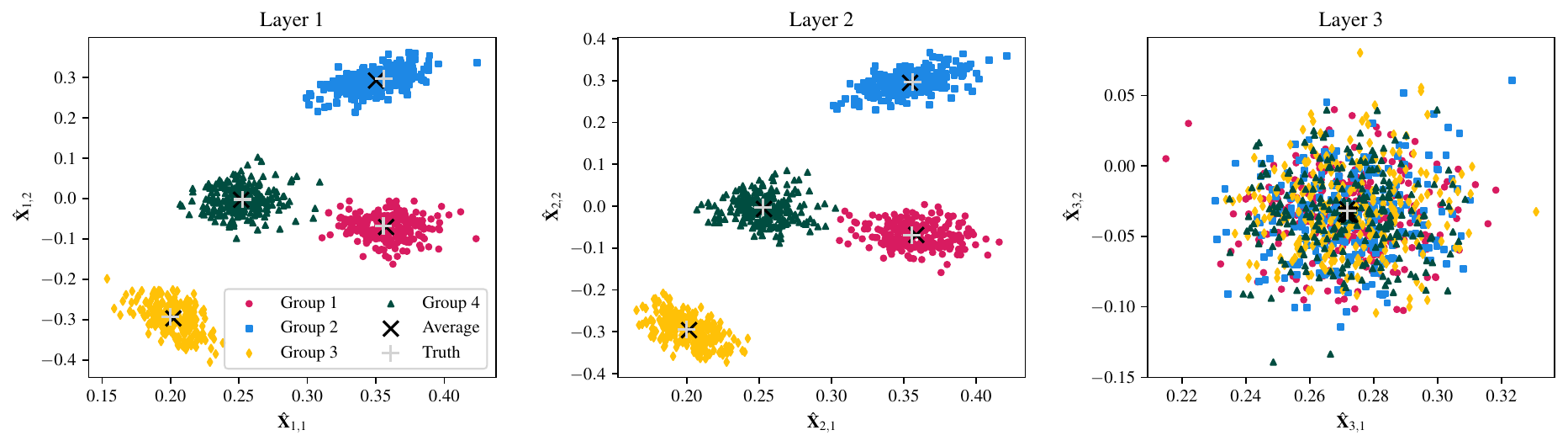}
    \label{fig:leftembeddings}
\end{subfigure}
\begin{subfigure}{\textwidth}
    \centering
    \caption{Right embedding}
    \includegraphics[width=.975\textwidth]{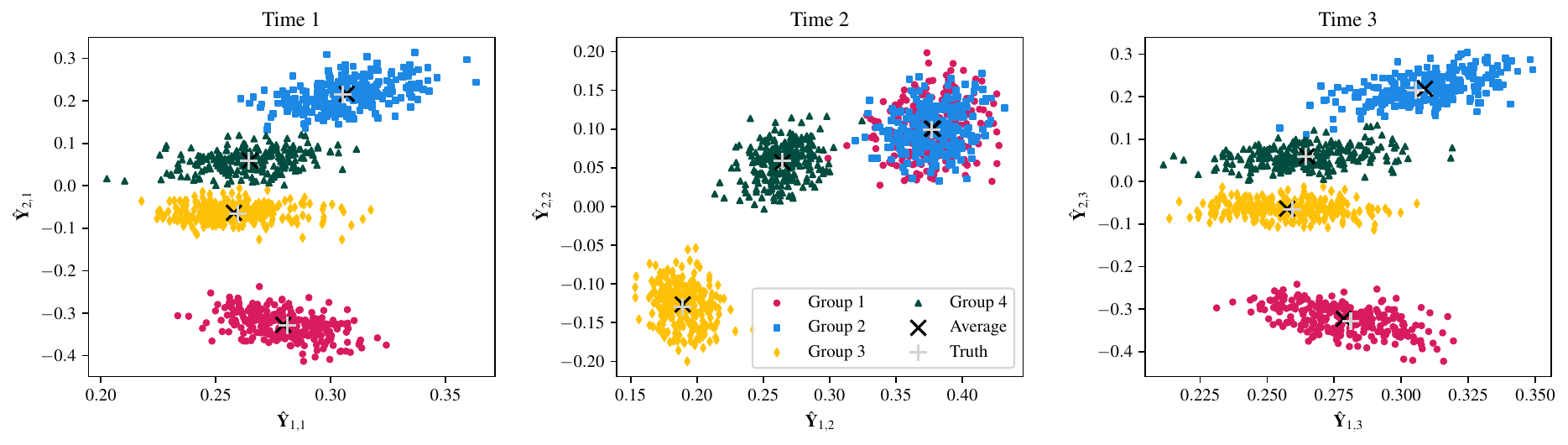}
    \label{fig:rightembeddings}
\end{subfigure}
\caption{Scatterplots of the first two dimensions of the left and right DUASE under a simulated DMPSBM with $n=1000$, $G=4$ groups of equal size, $\K=3$ and $\T=3$. Note that only the first two dimensions of the five-dimensional embedding are displayed.}
\label{fig:sbm}
\end{figure}

Figures \ref{fig:leftembeddings} and \ref{fig:rightembeddings} display scatterplots of the first two dimensions of the right and left embeddings obtained via DUASE applied on a realisation of the graph adjacency matrices under the DMPSBM, where the color of each point corresponds to community membership. Additionally, the average per-group and the true underlying latent position for the group are plotted. The theoretical latent positions were obtained by embedding the matrix $\mat B$ constructed as in \eqref{eq:B}, and performing orthogonal Procrustes alignment with the group means. 
Within each layer and time step, we observe clearly defined Gaussian clusters for each community centered at each of the true latent positions, as expected from Theorem~\ref{result:CLT}. Additionally, as expected, two communities overlap at time $t=2$, and all four communities overlap in layer $k=3$. The left DUASE embedding for layers $k=1$ and $k=2$ are comparable, and similarly for the right DUASE embeddings for time points $t=1$ and $t=3$. 
Overall, this simulation shows that DUASE has two desirable properties inherited from UASE: \textit{cross-sectional stability} and \textit{longitudinal stability} \citep{gallagher2021spectral}, occurring simultaneously on the different layers and time points. 

\subsection{Global change detection via the iso-mirror}
\label{sec:isomirror}

Consider the right DUASE embedding $\hYm=[\hYm^1\mid\cdots\mid\hYm^{\T}]\in\mathbb R^{nT \times d}$ and the associated sequence of time-specific embeddings $\hYm^1,\dots,\hYm^{\intercal}\in\mathbb R^{n\times d}$. 
 \cite{athreya2023euclidean} propose to calculate a $\T\times\T$ distance matrix $\hat{\mathcal D}^\varphi$, with entries
\begin{equation}
\hat{\mathcal D}^\varphi_{t,s} = \hat d_\mathrm{MV} (\hat{\mat Y}^t, \hat{\mat Y}^s) 
= \min_{\mat Q\in\mathbb O(d)} \frac{1}{\sqrt{n}} \norm{\hat{\mat Y}^t - \hat{\mat Y}^s\mat Q},\ t,s\in[\T],
\label{eq:isomirror}
\end{equation}
where $\mathbb O(d)$ is the orthogonal group with signature $d$. 
This dissimilarity metric is defined in \cite{athreya2023euclidean} within the context of a \textit{latent position process} (LPP) for a time series of a random dot product graphs, given by a map $\varphi$ that assigns time points $t$ to $d$-dimensional random variables $\varphi(t)=\zeta_t$ such that $\mathbb E[\zeta_t\zeta_t^\intercal]$ is finite and has full rank. 
If discrete samples from the latent position process $\varphi$ are taken for each node to construct the latent position matrices $\Ymat^t,\ t\in[T]$ for each component in the time series of RDPGs, and the latent positions are estimated via standard ASE, then the quantity $\hat d_\mathrm{MV} (\hat{\mat Y}^t, \hat{\mat Y}^s)$ in \eqref{eq:isomirror} consistently estimates the maximum directional metric 
\begin{equation}
    d_\mathrm{MV} \{\varphi(t), \varphi(s)\} = d_\mathrm{MV} (\zeta_t, \zeta_s) = \min_{\mat Q\in\mathbb O(d)} \norm{\mathbb E[(\zeta_t - \mat Q\zeta_s)(\zeta_t - \mat Q\zeta_s)^\intercal]}^{1/2}. 
\end{equation}
The components of the right DUASE are directly comparable (as shown, for example, in the simulation in Figure~\ref{fig:sbm}), implying that the Procrustes transformation in \eqref{eq:isomirror} is not necessary, and we can simply set $\hat{\mathcal D}^\varphi_{t,s} = n^{-1/2}\norm{\hYm^t - \hYm^s},\ t,s\in[\T]$. 
\cite{athreya2023euclidean}
apply classic multidimensional scaling \citep[CMDS; see, for example,][]{Borg05} on $\hat{\mathcal D}^\varphi$ to provide a consistent estimate $\hat\psi$ of a lower-dimensional Lipschitz continuous curve $\psi:[0,T]\to\mathbb R^c,\ c<d$, called a \textit{mirror}, which represents an Euclidean realisation of the manifold $\varphi([0,T])$. Additionally, \cite{athreya2023euclidean} also apply ISOMAP \citep{Tenenbaum00} to the points in $\mathrm{CMDS}(\hat{\mathcal D}^\varphi)=\{\hat\psi(t)\in \mathbb R^c,\ t=1,\dots, T\}$ to obtain a 1-dimensional curve, which can be plotted against the time indices $t=1,\dots,T$, called an \textit{iso-mirror}. When this procedure is applied to the sequence of time-specific embeddings $\hYm^1,\dots,\hYm^{\intercal}\in\mathbb R^{n\times d}$ obtained from DUASE, this yields a joint Euclidean mirror $\hat\psi(t),\ t\in[T]$ for dynamic multiplex networks, combined across layers. 
This can be used to identify global changepoints within the graph \citep[see, for example,][]{Chen23,Chen24}, affecting all nodes and layers simultaneously. 
Similarly, the same procedure could be used on the unstacked left DUASE embeddings $\hXm^1,\dots,\hXm^{\K}\in\mathbb R^{n\times d}$ to obtain a time-averaged Euclidean mirror $\hat{\psi}(k)$ based on CMDS applied to the $\K\times\K$ matrix with entries $\hat{\mathcal D}^{\varphi}_{k,h}=n^{-1/2}\norm{\hXm^{k} - \hXm^{h}},\ k,h\in[\K]$. This could be used to identify differences between layers. It should be remarked that we primarily utilize the iso-mirror method applied on the left DUASE only to identify which layers behave similarly or differently. For changepoint detection in the layers, it should be further assumed that the layers have a natural ordering. An example of a dynamic multiplex graph with ordered layers is the POLECAT network in Section~\ref{sec:applications}. 

\begin{figure}[t]
\centering
\begin{subfigure}{0.495\textwidth}
    \centering
    \caption{Iso-mirror for $\T=3$ time indices}
    \includegraphics[width=0.975\textwidth]{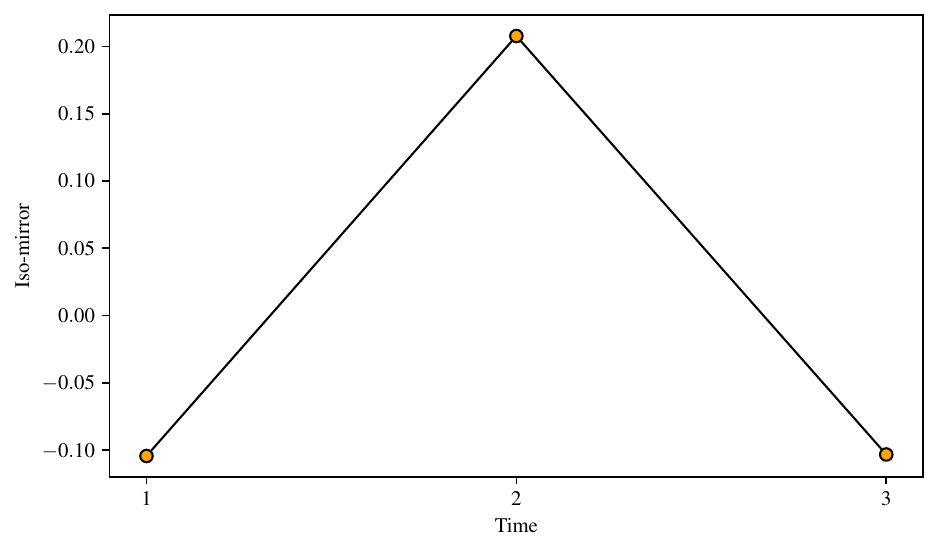}
    \label{fig:mirror_sbm_time}
\end{subfigure}
\begin{subfigure}{0.495\textwidth}
    \centering
    \caption{Iso-mirror for $\K=3$ layers}
    \includegraphics[width=0.975\textwidth]{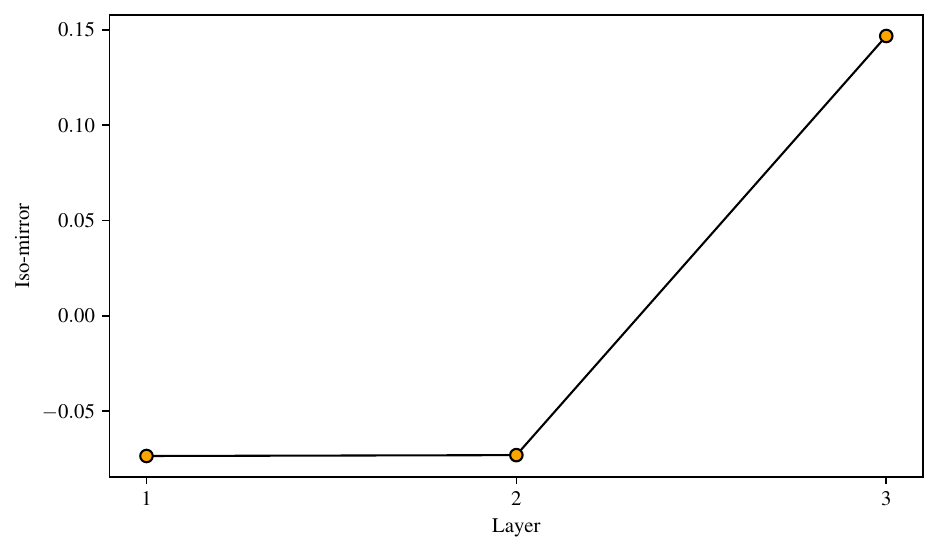}
    \label{fig:mirror_sbm_layer}
\end{subfigure}
\caption{Iso-mirrors calculated from 
DUASE 
on the 
DMPSBM 
in Figure~\ref{fig:sbm}.}
\label{fig:isomirror_sbm}
\end{figure}

In Figure~\ref{fig:isomirror_sbm}, we display the result of applying the iso-mirror method directly on the left and right DUASE embedding calculated from the DMPSBM simulated in Section~\ref{sec:sbm}. Since the same connection probability matrices were used for simulating the graph at times $t = 1$ and $t = 3$, the iso-mirror values at these indices are close (\textit{cf}. Figure~\ref{fig:mirror_sbm_time}). On the other hand $t=2$ exhibits a clear difference, since a different connection probability matrix is used. Similarly, Figure~\ref{fig:mirror_sbm_layer} shows that layers $k=1$ and $k=2$, identical by design in their connectivity matrices, also have similar iso-mirror values. On the other hand, the iso-mirror suggests that the third layer is characterised by a different connectivity structure from the first two, as designed in the simulation.

\section{Application on real-world networks}
\label{sec:applications}

To demonstrate practical uses of the DUASE embedding method for recovering latent position estimates, we analyse two real-world knowledge graphs: the POLECAT dataset of geopolitical interactions \citep{Halterman23}, and FinDKG \citep{Li24}, a dynamic knowledge graph extracted from financial news. 

\subsection{International relations}

\begin{figure}[!t]
\centering
\begin{subfigure}[t]{0.495\textwidth}
\caption{Iso-mirror on right DUASE $\hat\Ymat$}
\includegraphics[width=0.975\textwidth]{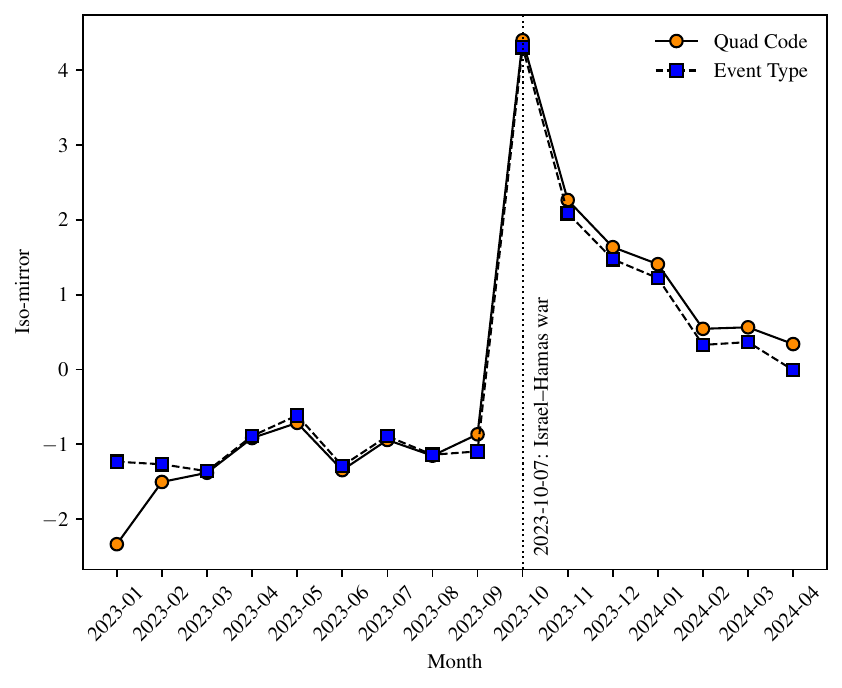}
\label{fig:im1}
\end{subfigure}
\hfill
\begin{subfigure}[t]{0.495\textwidth}
\centering
\caption{Iso-mirror on 
$\hat\Xmat$, grouped by event type}
\includegraphics[width=0.975\textwidth]{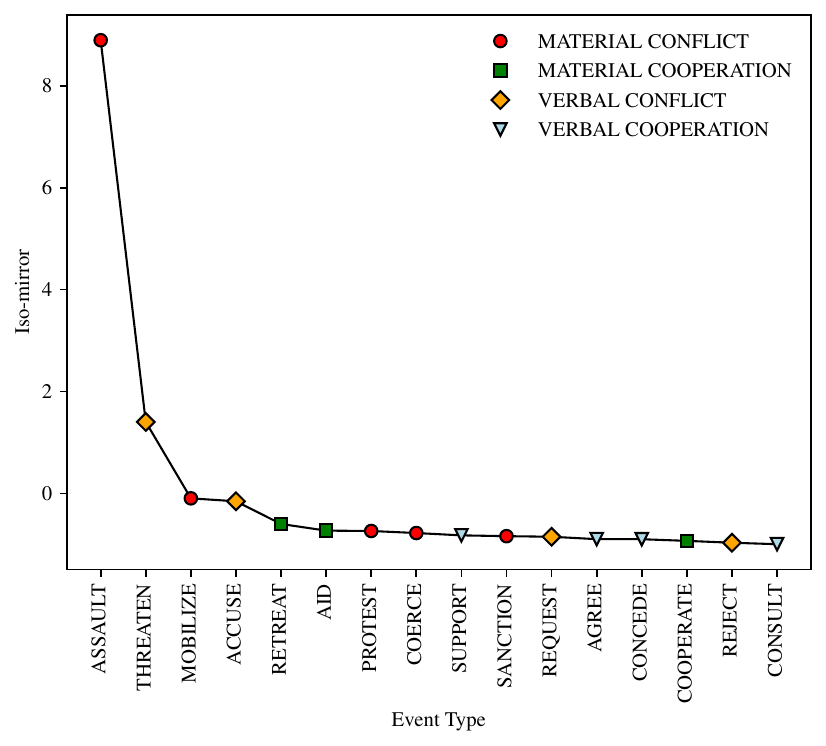}
\label{fig:im2}
\end{subfigure}
\\
\begin{subfigure}[t]{0.495\textwidth}
\centering
\caption{Iso-mirror on 
$\hat\Xmat$, grouped by quad code}
\includegraphics[width=0.975\textwidth]{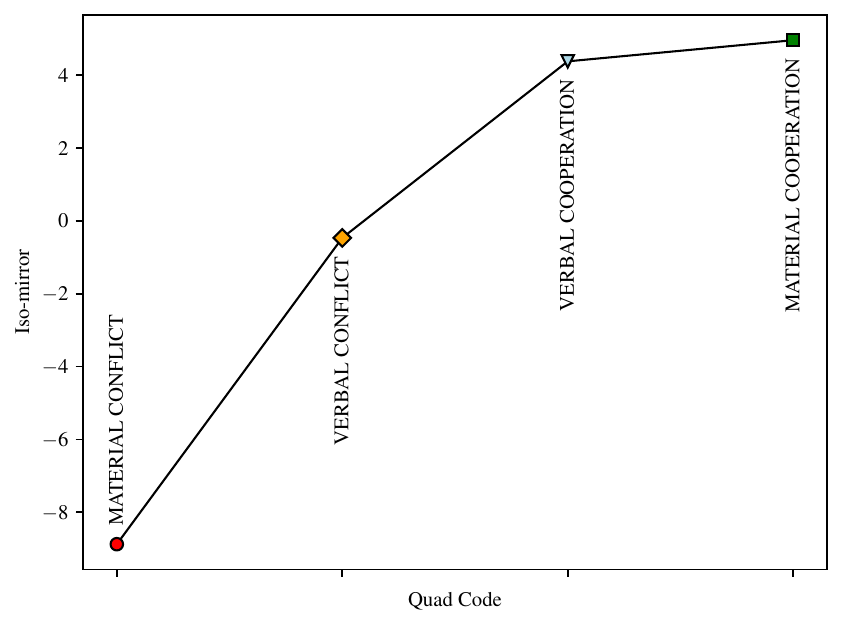}
\label{fig:im3}
\end{subfigure}
\hfill
\begin{subfigure}[t]{0.495\textwidth}
\centering
\caption{Iso-mirror on 
$\hat\Xmat$ vs. PLOVER intensity}
\includegraphics[width=0.975\textwidth]{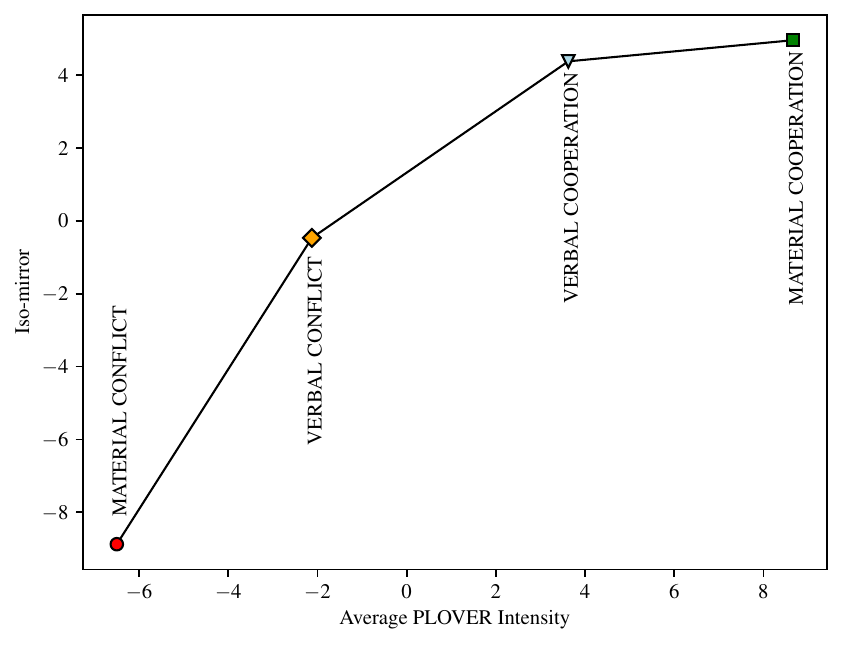}
\label{fig:im4}
\end{subfigure}
\caption{Iso-mirror across time and event types on the POLECAT data.}
\label{fig:isomap_polecat}
\end{figure}

 First, we study the POLECAT data of global political events \citep{Halterman23}, an earlier version of which, called ICEWS, has previously been studied in the dynamic multilayer graphs literature \citep[for example,][]{Loyal23}. 
 The dataset, after preprocessing, contains $\numprint{624888}$ political interaction events between $n=104$ countries. We group the events across $T=16$ months ranging between January 2023 and April 2024. Also, each event is associated with one of $\K=16$ event types based on the Political Language Ontology for Verifiable Event Records (PLOVER) categories. Each of the event types is further grouped into $K^\ast=4$ macro-groups called \textit{quad categories}: material cooperation, verbal cooperation, verbal conflict, and material conflict. The event types and corresponding quad macro-categories are summarised in Table~\ref{tab:plover}. We construct $\K\times T$ adjacency matrices for each month for each PLOVER event category, and we repeat the analysis considering the quad codes only, resulting in $K^\ast\times T$ adjacency matrices. To identify global structural changes in the graph, we run the iso-mirror procedure on the left and right DUASE as described in Section~\ref{sec:isomirror}. In this example, layers have a natural ordering, implied by a PLOVER intensity score for each category. The embedding dimension was selected using the scree-plot criterion of \cite{Zhu06}, resulting in $d=4$ for both graphs constructed via the event types or quad codes. The value of $c=2$ is chosen for CMDS via the scree-plot method, and the nearest neighbour graph for ISOMAP is constructed choosing the minimum threshold giving a connected graph. The results are plotted in Figure~\ref{fig:isomap_polecat}.

\begin{table}[t]
\centering
\scalebox{0.9}{
\begin{tabular}{ll}
\toprule
\textbf{Quad macro-category} & \textbf{PLOVER event categories} \\ \midrule
Verbal cooperation & AGREE, CONCEDE, CONSULT, SUPPORT \\ \hline
Material cooperation & AID, COOPERATE, RETREAT \\ \hline
Verbal conflict & ACCUSE, REJECT, REQUEST, THREATEN \\ \hline
Material conflict & ASSAULT, COERCE, MOBILIZE, PROTEST, SANCTION \\ \bottomrule
\end{tabular}
}
\caption{Quad categories and corresponding PLOVER event types.}
\label{tab:plover}
\end{table}

From Figure~\ref{fig:im1}, we identify a clear changepoint in the iso-mirror coinciding with the Israel-Hamas conflict, started on 7th October 2023 with the attacks led by Hamas on the Gaza envelope in southern Israel. In the same month, Isreal launched a bombing campaign targeting Hamas, and invaded the Gaza Strip on 27th October 2023. As expected, such a major shift in the geopolitical landscape leads to a clear distortion in the iso-mirror curves obtained via DUASE, both using the event types or the quad codes as layers.

Since DUASE yields a layer-specific embedding in addition to the time-specific embedding, it is also possible to obtain an iso-mirror representation for the event categories. Figures~\ref{fig:im2} and~\ref{fig:im3} display the iso-mirror scores for each layer, ranked in decreasing order. 
From Figure~\ref{fig:im2}, it also appears that ASSAULT and THREATEN event types result in largely different connectivity compared to other event types.
It appears that the event categories are naturally ordered, with event types related to conflict being at opposite ends of the ranking compared to cooperation categories. This is particularly evident in Figure~\ref{fig:im3}, where a clear transition from VERBAL COOPERATION to MATERIAL CONFLICT is observed across the iso-mirror scores. 
To further confirm this, Figure~\ref{fig:im4} displays a scatterplot between the average PLOVER intensity, calculated from the POLECAT Data Dictionary v5.8, and the iso-mirror scores, confirming a similar structure to Figure~\ref{fig:im3}. 

\subsection{Financial news}

\begin{figure}[t]
\centering
\begin{subfigure}[t]{0.495\textwidth}
\caption{Iso-mirror for right DUASE $\hat \Ymat$}
\includegraphics[width=0.975\textwidth]{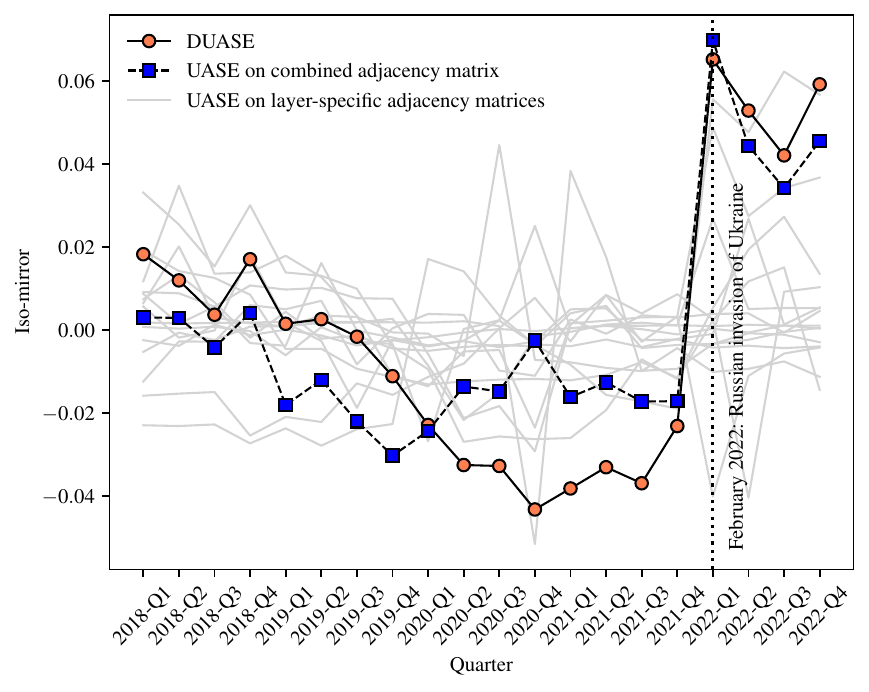}
\label{fig:imf1}
\end{subfigure}
\hfill
\begin{subfigure}[t]{0.495\textwidth}
\centering
\caption{Iso-mirror for left DUASE $\hat \Xmat$}\includegraphics[width=0.975\textwidth]{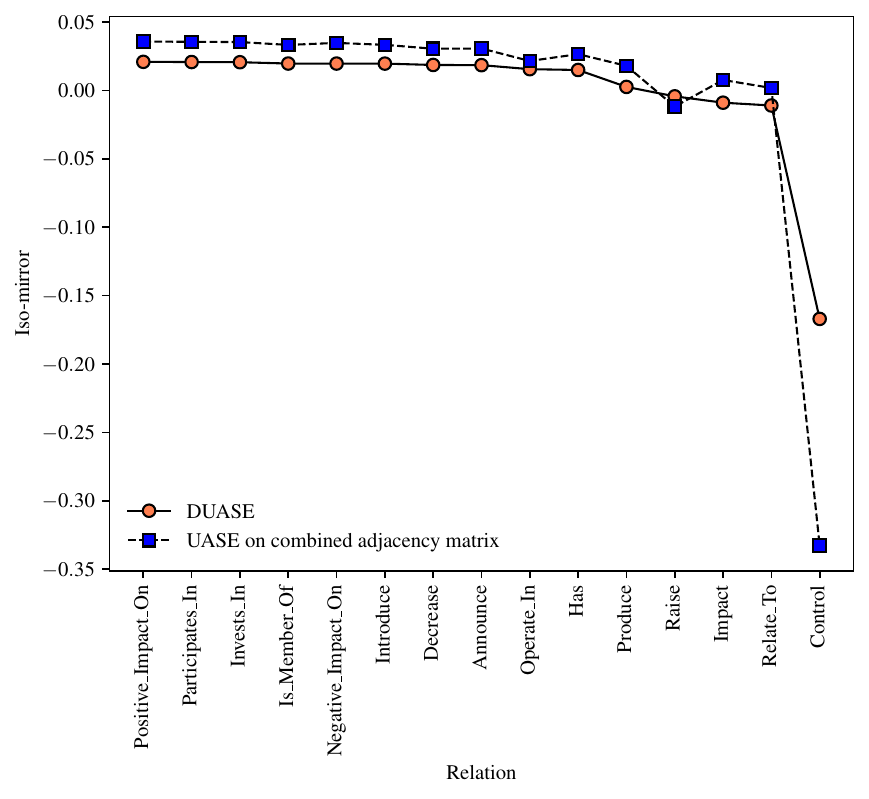}
\label{fig:imf2}
\end{subfigure}
\caption{Iso-mirror across time and event types on the FinDKG data.}
\label{fig:isomap_findkg}
\end{figure}

In a second example, we apply the DUASE algorithm to the FinDKG dataset \cite{Li24}, which contains events obtained from financial news articles. 
The graph has a total of $\numprint{241948}$ edges between $n=\numprint{13637}$ nodes, with $K=15$ different connection types related to financial concepts, such as ``Raise'', ``Invests\_In'' or ``Produce''. Nodes represent financial institutions, politicians, businessmen, countries, financial concepts, and commodities. We grouped the observations into $T=20$ quarters, ranging from January 2018 until December 2022. 
Similarly to the previous part, the embedding dimension $d=9$ is chosen via the scree-plot criterion \citep{Zhu06}, with $c=2$ for CMDS, and a choice of the minimum number of nearest neighbours for ISOMAP returning a connected graph. Additionally, we also compare DUASE with UASE \citep{jones2021multilayer} on individual layers of the graph, and with UASE on the averaged adjacency matrix across layers over time, and across time over layers. It must be remarked that DUASE returns a left and right embedding suitable for running the iso-mirror routine \textit{simultaneously}, which is computationally convenient compared to the two separate UASE procedures needed to obtain an iso-mirror across time, and an iso-mirror across layers. 
This is a relevant advantage of DUASE over alternative embedding methods. 
The results are 
in Figure~\ref{fig:isomap_findkg}.

In Figure~\ref{fig:imf1}, the ISOMAP curve for the right DUASE embedding displays a distinct inflection point corresponding to the Russian invasion of Ukraine, which occurred in February 2022 (Q1-2022). 
A similar changepoint is detected via the iso-mirror applied to the right embedding calculated via UASE on an averaged adjacency matrix that does not consider the layers. 
The changepoint is not present in most of the iso-mirrors calculated via UASE applied to layer-specific adjacency matrices, suggesting that the sharp change to connectivity is only visible when information from all layers is considered simultaneously.

Figure~\ref{fig:imf2} instead displays the iso-mirrors calculated from the left DUASE embedding and from UASE applied on layer-specific adjacency matrices, combined over time. In this case, layers do not have a natural ordering, so we primarily use the iso-mirror methods to idenfity groups of layers behaving similarly. We observe that the iso-mirror scores for most relation types are similar, with the notable exception of the ``Control'' relation. This may be indicative of the fact that the other relations are generally related to economic cooperation while ``Control'' is generally adversarial. It is also possible that this is an artifact of the Russia-Ukraine war as control of military targets as well as oil and gas supply lines have become 
 a central theme in news reporting. 

\section{Conclusion and discussion}
\label{sec:conclusion}

In this work we have introduced the dynamic multiplex random dot product graph (DMRDPG), a model for dynamic graphs with multiple connection types. Additionally, we proposed doubly unfolded adjacency spectral embedding (DUASE), a computationally cheap spectral embedding method, which is able to estimate the parameters of a DMPRDPG from a set of observed adjacency matrices. In Theorems~\ref{result:TwotoInfNorm} and~\ref{result:CLT}, we have shown that the DUASE estimates are both consistent as well as asymptotically normally distributed. In the context of a dynamic multilayer SBM, which is a special case of the DMPRDPG, this provides a theoretical justification for the use of Gaussian mixture modelling for clustering tasks in the embedding space. DUASE also ties in with recent developments in the literature. In particular, we highlight the use of the iso-mirror algorithm \citep{athreya2023euclidean} in conjunction with DUASE as a way to efficiently capture the time-specific or layer-specific trends and inflection points of a network. In two applications of the iso-mirror algorithm with DUASE on real-world networks, we find that this method detects change points over time related to major geopolitical shifts, and also provides insights into time-invariant relationships between layers.

\section*{Code}

Data and code to implement the methods proposed in this work, and reproduce the experiments and real data examples, are available in the Github repository \href{https://github.com/mjbaum/dmprdpg}{\texttt{mjbaum/dmprdpg}}. 

\section*{Acknowledgements}

Maximilian Baum acknowledges funding from the Department of Mathematics at Imperial College London. Francesco Sanna Passino acknowledges funding from the EPSRC, grant number
EP/Y002113/1. The authors thank Dr Anna Calissano, Professor Nick Heard, and Dr Alexander Modell for helpful discussions. 

\bibliographystyle{rss}
\singlespacing
\bibliography{references}

\begin{thebibliography}{6}
\expandafter\ifx\csname natexlab\endcsname\relax\def\natexlab#1{#1}\fi
\expandafter\ifx\csname url\endcsname\relax
  \def\url#1{\texttt{#1}}\fi
\expandafter\ifx\csname urlprefix\endcsname\relax\def\urlprefix{URL: }\fi

\bibitem[{Chen et~al.(2021)Chen, Chi, Fan and Ma}]{SpectralMethodsforDS}
Chen, Y., Chi, Y., Fan, J., and Ma, C. (2021) Spectral methods for data
  science: A statistical perspective.
\newblock \textit{Foundations and Trends{\textregistered} in Machine Learning},
  \textbf{14}, 566–806.

\bibitem[{Horn and Johnson(2012)}]{horn2012matrix}
Horn, R.~A. and Johnson, C.~R. (2012) \textit{Matrix analysis}.
\newblock Cambridge University Press.

\bibitem[{Jones and Rubin-Delanchy(2021)}]{jones2021multilayer}
Jones, A. and Rubin-Delanchy, P. (2021) The multilayer random dot product
  graph.
\newblock \textit{arXiv e-print arXiv:2007.10455}.

\bibitem[{Sch{\"o}nemann(1966)}]{schonemann1966generalized}
Sch{\"o}nemann, P.~H. (1966) A generalized solution of the orthogonal
  {Procrustes} problem.
\newblock \textit{Psychometrika}, \textbf{31}.

\bibitem[{Tropp et~al.(2015)}]{tropp2015introduction}
Tropp, J.~A. et~al. (2015) An introduction to matrix concentration
  inequalities.
\newblock \textit{Foundations and Trends{\textregistered} in Machine Learning},
  \textbf{8}, 1--230.

\bibitem[{Yu et~al.(2014)Yu, Wang and Samworth}]{Yu14}
Yu, Y., Wang, T., and Samworth, R.~J. (2014) {A useful variant of the
  Davis--Kahan theorem for statisticians}.
\newblock \textit{Biometrika}, \textbf{102}, 315--323.

\end{thebibliography}


\begin{thebibliography}{48}
\expandafter\ifx\csname natexlab\endcsname\relax\def\natexlab#1{#1}\fi
\expandafter\ifx\csname url\endcsname\relax
  \def\url#1{\texttt{#1}}\fi
\expandafter\ifx\csname urlprefix\endcsname\relax\def\urlprefix{URL: }\fi

\bibitem[{Arroyo et~al.(2021)Arroyo, Athreya, Cape, Chen, Priebe and
  Vogelstein}]{Arroyo21}
Arroyo, J., Athreya, A., Cape, J., et~al. (2021) Inference for multiple
  heterogeneous networks with a common invariant subspace.
\newblock \textit{Journal of Machine Learning Research}, \textbf{22}, 1--49.

\bibitem[{Athreya et~al.(2018)Athreya, Fishkind, Tang, Priebe, Park,
  Vogelstein, Levin, Lyzinski, Qin and Sussman}]{athreya2018statistical}
Athreya, A., Fishkind, D.~E., Tang, M., et~al. (2018) Statistical inference on
  random dot product graphs: a survey.
\newblock \textit{Journal of Machine Learning Research}, \textbf{18}, 1--92.

\bibitem[{Athreya et~al.(2025)Athreya, Lubberts, Park and
  Priebe}]{athreya2023euclidean}
Athreya, A., Lubberts, Z., Park, Y., and Priebe, C.~E. (2025) Euclidean mirrors
  and dynamics in network time series.
\newblock \textit{Journal of the American Statistical Association},
  \textbf{120}, 1025–1036.

\bibitem[{Athreya et~al.(2016)Athreya, Priebe, Tang, Lyzinski, Marchette and
  Sussman}]{athreya2016limit}
Athreya, A., Priebe, C.~E., Tang, M., et~al. (2016) A limit theorem for scaled
  eigenvectors of random dot product graphs.
\newblock \textit{Sankhya A}, \textbf{78}, 1--18.

\bibitem[{{Baltodano L{\'o}pez} and {Casarin}(2022)}]{Lopez22}
{Baltodano L{\'o}pez}, O. and {Casarin}, R. (2022) {A Dynamic Stochastic Block
  Model for Multi-Layer Networks}.
\newblock \textit{arXiv e-prints}, arXiv:2209.09354.

\bibitem[{Billio et~al.(2024)Billio, Casarin and Iacopini}]{Billio24}
Billio, M., Casarin, R., and Iacopini, M. (2024) {Bayesian Markov-Switching
  Tensor Regression for Time-Varying Networks}.
\newblock \textit{Journal of the American Statistical Association},
  \textbf{119}, 109--121.

\bibitem[{Bollob\'{a}s and Riordan(2009)}]{Bollobas09}
Bollob\'{a}s, B. and Riordan, O. (2009) Metrics for sparse graphs.
\newblock \textit{Surveys in Combinatorics, LMS Lecture Notes Series 365, CUP
  2009}, 211--287.

\bibitem[{Borg and Groenen(2005)}]{Borg05}
Borg, I. and Groenen, P. (2005) \textit{Modern Multidimensional Scaling: Theory
  and Applications}.
\newblock Springer Series in Statistics. Springer New York.

\bibitem[{Cape et~al.(2019)Cape, Tang and Priebe}]{cape19two}
Cape, J., Tang, M., and Priebe, C.~E. (2019) {The two-to-infinity norm and
  singular subspace geometry with applications to high-dimensional statistics}.
\newblock \textit{The Annals of Statistics}, \textbf{47}, 2405 -- 2439.

\bibitem[{{Chen} et~al.(2024){Chen}, {Lubberts}, {Athreya}, {Park} and
  {Priebe}}]{Chen24}
{Chen}, T., {Lubberts}, Z., {Athreya}, A., {Park}, Y., and {Priebe}, C.~E.
  (2024) {Euclidean mirrors and first-order changepoints in network time
  series}.
\newblock \textit{arXiv e-prints}, arXiv:2405.11111.

\bibitem[{{Chen} et~al.(2023){Chen}, {Park}, {Saad-Eldin}, {Lubberts},
  {Athreya}, {Pedigo}, {Vogelstein}, {Puppo}, {Silva}, {Muotri}, {Yang},
  {White} and {Priebe}}]{Chen23}
{Chen}, T., {Park}, Y., {Saad-Eldin}, A., et~al. (2023) {Discovering a change
  point and piecewise linear structure in a time series of organoid networks
  via the iso-mirror}.
\newblock \textit{arXiv e-prints}, arXiv:2303.04871.

\bibitem[{Chen et~al.(2021)Chen, Chi, Fan and Ma}]{SpectralMethodsforDS}
Chen, Y., Chi, Y., Fan, J., and Ma, C. (2021) Spectral methods for data
  science: A statistical perspective.
\newblock \textit{Foundations and Trends{\textregistered} in Machine Learning},
  \textbf{14}, 566–806.

\bibitem[{Corneck et~al.(2026)Corneck, Cohen and
  Sanna~Passino}]{corneck2026spectral}
Corneck, J., Cohen, E.~A., and Sanna~Passino, F. (2026) Spectral embedding of
  inhomogeneous {Poisson} processes on multiplex networks.
\newblock \textit{arXiv preprint arXiv:2601.16784}.

\bibitem[{De~Domenico et~al.(2013)De~Domenico, Sol\'e-Ribalta, Cozzo, Kivel\"a,
  Moreno, Porter, G\'omez and Arenas}]{DeDomenico13}
De~Domenico, M., Sol\'e-Ribalta, A., Cozzo, E., et~al. (2013) Mathematical
  formulation of multilayer networks.
\newblock \textit{Physical Review X}, \textbf{3}, 041022.

\bibitem[{Durante et~al.(2017)Durante, Mukherjee and Steorts}]{Durante17}
Durante, D., Mukherjee, N., and Steorts, R.~C. (2017) Bayesian learning of
  dynamic multilayer networks.
\newblock \textit{Journal of Machine Learning Research}, \textbf{18}, 1--29.

\bibitem[{Fensel et~al.(2020)Fensel, {\c{S}}im{\c{s}}ek, Angele, Huaman,
  K{\"a}rle, Panasiuk, Toma, Umbrich and Wahler}]{Fensel2020}
Fensel, D., {\c{S}}im{\c{s}}ek, U., Angele, K., et~al. (2020)
  \textit{Introduction: What Is a Knowledge Graph?}, 1--10.
\newblock Cham: Springer International Publishing.

\bibitem[{Gallagher et~al.(2021)Gallagher, Jones and
  Rubin-Delanchy}]{gallagher2021spectral}
Gallagher, I., Jones, A., and Rubin-Delanchy, P. (2021) Spectral embedding for
  dynamic networks with stability guarantees.
\newblock \textit{Advances in Neural Information Processing Systems},
  \textbf{34}, 10158--10170.

\bibitem[{Grover and Leskovec(2016)}]{grover2016node2vec}
Grover, A. and Leskovec, J. (2016) node2vec: Scalable feature learning for
  networks.
\newblock In \textit{Proceedings of the 22nd ACM SIGKDD International
  Conference on Knowledge Discovery and Data Mining}, 855--864.

\bibitem[{Halterman et~al.(2023)Halterman, Bagozzi, Beger, Schrodt and
  Scraborough}]{Halterman23}
Halterman, A., Bagozzi, B.~E., Beger, A., Schrodt, P., and Scraborough, G.
  (2023) {PLOVER} and {POLECAT}: A new political event ontology and dataset.
\newblock \textit{Tech. rep.}

\bibitem[{Han et~al.(2015)Han, Xu and Airoldi}]{Han15}
Han, Q., Xu, K.~S., and Airoldi, E.~M. (2015) Consistent estimation of dynamic
  and multi-layer block models.
\newblock In \textit{Proceedings of the 32nd International Conference on
  International Conference on Machine Learning}, vol.~37 of \textit{ICML'15},
  1511–1520.

\bibitem[{Hoff(2015)}]{Hoff15}
Hoff, P.~D. (2015) {Multilinear tensor regression for longitudinal relational
  data}.
\newblock \textit{The Annals of Applied Statistics}, \textbf{9}, 1169 -- 1193.

\bibitem[{Hoff et~al.(2002)Hoff, Raftery and Handcock}]{hoff2002latent}
Hoff, P.~D., Raftery, A.~E., and Handcock, M.~S. (2002) Latent space approaches
  to social network analysis.
\newblock \textit{Journal of the American Statistical Association},
  \textbf{97}, 1090--1098.

\bibitem[{Holland et~al.(1983)Holland, Laskey and
  Leinhardt}]{holland1983stochastic}
Holland, P.~W., Laskey, K.~B., and Leinhardt, S. (1983) Stochastic blockmodels:
  First steps.
\newblock \textit{Social networks}, \textbf{5}, 109--137.

\bibitem[{Horn and Johnson(2012)}]{horn2012matrix}
Horn, R.~A. and Johnson, C.~R. (2012) \textit{Matrix analysis}.
\newblock Cambridge University Press.

\bibitem[{Huang et~al.(2023)Huang, Weng and Feng}]{Huang23}
Huang, S., Weng, H., and Feng, Y. (2023) Spectral clustering via adaptive layer
  aggregation for multi-layer networks.
\newblock \textit{Journal of Computational and Graphical Statistics},
  \textbf{32}, 1170--1184.

\bibitem[{Jones and Rubin-Delanchy(2021)}]{jones2021multilayer}
Jones, A. and Rubin-Delanchy, P. (2021) The multilayer random dot product
  graph.
\newblock \textit{arXiv e-print arXiv:2007.10455}.

\bibitem[{Ke et~al.(2019)Ke, Shi and Xia}]{ke2019community}
Ke, Z.~T., Shi, F., and Xia, D. (2019) Community detection for hypergraph
  networks via regularized tensor power iteration.
\newblock \textit{arXiv e-print arXiv:1909.06503}.

\bibitem[{Lei and Lin(2023)}]{Lei23}
Lei, J. and Lin, K.~Z. (2023) Bias-adjusted spectral clustering in multi-layer
  stochastic block models.
\newblock \textit{Journal of the American Statistical Association},
  \textbf{118}, 2433--2445.

\bibitem[{Levin et~al.(2017)Levin, Athreya, Tang, Lyzinski, Park and
  Priebe}]{levin2017central}
Levin, K., Athreya, A., Tang, M., et~al. (2017) A central limit theorem for an
  omnibus embedding of multiple random graphs and implications for multiscale
  network inference.
\newblock \textit{arXiv e-prints, arXiv:1705.09355}.

\bibitem[{Li and Sanna~Passino(2024)}]{Li24}
Li, X.~V. and Sanna~Passino, F. (2024) {FinDKG}: Dynamic knowledge graphs with
  large language models for detecting global trends in financial markets.
\newblock In \textit{Proceedings of the 5th ACM International Conference on AI
  in Finance}, 573--581.

\bibitem[{Loyal and Chen(2023)}]{Loyal23}
Loyal, J.~D. and Chen, Y. (2023) An eigenmodel for dynamic multilayer networks.
\newblock \textit{Journal of Machine Learning Research}, \textbf{24}, 1--69.

\bibitem[{Lyzinski et~al.(2016)Lyzinski, Tang, Athreya, Park and
  Priebe}]{lyzinski2016community}
Lyzinski, V., Tang, M., Athreya, A., Park, Y., and Priebe, C.~E. (2016)
  Community detection and classification in hierarchical stochastic
  blockmodels.
\newblock \textit{IEEE Transactions on Network Science and Engineering},
  \textbf{4}, 13--26.

\bibitem[{Malik et~al.(2021)Malik, Ubaru, Horesh, Kilmer and
  Avron}]{malik2021dynamic}
Malik, O.~A., Ubaru, S., Horesh, L., Kilmer, M.~E., and Avron, H. (2021)
  {Dynamic graph convolutional networks using the tensor M-product}.
\newblock In \textit{Proceedings of the 2021 SIAM International Conference on
  Data Mining (SDM)}, 729--737. SIAM.

\bibitem[{Oselio et~al.(2014)Oselio, Kulesza and Hero}]{Oselio14}
Oselio, B., Kulesza, A., and Hero, A.~O. (2014) Multi-layer graph analysis for
  dynamic social networks.
\newblock \textit{IEEE Journal of Selected Topics in Signal Processing},
  \textbf{8}, 514--523.

\bibitem[{Rohe et~al.(2016)Rohe, Qin and Yu}]{Rohe16}
Rohe, K., Qin, T., and Yu, B. (2016) Co-clustering directed graphs to discover
  asymmetries and directional communities.
\newblock \textit{Proceedings of the National Academy of Sciences},
  \textbf{113}, 12679--12684.

\bibitem[{Rubin-Delanchy(2020)}]{RubinDelanchy20}
Rubin-Delanchy, P. (2020) Manifold structure in graph embeddings.
\newblock In \textit{Advances in Neural Information Processing Systems} (eds.
  H.~Larochelle, M.~Ranzato, R.~Hadsell, M.~Balcan and H.~Lin), vol.~33,
  11687--11699. Curran Associates, Inc.

\bibitem[{Rubin-Delanchy et~al.(2022)Rubin-Delanchy, Cape, Tang and
  Priebe}]{RubinDelanchy22}
Rubin-Delanchy, P., Cape, J., Tang, M., and Priebe, C.~E. (2022) {A Statistical
  Interpretation of Spectral Embedding: The Generalised Random Dot Product
  Graph}.
\newblock \textit{Journal of the Royal Statistical Society Series B:
  Statistical Methodology}, \textbf{84}, 1446--1473.

\bibitem[{Sanna~Passino and Heard(2020)}]{SannaPassino20}
Sanna~Passino, F. and Heard, N.~A. (2020) Bayesian estimation of the latent
  dimension and communities in stochastic blockmodels.
\newblock \textit{Statistics and Computing}, \textbf{30}, 1291--1307.

\bibitem[{Sewell and Chen(2015)}]{sewell2015latent}
Sewell, D.~K. and Chen, Y. (2015) Latent space models for dynamic networks.
\newblock \textit{Journal of the American Statistical Association},
  \textbf{110}, 1646--1657.

\bibitem[{Sosa and Betancourt(2022)}]{Sosa22}
Sosa, J. and Betancourt, B. (2022) A latent space model for multilayer network
  data.
\newblock \textit{Computational Statistics \& Data Analysis}, \textbf{169},
  107432.

\bibitem[{Sussman et~al.(2012)Sussman, Tang, Fishkind and
  Priebe}]{sussman2012consistent}
Sussman, D.~L., Tang, M., Fishkind, D.~E., and Priebe, C.~E. (2012) A
  consistent adjacency spectral embedding for stochastic blockmodel graphs.
\newblock \textit{Journal of the American Statistical Association},
  \textbf{107}, 1119--1128.

\bibitem[{Sussman et~al.(2013)Sussman, Tang and Priebe}]{sussman2013consistent}
Sussman, D.~L., Tang, M., and Priebe, C.~E. (2013) Consistent latent position
  estimation and vertex classification for random dot product graphs.
\newblock \textit{IEEE Transactions on Pattern Analysis and Machine
  Intelligence}, \textbf{36}, 48--57.

\bibitem[{Tao and Vu(2010)}]{tao2010random}
Tao, T. and Vu, V. (2010) Random matrices: Universality of local eigenvalue
  statistics up to the edge.
\newblock \textit{Communications in Mathematical Physics}, \textbf{298},
  549--572.

\bibitem[{Tenenbaum et~al.(2000)Tenenbaum, de~Silva and Langford}]{Tenenbaum00}
Tenenbaum, J.~B., de~Silva, V., and Langford, J.~C. (2000) A global geometric
  framework for nonlinear dimensionality reduction.
\newblock \textit{Science}, \textbf{290}, 2319--2323.

\bibitem[{Wang et~al.(2026)Wang, Li, Madrid~Padilla, Yu and Rinaldo}]{Wang23}
Wang, F., Li, W., Madrid~Padilla, O.~H., Yu, Y., and Rinaldo, A. (2026)
  Multilayer random dot product graphs: Estimation and online change point
  detection.
\newblock \textit{Journal of the Royal Statistical Society Series B:
  Statistical Methodology}, \textbf{88}, 282--312.

\bibitem[{Yang et~al.(2021)Yang, Priebe, Park and Marchette}]{Yang20}
Yang, C., Priebe, C.~E., Park, Y., and Marchette, D.~J. (2021) Simultaneous
  dimensionality and complexity model selection for spectral graph clustering.
\newblock \textit{Journal of Computational and Graphical Statistics},
  \textbf{30}, 422--441.

\bibitem[{Zhen and Wang(2023)}]{zhen2023community}
Zhen, Y. and Wang, J. (2023) Community detection in general hypergraph via
  graph embedding.
\newblock \textit{Journal of the American Statistical Association},
  \textbf{118}, 1620--1629.

\bibitem[{Zhu and Ghodsi(2006)}]{Zhu06}
Zhu, M. and Ghodsi, A. (2006) Automatic dimensionality selection from the scree
  plot via the use of profile likelihood.
\newblock \textit{Computational Statistics \& Data Analysis}, \textbf{51},
  918--930.

\end{thebibliography}


\begin{center}
{\LARGE\textbf{ 
SUPPLEMENTARY MATERIAL
}}
\end{center}

The supplementary material contains the proofs supporting the two main results in our work, Theorems~\ref{result:TwotoInfNorm} and~\ref{result:CLT}. The proofs are based on adaptations of the results in the literature on random dot product graphs \citep[see, for example,][]{athreya2018statistical} and multilayer RDPGs \citep{jones2021multilayer}. The results in Section \ref{sec:intermediate_proofs} are derived under Assumptions~\ref{ass:bounded-latent-space}--\ref{ass:second-moment-convergence}. 

\section{Intermediate results and proofs} \label{sec:intermediate_proofs}

\begin{proposition}\label{prop:XTXminDeltaRate}
Let $(\Amat, \Xmat, \Ymat) \sim \mathrm{DMPRDPG}(F_{\rho_n})$ with $\Kn$ layers and $\Tn$ time points, defined as in Definition~\ref{def:DMPRDPG}. Let the second moment matrices of the latent position distributions be $\bDelta_{X,k} = \mathbb{E} [\boldsymbol{\xi}^k\boldsymbol{\xi}^{k\intercal}], k\in[\Kn]$, and $\bDelta_{Y,t} = \mathbb{E} [\boldsymbol{\nu}^t\boldsymbol{\nu}^{t\intercal}],\ t\in[\Tn]$. Then: 
\begin{align}
\norm{{\Xmat}^{k\intercal} {\Xmat}^k - \rho_n n \bDelta_{X,k}} = \Op \{n^{1/2}\log^{1/2}(n)\}, & & 
\norm{{\Ymat}^{t\intercal} {\Ymat}^t -\rho_n n\bDelta_{Y,t}} = \Op \{n^{1/2}\log^{1/2}(n)\}.
\end{align}
\end{proposition}

\begin{proof}
    By Assumptions \ref{ass:bounded-latent-space} and \ref{ass:independence} each $\Xmat^{k \intercal} \Xmat^k,\ k\in[\Kn]$, and $\Ymat^{t \intercal} \Ymat^t,\ t\in[\Tn]$, is the sum of $n$ independent and identically distributed random variables bounded by a constant $c$ with expectation $\rho_n \bDelta_{X,k}$ and $\rho_n \bDelta_{Y,t}$ respectively. Therefore, we can apply Hoeffding's inequality, which gives
    \begin{equation}
        \mathbb{P} \left(\frob{\Xmat^{k \intercal} \Xmat^k - \rho_n n\bDelta_{X,k}} \geq \tau \right) \leq \exp \left(- \frac{2\tau^2}{n c^2} \right).
    \end{equation}
    Hence, for any $\alpha > 0$ we can set $\tau = C_\alpha n^{1/2} \log^{1/2}(n)$ for some $C_\alpha$, to show that
    \begin{equation}
    \mathbb{P} \left(\frob{\Xmat^{k \intercal} \Xmat^k - \rho_n n\bDelta_{X,k}} \geq C_\alpha n^{1/2} \log^{1/2}(n) \right) \leq \exp \left(- \frac{C_\alpha^2 \log(n)}{c^2} \right).
    \end{equation}
    By choosing $C_\alpha = \sqrt{c^2 \alpha}$, we find 
    \begin{equation}
        \mathbb{P} \left(\frob{\Xmat^{k \intercal} \Xmat^k - \rho_n n\bDelta_{X,k}} \geq C_\alpha n^{1/2} \log^{1/2}(n) \right) \leq n^{-\alpha}.
    \end{equation}
Therefore, by the definition of the $\Op$ notation, it follows that: 
\begin{equation}
    \frob{{\Xmat}^{k\intercal} {\Xmat}^k -\rho_n n\bDelta_{X,k}} = \Op \{n^{1/2}\log^{1/2}(n)\}.
\end{equation}
For any matrix $\mat{A}$, 
$\frob{\mat{A}} \geq \norm{\mat{A}}$, which gives the desired result. 
A similar argument holds for the statement $\norm{{\Ymat}^{t\intercal} {\Ymat}^t -\rho_n n\bDelta_{Y,t}} = \Op \{n^{1/2}\log^{1/2}(n)\}$.
\end{proof}



Next, we introduce Proposition~\ref{prop:Psingvalorder}  which provides a control on the singular values of $\Pmat$. The proof uses Proposition~\ref{prop:XTXminDeltaRate}. 

\begin{restatable}[Singular values of $\Pmat$]{proposition}{controlsvals}\label{prop:Psingvalorder}
Let $(\Amat, \Xmat, \Ymat) \sim \mathrm{DMPRDPG}(F_\rho)$  and define $\Pmat = \Xmat \Ymat^\intercal$. Furthermore, let $\sigma_\ell(\Pmat)$ denote the $\ell$-th non-zero singular value of $\Pmat$ for $\ell \in [d]$ and let the second moment matrices of the latent position distributions be $\bDelta_{X,k} = \mathbb{E} [\boldsymbol{\xi}^k\boldsymbol{\xi}^{k\intercal}],\ k\in[\Kn]$, and $\bDelta_{Y,t} = \mathbb{E} [\boldsymbol{\nu}^t\boldsymbol{\nu}^{t\intercal}],\ t\in[\Tn]$, and define the matrices $\tDelta_X = \lim_{n\to\infty}\Kn^{-1} \sum_{k=1}^{\Kn}\bDelta_{X,k} $ and $\tDelta_Y = \lim_{n\to\infty}\Tn^{-1}\sum_{t= 1}^{\Tn} \bDelta_{Y,t}$. Then, for $n\to\infty$: 
\begin{equation}
\frac{\sigma_\ell(\Pmat)}{\rho_n n \Tn^{1/2} \Kn^{1/2}} \to \sqrt{\lambda_\ell\left(\tilde{\boldsymbol\Delta}_X \tilde{\boldsymbol\Delta}_Y\right)},
\end{equation}
with overwhelming probability. Consequently:
\begin{enumerate}[label=\roman*.]
    \item $\sigma_\ell(\Pmat) = \Op(\rho_n n\Kn^{1/2} \Tn^{1/2})$ with overwhelming probability, 
    \item $\sigma_\ell(\Pmat) = \OmegaP(\rho_n n\Kn^{1/2} \Tn^{1/2})$ with overwhelming probability.
\end{enumerate}
\end{restatable}

\begin{proof} 
The non-zero eigenvalues of a product of matrices are invariant under cyclic permutations, hence we can write:
\begin{equation}
\sigma_\ell(\Pmat) = \sqrt{{\lambda_i(\Xmat\Ymat_{\vphantom{0}}^\intercal \Ymat\Xmat_{\vphantom{0}}^\intercal)}} = \sqrt{\lambda_i(\Xmat_{\vphantom{0}}^{\intercal}\Xmat\Ymat_{\vphantom{0}}^\intercal{\Ymat})}, 
\end{equation}
directly following from the definition $\Pmat = {\Xmat\Ymat}^{\intercal}$.
From Proposition \ref{prop:XTXminDeltaRate}, the following rates hold mutually with overwhelming probability for each $k \in [\Kn], t \in[\Tn]$:
\begin{align}
\norm{{\Xmat}^{k\intercal} {\Xmat}^k - \rho_n n\bDelta_{X,k}} = \Op \{n^{1/2}\log^{1/2}(n)\},\ & & \norm{{\Ymat}^{t\intercal} {\Ymat}^t - \rho_n n\bDelta_{Y,t}} = \Op \{n^{1/2}\log^{1/2}(n)\}.
\end{align}
Let $\bDelta_{X,n}^\ast = \Kn^{-1}\sum_{k=1}^{\Kn} \bDelta_{X,k}$ and $\bDelta_{Y,n}^\ast = \Tn^{-1}\sum_{t=1}^{\Tn} \bDelta_{Y,t}$. Because 
\if1\arxiv
$\norm{{\Xmat}^{\intercal} {\Xmat} - \rho_n n \Kn \bDelta_{X,n}^\ast} \leq \sum_{k=1}^{\Kn} \\ \norm{{\Xmat}^{k\intercal} {\Xmat}^k - \rho_n n\bDelta_{X,k}} $
\else
$\norm{{\Xmat}^{\intercal} {\Xmat} - \rho_n n \Kn \bDelta_{X,n}^\ast} \leq \sum_{k=1}^{\Kn} \norm{{\Xmat}^{k\intercal} {\Xmat}^k - \rho_n n\bDelta_{X,k}} $
\fi
and $\norm{{\Ymat}^{\intercal} {\Ymat} - \rho_n n \Tn \bDelta_{Y,n}^\ast} \leq \sum_{t=1}^{\Tn} \norm{{\Ymat}^{t\intercal} {\Ymat}^t - \rho_n n\bDelta_{Y,t}}$ and by Assumption \ref{ass:KT-growth} both $\Tn$ and $\Kn$ are $O\{\log(n)\}$ we have 
\begin{align}
    \norm{{\Xmat}^{\intercal} {\Xmat} - \rho_n n \Kn \bDelta_{X,n}^\ast} = \Op \{n^{1/2}\log^{3/2}(n)\}, & & 
    \norm{{\Ymat}^{\intercal} {\Ymat} - \rho_n n \Tn \bDelta_{Y,n}^\ast} = \Op \{n^{1/2}\log^{3/2}(n)\}.
    \label{eq:bounds_xx_yy}
\end{align}
The triangle inequality yields
\begin{equation}
    \norm{{\Ymat}^\intercal {\Ymat}} \leq \rho_n n \Tn\norm{\bDelta_{Y,n}^\ast} + \norm{{\Ymat}^\intercal {\Ymat} - \rho_n n \Tn \bDelta_{Y,n}^\ast} = \Op(\rho_n n \Tn ).
    \label{eq:bound_yy}
\end{equation}
Furthermore, we can write: 
\begin{equation}
{\Xmat}^\intercal \Xmat {\Ymat}^\intercal {\Ymat} - \rho_n^2 n^2 \Tn \Kn \bDelta_{X,n}^\ast \bDelta_{Y,n}^\ast = ({\Xmat}^\intercal {\Xmat} - \rho_n n \Kn \bDelta_{X,n}^\ast){\Ymat}^\intercal {\Ymat} + \rho_n n \Kn \bDelta_{X,n}^\ast({\Ymat}^\intercal {\Ymat} - \rho_n n \Tn \bDelta_{Y,n}^\ast).
\end{equation}
Therefore, applying the triangle inequality gives:
\begin{multline}
\norm{{\Xmat}^\intercal \Xmat {\Ymat}^\intercal {\Ymat} - \rho_n^2 n^2 \Tn \Kn \bDelta_{X,n}^\ast \bDelta_{Y,n}^\ast} \leq \\ \norm{{\Xmat}^\intercal {\Xmat} - \rho_n n \Kn \bDelta_{X,n}^\ast}~\norm{{\Ymat}^\intercal {\Ymat}} + \rho_n n \Kn \norm{\bDelta_{X,n}^\ast}~\norm{{\Ymat}^\intercal {\Ymat} - \rho_n n \Tn \bDelta_{Y,n}^\ast}.
\end{multline}
Using Equations \eqref{eq:bounds_xx_yy} and \eqref{eq:bound_yy}, we  get: 
\begin{equation}
\norm{{\Xmat}^\intercal \Xmat {\Ymat}^\intercal {\Ymat} - \rho_n^2 n^2 \Tn \Kn \bDelta_{X,n}^\ast \bDelta_{Y,n}^\ast} = \Op \{\rho_n^2 \Kn n^{3/2}\log^{3/2}(n)\}.
\end{equation}
It follows that $\rho_n^{-2}n^{-2} \Tn^{-1} \Kn^{-1} {\Xmat}_{\vphantom{0}}^\intercal \Xmat {\Ymat}_{\vphantom{0}}^\intercal {\Ymat}$ converges to $\tilde \bDelta_X \tilde \bDelta_Y$ in spectral norm since
\begin{equation}\label{eq:xtxyty-bound}
\bignorm{\frac{{\Xmat}^\intercal \Xmat {\Ymat}^\intercal {\Ymat}}{\rho_n^2 n^2 \Tn \Kn} - \bDelta_{X,n}^\ast \bDelta_{Y,n}^\ast} = \Op \left\{ \frac{\rho_n^2 \Kn n^{3/2} \log^{3/2}(n)}{\rho_n^2 n^2 \Tn \Kn}\right\}
\end{equation}
and 
\begin{multline}
    \bignorm{\frac{{\Xmat}^\intercal \Xmat {\Ymat}^\intercal {\Ymat}}{\rho_n^2 n^2 \Tn \Kn} - \tilde \bDelta_X \tilde \bDelta_Y} \leq \bignorm{\frac{{\Xmat}^\intercal \Xmat {\Ymat}^\intercal {\Ymat}}{\rho_n^2 n^2 \Tn \Kn} - \bDelta_{X,n}^\ast \bDelta_{Y,n}^\ast} + \bignorm{\bDelta_{X,n}^\ast \bDelta_{Y,n}^\ast - \tilde \bDelta_X \tilde \bDelta_Y} \\ \leq \bignorm{\frac{{\Xmat}^\intercal \Xmat {\Ymat}^\intercal {\Ymat}}{\rho_n^2 n^2 \Tn \Kn} - \bDelta_{X,n}^\ast \bDelta_{Y,n}^\ast} + \bignorm{\bDelta_{X,n}^\ast (\bDelta_{Y,n}^\ast - \tilde \bDelta_Y)} + \bignorm{ (\bDelta_{X,n}^\ast - \tilde \bDelta_X) \tilde \bDelta_Y} \\ \leq \bignorm{\frac{{\Xmat}^\intercal \Xmat {\Ymat}^\intercal {\Ymat}}{\rho_n^2 n^2 \Tn \Kn} - \bDelta_{X,n}^\ast \bDelta_{Y,n}^\ast} + \bignorm{\bDelta_{X,n}^\ast} \bignorm{\bDelta_{Y,n}^\ast - \tilde \bDelta_Y} + \bignorm{\bDelta_{X,n}^\ast - \tilde \bDelta_X} \bignorm{\tilde \bDelta_Y}.
\end{multline}
The first term on the right-hand side is bounded by Equation \eqref{eq:xtxyty-bound} while the remaining terms converge to $0$ by Assumption \ref{ass:second-moment-convergence}. Convergence in the spectral norm implies element-wise convergence, which along with the continuity of the characteristic polynomial implies convergence of the singular values. Hence, for $n \to \infty$: 
\begin{equation}
\frac{\sigma_\ell \left( \Pmat \right)}{\rho_n n \Tn^{1/2} \Kn^{1/2}} = \sqrt{\frac{\lambda_\ell \left(\Xmat \Ymat_{\vphantom{0}}^\intercal \Ymat \Xmat_{\vphantom{0}}^\intercal \right)}{\rho_n^2 n^2 \Tn \Kn}} \to \sqrt{\lambda_\ell \left( \tilde \bDelta_X \tilde \bDelta_Y \right)}.
\end{equation}
By Assumption \ref{ass:second-moment-convergence}, the matrices $\tilde \bDelta_X$ and $\tilde \bDelta_Y$ are full rank and fixed. Hence, the first $d$ eigenvalues of $\tilde \bDelta_X \tilde \bDelta_Y$ are all non-zero constants. It follows that the asymptotic growth rate of each $\sigma_\ell\left( \Pmat \right)$ is exactly $\rho_n n \Tn^{1/2} \Kn^{1/2}$.
\end{proof}


The core idea that is used to prove Theorems $\ref{result:TwotoInfNorm}$ and $\ref{result:CLT}$ is to show that the right and left DUASE embeddings of the matrix $\Pmat$ are, up to an invertible linear transformation, equal to the true latent position matrices $\Xmat$ and $\Ymat$. The matrix $\Amat$ is then regarded as a randomly perturbed version of $\Pmat$. Hence, a critical bound for the proofs that follow is a bound on the norm of this random perturbation which we show below.
\begin{restatable}[Bound for ``variance'' of $\Amat$]{proposition}{variancebound}\label{prop:AminPorder}
Let $(\Amat, \Xmat, \Ymat) \sim \mathrm{DMPRDPG}(F_\rho)$  and define $\Pmat = \Xmat \Ymat^\intercal$. Then,
\begin{equation}
    ||\Amat-\Pmat|| = \Op \left\{\rho_n^{1/2} \maxKT^{1/2}n^{1/2}\log^{1/2}(n)\right\}.
\end{equation}
\end{restatable}
\begin{proof}
We begin by conditioning on a fixed connection probability matrix $\Pmat^\ast$ and decomposing the matrix $(\Amat - \Pmat^\ast)$ as $(\Amat - \Pmat^\ast) = \mat{M} + \mat{P}_{0}$ where $\mat{M} \in \mathbb{R}^{n \Kn \times n \Tn}$ is the double unfolding of the matrices $\Amat^{k,t} - \Pmat^{\ast k,t},\ k\in[\Kn],\ t\in[\Tn]$, with the diagonal terms set to $0$, whereas $\mat{P}_{0}$ contains only the diagonal terms. We proceed by establishing bounds on the spectral norms of $\mat{M}$ and $\mat{P}_{0}$ individually. For $\mat{P}_{0}$, bounding the spectral norm by the Frobenius norm yields 
\begin{equation}
\norm{\mat{P}_{0}} \leq \frob{\mat{P}_{0}} = \sqrt{\sum_{k= 1}^{\Kn} \sum_{t = 1}^{\Tn} \sum_{i=1}^n \left(\Pmat^{\ast k,t}_{i,i} \right)^2} \leq \rho_n n^{1/2} \Kn^{1/2} \Tn^{1/2}
\end{equation}
with probability one. 
In order to bound $\mat{M}$ with overwhelming probability, we make use of the \emph{matrix Bernstein’s inequality} \citepSM[see, for example, Theorem 1.6.2 in][]{tropp2015introduction}. 
In order to apply the 
inequality, we must first bound the matrix variance statistic $v(\Mmat) = \max(\norm{\mathbb{E} \left[\Mmat \Mmat^\intercal \right]}, \norm{\mathbb{E} \left[\Mmat^\intercal \Mmat \right]})$. We break down $\mat{M}$ into $n \times n$ sub-matrices where each $\mat{M}^{k,t}$ corresponds to the quantity $\Amat^{k,t} - \Pmat^{\ast k,t}$ with the diagonal terms set to $0$. Now, for each $\mat{M}^{k,t},\ k\in[\Kn],\ t\in[\Tn]$, we can write:  
\begin{equation}
    \left[ \mat{M}^{k,t} {\mat{M}^{k,t}}^\intercal \right]_{i,j} = \sum_{l \neq i,j} (\Amat^{k,t}_{l,i} -\Pmat^{\ast k,t}_{l,i})(\Amat^{k,t}_{l,j} -\Pmat^{\ast k,t}_{l,j}).
\end{equation}
Since $\mathbb{E}[\Amat^{k,t}_{l,i}]=\Pmat^{\ast k,t}_{l,i}$, we obtain:
\begin{equation}
\mathbb{E}\left[ \mat{M}^{k,t} {\mat{M}^{k,t}}^\intercal \right]_{i,j} = 
\begin{cases}
    \sum_{l \neq i} \Pmat^{\ast k,t}_{l,i} (1- \Pmat^{\ast k,t}_{l,i}), & \text{if}\ i=j,\\
    0, & \text{if}\ i \neq j.
  \end{cases}
\end{equation}
The matrix $\mathbb{E}[\mat{M}^{k,t} {\mat{M}^{k,t}}^\intercal]$ is diagonal, with diagonal entries less than $ \rho_n n$ since $\Pmat^{\ast k,t}_{i,j} \leq \rho_n$. Therefore, for all $k\in[\Kn],\ t\in[\Tn]$, we have:
\begin{equation}
\bignorm{\mathbb{E}\left[ \mat{M}^{k,t} {\mat{M}^{k,t}}^\intercal \right]} \leq \rho_n n.
\end{equation}
By Definition \ref{def:DMPRDPG} the 
noise at each entry of every layer and time point is independent, hence 
\if1\arxiv
$\mathbb{E}[ \mat{M}^{k,t} {\mat{M}^{h,s}}^\intercal] \\ = 0$ 
\else
$\mathbb{E}[ \mat{M}^{k,t} {\mat{M}^{h,s}}^\intercal] = 0$ 
\fi
for $k,h\in[\Kn],\ t,s\in[\Tn],\ (k,t) \neq (h,s)$.
If $\mathbb{E}[\Mmat \Mmat^\intercal]$ is further divided into $\Kn^2$ sub-matrices, each of dimension $n\times n$, denoted $\mat{R}_{k,h}$, $k,h\in[\Kn]$, we can write:
\begin{equation}
\mat{R}_{k,h} = \sum_{t=1}^{\Tn} \mathbb{E}\left[ \mat{M}^{k,t} {\mat{M}^{h,t}}^\intercal \right] = \begin{cases}
    \sum_{t=1}^{\Tn}  \mathbb{E}\left[ \mat{M}^{k,t} {\mat{M}^{k,t}}^\intercal \right] & \text{if}\ k=h,\\
    \mat{0}_{n\times n} & \text{if}\ k \neq h.
  \end{cases}
\end{equation}
 Hence, $\mathbb{E}[\mat{M} \mat{M}^\intercal]$ is a diagonal matrix in $\mathbb{R}^{n\Kn \times n\Kn}$, with diagonal entries less than $\rho_n n \Tn$. Therefore, $\norm{\mathbb{E}[\mat{M} \mat{M}^\intercal]} \leq \rho_n n \Tn$. 
 An identical argument can be used to show that $\norm{\mathbb{E}[\mat{M}^\intercal \mat{M}]} \leq \rho_n n \Kn$. Hence, $v(\mat{M}) \leq \rho_n n \maxKT$. Plugging this value into the matrix Bernstien's inequality yields:
 \begin{equation}
\mathbb{P} \left( \norm{\mat{M}} \geq \tau \right) \leq n(\Tn+\Kn)\exp \left(-\frac{3\tau^2}{6 \rho_n n \maxKT + 2\tau} \right)
\end{equation}
for any $\tau\geq0$. For any $\alpha>0$, we define a constant $C_\alpha = \sqrt{7(\alpha + 2)/3}$ dependent on $\alpha$,   and select $\tau= C_\alpha \rho_n^{1/2}n^{1/2}\maxKT^{1/2}\log^{1/2}(n)$, which gives:
\begin{align}
\mathbb{P} &\left( \norm{\mat{M}} \geq 
C_\alpha 
\rho_n^{1/2}\maxKT^{1/2}n^{1/2}\log^{1/2}(n) \right) \\ &\leq n(\Tn+\Kn)\exp \left(\frac{-7(\alpha + 2) \rho_n \maxKT n \log(n)}{6 \rho_n n \maxKT + 2\sqrt{7(\alpha + 2)/3}\rho_n^{1/2}n^{1/2}\maxKT^{1/2}\log^{1/2}(n)} \right).
\end{align}
Let $n_1 = \mathrm{inf}\{ n>0:  \rho_n \maxKT n \geq 2\sqrt{7(\alpha + 2)/3}~\rho_n^{1/2}n^{1/2}\maxKT^{1/2}\log^{1/2}(n)\}$ and let $n_2 = \mathrm{inf}\{ n>0:  \Kn+\Tn \leq n\}$. 
Both $n_1$ and $n_2$ must be finite by our assumptions on the growth rates of $\rho_n$, $\Kn$ and $\Tn$ (Assumptions \ref{ass:rho-growth} and \ref{ass:KT-growth}). Define $ n^* = \max(n_1, n_2)$. For $n \geq n^*$, we then have the following bound:
\begin{align}
\mathbb{P} &\left( \norm{\mat{M}} \geq 
C_\alpha \rho_n^{1/2}n^{1/2}\maxKT^{1/2}\log^{1/2}(n) \right) \\ 
&\leq n(\Tn+\Kn)\exp \left(\frac{-7(\alpha + 2) \rho_n \maxKT n \log(n)}{6 \rho_n n \maxKT + 2 \sqrt{7(\alpha + 2)/3}~\rho_n^{1/2}n^{1/2}\maxKT^{1/2}\log^{1/2}(n)} \right) \\ 
&\leq n(\Tn+\Kn)\exp \left(\frac{-7(\alpha + 2) \rho_n n \maxKT \log(n)}{7 \rho_n n \maxKT} \right) \\ 
&= n(\Tn+\Kn)\exp \left\{-(\alpha + 2) \log(n) \right\} \\ &\leq n^2\exp \left\{-(\alpha + 2) \log(n) \right\} = n^{-\alpha}. 
\end{align}
It follows that 
\begin{equation}
    \mathbb{P} \left( \norm{\mat{M}} \leq C_\alpha\rho_n^{1/2}n^{1/2}\maxKT^{1/2}\log^{1/2}(n) \right) \geq 1- n^{-\alpha},\ \quad n>n^\ast,
\end{equation}
for $C_\alpha=\sqrt{7(\alpha + 2)/3}$. By the definition of $\Op$, this gives: 
\begin{equation}
    \norm{\mat{M}} = \Op \left\{\rho_n^{1/2}n^{1/2}\maxKT^{1/2}\log^{1/2}(n) \right\}.
\end{equation}
By Assumption \ref{ass:KT-growth}, $\Tn$ and $\Kn$ scale as $O\{\log (n)\}$, and therefore the spectral norm of $\mat{M}$ dominates that of $\mat{P}_0$ asymptotically. Hence  
\begin{equation}
    \norm{\Amat-\Pmat^\ast } \leq \norm{\mat{M}} + \norm{\mat{P}_{0}} = \Op \left\{ \rho_n^{1/2} n^{1/2} \maxKT^{1/2}\log^{1/2}(n) \right\}.
\end{equation}
This means that for every $\Pmat^\ast$ and for every $\alpha > 0 $ setting $C_{\alpha}=\sqrt{7(\alpha + 2)/3}$ yields 
\begin{equation}
    \mathbb{P} \left(\norm{\Amat - \Pmat^\ast} \leq C_{\alpha} \rho_n^{1/2} n^{1/2} \maxKT^{1/2}\log^{1/2}(n) \right) \geq 1-n^{- \alpha}
\end{equation}
for all $n > n^\ast$. Because this rate holds for any $\Pmat^\ast$ we conclude that 
\begin{equation}
     ||\Amat-\Pmat|| = \Op \left\{\rho_n^{1/2} \maxKT^{1/2}n^{1/2}\log^{1/2}(n)\right\}
\end{equation}
which is the required result. 
\end{proof}
Using the two previous results we can now establish a bound on the singular values of the observed adjacency matrix $\Amat$.
\begin{proposition}\label{prop:ASingValOrder}
Let $(\Amat, \Xmat, \Ymat) \sim \mathrm{DMPRDPG}(F_\rho)$. 
The non-zero singular values $\sigma_\ell(\Amat)$ for $\ell\in \{1,\dots, d\}$ satisfy:

\begin{enumerate}[label=\roman*.]
    \item $\sigma_\ell(\Amat) = \Op (\rho_n n \Kn^{1/2} \Tn^{1/2})$; 
    \item $\sigma_\ell(\Amat) = \OmegaP (\rho_n n \Kn^{1/2} \Tn^{1/2})$.
\end{enumerate}
\end{proposition}
\begin{proof}
We make use of Corollary 7.3.5 from \citeSM{horn2012matrix},
 which states that for for any two matrices $\mat{M}_1$ and $\mat{M}_2$ with the same dimension, the following inequalities hold: 
 \begin{equation}
 \sigma_\ell(\mat{M}_1)- \norm{\mat{M}_2- \mat{M}_1} \leq \sigma_\ell(\mat{M}_2) \leq \sigma_\ell(\mat{M}_2) - \norm{\mat{M}_1 - \mat{M}_2}.
 \end{equation}
 Therefore, setting $\mat M_1=\Pmat$ and $\mat M_2=\Amat$ gives:
 \begin{equation}
 \sigma_\ell(\Pmat) - \norm{\mat{A} - \Pmat} \leq \sigma_\ell(\mat{A}) \leq \sigma_\ell(\Pmat) - \norm{\mat{A}- \Pmat}.
 \end{equation}
 Applying Proposition \ref{prop:Psingvalorder} for $\sigma_\ell(\Pmat)$ and Proposition \ref{prop:AminPorder} for $\norm{\Amat-\Pmat}$, we see that both the upper bound and lower bound for $\sigma_\ell(\mat{A})$ are $\Op (\rho_n n \Kn^{1/2} \Tn^{1/2}).$
 \end{proof}

\begin{proposition}\label{prop:UpAminPVporder}
Let $(\Amat, \Xmat, \Ymat) \sim \mathrm{DMPRDPG}(F_{\rho_n})$ with $\Kn$ layers and $\Tn$ time points, defined as in Definition~\ref{def:DMPRDPG} where $\Amat$ has a singular value decomposition $\Amat = \UmatA\DmatA\VmatA^\intercal + \mat{U}_{\mat{A} \perp} \mat{D}_{\mat{A} \perp} \mat{V}_{\mat{A} \perp}^\intercal$. Let $\Pmat=\Xmat\Ymat^\intercal$, with singular value decomposition $\Pmat = \UmatP\DmatP\VmatP^\intercal$. Then: 
\begin{equation}
\frob{\UmatP^\intercal(\Amat-\Pmat)\VmatP} = \Op \{\log^{1/2}(n)\}.
\end{equation}
\end{proposition}
\begin{proof}
Once again, we begin by conditioning on fixed latent positions. For any $p,q\in [d]$, $k\in[\Kn]$ and $t\in [\Tn]$, define $\UmatP^k$ to be the $[n(k-1)+1]$-th through $nk$-th rows of $\UmatP$ and define $\VmatP^t$ to be the $[n(t-1)+1]$-th through $nt$-th rows of $\VmatP$. Let $\vec u^{k,p}=(u^{k,p}_1,\dots,u^{k,p}_n)\in\mathbb R^n$ and $\vec v^{t,q}=(v^{t,q}_1,\dots,v^{t,q}_n)\in\mathbb R^n$ denote the $p$-th and $q$-th columns of $\UmatP^k$ and $\VmatP^t$ respectively, and define 
\begin{equation}
    E_{p,q}^{k,t} = \sum_{i=1}^n \sum_{j=1,i \neq j}^n u_i^{k,p} v_j^{t,q}\left( \Amat^{k,t}_{i,j} - \Pmat^{k,t}_{i,j}\right) - \sum_{i=1}^n u_i^{k,p} v_i^{t,q}\Pmat^{k,t}_{i,i}.
    \label{eq:epq}
\end{equation}
Using $E_{p,q}^{k,t}$, we can write: 
\begin{equation}
[\UmatP^\intercal(\Amat-\Pmat)\VmatP]_{p,q} = \sum_{k=1}^{\Kn} \sum_{t=1}^{\Tn} E^{k,t}_{p,q}.
\end{equation}
The second term in the definition of $E_{p,q}^{k,t}$ in \eqref{eq:epq} 
is not relevant for the asymptotic analysis of the entry $[\UmatP^\intercal(\Amat-\Pmat)\VmatP]_{p,q}$, since  
\begin{equation}
    \bigabs{\sum_{k=1}^{\Kn} \sum_{t=1}^{\Tn} \sum_{i=1}^n u_i^{k,p} v^{t,q}_i\Pmat^{k,t}_{i,i}} \leq \left( \sum_{k=1}^{\Kn} \sum_{i=1}^n \abs{u_i^{k,p}}^2 \right)^{1/2} \left( \sum_{t=1}^{\Tn} \sum_{i=1}^n \abs{v^{t,q}_i}^2 \right)^{1/2} = \Op \left( \rho_n \right)
\end{equation}
by Cauchy-Schwarz. 
On the other hand, for the first part of \eqref{eq:epq}, 
we note that each $E^{k,t}_{p,q}$ is the sum of independent random variables with 0 mean, bounded in absolute value by $\abs{u_i^{k,p} v^{t,q}_j}$. Therefore, we can apply Hoeffding's inequality to find:
\begin{equation}
\mathbb{P} \left( \bigabs{\sum_{k=1}^{\Kn} \sum_{t=1}^{\Tn} E^{k,t}_{p,q}} \geq \tau \right) \leq 2 \exp\left\{ -\frac{\tau^2}{2}\left(\sum_{k=1}^{\Kn}\sum_{t=1}^{\Tn} \sum_{i=1}^n \sum_{j=1,j\neq i}^n  \bigabsnogap{u_i^{k,p} v^{t,q}_j}^2\right)^{-1}\right\}.\end{equation}
Selecting $\tau = C_\alpha^{1/2}\log^{1/2}(n)$ yields
\begin{equation}
\mathbb{P}\left( \bigabs{\sum_{k=1}^{\Kn} \sum_{t=1}^{\Tn} E^{t,k}_{p,q}} \geq C_\alpha^{1/2}\log^{1/2}(n) \right) \leq 2 \exp\left\{-\frac{C_\alpha\log(n)}{2} \left(\sum_{k=1}^{\Kn}\sum_{t=1}^{\Tn} \sum_{i=1}^n \sum_{j=1,j\neq i}^n  \bigabsnogap{u_i^{k,p} v^{t,q}_j}^2\right)^{-1}\right\}.
\end{equation}
By definition $\vec u^{k,p} $ and $\vec v^{t,q}$ each partition columns of the unitary matrices $\UmatP$ and $\VmatP$; hence, we can conclude that $\sum_{k=1}^{\Kn}\sum_{t=1}^{\Tn} \sum_{i=1}^n \sum_{j=1,j\neq i}^n \abs{u_i^{k,p} v^{t,q}_j}^2 \leq 1$.  Therefore, $\abs{\sum_{k=1}^{\Kn} \sum_{t=1}^{\Tn} E^{k,t}_{p,q}\ } < C_\alpha^{1/2} \log^{1/2}(n)$ with probability 
\begin{equation}
1 - 2 \exp\left\{-\frac{C_\alpha\log(n)}{2} \left(\sum_{k=1}^{\Kn}\sum_{t=1}^{\Tn} \sum_{i=1}^n \sum_{j=1,j\neq i}^n  \bigabsnogap{u_i^{k,p} v^{t,q}_j}^2\right)^{-1}\right\} \geq 1-
\exp\left\{-\frac{C_\alpha\log(n)}{2} + \log(2) \right\}.
\end{equation}
By selecting $C_\alpha = 2\alpha + 1$ it follows that
\begin{equation}\left[\UmatP^\intercal(\Amat-\Pmat)\VmatP\right]_{p,q} = \Op \left\{\log^{1/2}(n)\right\},
\end{equation}
for any $p,q\in[d]$. Because each of the $d^2$ elements of $\UmatP^\intercal(\Amat-\Pmat)\VmatP$ scale as $\Op\{\log^{1/2}(n)\}$ it follows that the Frobenius norm of the entire matrix is also $\Op\{\log^{1/2}(n)\}$.
Because this rate holds for any choice of latent positions we conclude that 
\begin{equation}
\frob{\UmatP^\intercal(\Amat-\Pmat)\VmatP} = \Op \{\log^{1/2}(n)\}.
\end{equation}
\end{proof}

\begin{proposition}\label{prop:RatesforVariousTerms1} 
Let $(\Amat, \Xmat, \Ymat) \sim \mathrm{DMPRDPG}(F_{\rho_n})$ with $\Kn$ layers and $\Tn$ time points, defined as in Definition~\ref{def:DMPRDPG}. Let $\Pmat=\Xmat\Ymat^\intercal$, with singular value decomposition $\Pmat = \UmatP\DmatP\VmatP^\intercal$. Then the following results hold:
\begin{enumerate}
  \item 
  \begin{enumerate}
      \item $\norm{\UmatA \UmatA^\intercal - \UmatP \UmatP^{\intercal}} = \Op\left\{ \Kn^{-1/2} \Tn^{-1/2} \maxKT^{1/2} \rho_n^{-1/2} n^{-1/2}\log^{1/2}(n)\right\}$,
      \item $\norm{\VmatA \VmatA^\intercal - \VmatP \VmatP^\intercal} = \Op\left\{\Kn^{-1/2} \Tn^{-1/2} \maxKT^{1/2} \rho_n^{-1/2} n^{-1/2}\log^{1/2}(n)\right\}$.
  \end{enumerate}
  \item 
  \begin{enumerate}
  \item $\frob{\UmatA  - \UmatP \UmatP^\intercal \UmatA} = \Op\left\{ \Kn^{-1/2} \Tn^{-1/2}  \maxKT^{1/2}\rho_n^{-1/2} n^{-1/2}\log^{1/2}(n)\right\}$,
  \item $\frob{\VmatA - \VmatP \VmatP^\intercal \VmatA} = \Op\left\{ \Kn^{-1/2} \Tn^{-1/2}  \maxKT^{1/2} \rho_n^{-1/2} n^{-1/2}\log^{1/2}(n)\right\}.$
  \end{enumerate}
  \item 
  \begin{enumerate}
  \item $\frob{\UmatP^\intercal \UmatA \DmatA  - \DmatP \VmatP^\intercal \VmatA} = \Op\left\{ \Kn^{-1/2} \Tn^{-1/2} \maxKT \log(n) \right\}$, 
  \item $\frob{\DmatP \UmatP^\intercal \UmatA  - \VmatP^\intercal \VmatA \DmatA} = \Op\left\{ \Kn^{-1/2} \Tn^{-1/2} \maxKT \log(n) \right\}$.
  \end{enumerate}
  \item $\frob{\UmatP^\intercal \UmatA - \VmatP^\intercal \VmatA} = \Op \left\{\Kn^{-1} \Tn^{-1} \maxKT \rho_n^{-1} n^{-1} \log(n)\right\}$. 
\end{enumerate}
\end{proposition}
\begin{proof} This proof is divided into four parts, corresponding to the four statements in the result. 
\begin{enumerate}
  \item Define $\sigma_1, \dots \sigma_d$ to be the singular values of the matrix $\UmatP^\intercal \UmatA$ and $\theta_\ell = \cos^{-1}(\sigma_\ell)$ to be the principal angles. From Lemma 2.4 in \citeSM{SpectralMethodsforDS}, we know that the non-zero eigenvalues of $\UmatA \UmatA^\intercal - \UmatP \UmatP^\intercal$ are equal to $\sin(\theta_\ell)$. By invoking a variant of the Davis-Kahan theorem \citepSM[see][Theorem 12]{Yu14,jones2021multilayer} we find that
    \begin{equation}
      \norm{\UmatA \UmatA^\intercal - \UmatP \UmatP^{\intercal}} = \max_{i \in \{1, \dots d \}} \abs{\sin(\theta_i)} \leq \frac{2 \sqrt{d}\norm{\Amat  - \Pmat}[2\sigma_1(\Pmat) + \norm{\Amat - \Pmat}]}{\sigma_d(\Pmat)^2}.
    \end{equation} 
for large $n$. By Propositions \ref{prop:Psingvalorder} and \ref{prop:AminPorder}, the order of the right-hand side is 
\begin{equation} \Op\left\{\frac{\maxKT^{1/2} \log^{1/2}(n)}{\Kn^{1/2} \Tn^{1/2}\rho_n^{1/2} n^{1/2}}\right\}.
\end{equation}
  An analogous argument can be used to attain the same rate for $\norm{\VmatA \VmatA^\intercal - \VmatP \VmatP^{\intercal}}$.

  \item Using the rate derived above in Part 1, and the fact the fact that $\UmatA$ is a truncated unitary matrix, and hence multiplication by $\UmatA$ does not increase the growth rate of the norm, we see that
  \begin{multline*}
  \frob{\UmatA - \UmatP \UmatP^\intercal \UmatA} = \frob{(\UmatA \UmatA^\intercal - \UmatP \UmatP^\intercal)\UmatA} \\ \leq d \cdot\frob{\UmatA \UmatA^\intercal - \UmatP \UmatP^\intercal} = \Op\left\{\frac{\maxKT^{1/2} \log^{1/2}(n)}{\Kn^{1/2} \Tn^{1/2}\rho_n^{1/2} n^{1/2}}\right\}.
  \end{multline*}
  Again, the same argument can be used to show that $\frob{\VmatA - \VmatP \VmatA^\intercal \VmatA}$. 

  \item Algebraic manipulation shows that 
  \begin{multline}
      \UmatP^\intercal \UmatA \DmatA  - \DmatP \VmatP^\intercal \VmatA = \UmatP^\intercal (\Amat - \Pmat)\VmatA \\ = \UmatP^\intercal (\Amat - \Pmat )(\VmatA - \VmatP \VmatP^\intercal \VmatA) + \UmatP^\intercal (\Amat - \Pmat )\VmatP \VmatP^\intercal \VmatA.
    \end{multline}
  Analysing these terms separately and applying the bound from Part 2, combined with the results in Propositions~\ref{prop:AminPorder} and \ref{prop:UpAminPVporder}, we find that
  \begin{equation}
  \frob{\UmatP^\intercal (\Amat - \Pmat )(\VmatA - \VmatP \VmatP^\intercal \VmatA)} = \Op\left\{\Kn^{-1/2} \Tn^{-1/2} \maxKT \log(n)\right\},
  \end{equation}
  and 
  \begin{equation}
  \frob{\UmatP^\intercal (\Amat - \Pmat )\VmatP \VmatP^\intercal \VmatA} = \Op\left\{\log^{1/2}(n)\right\}.
  \end{equation}
Combining the two rates gives the required result. 
  The same rate can be attained analogously for $\frob{\DmatP \UmatP^\intercal \UmatA  - \VmatP^\intercal \VmatA \DmatA}$.
  \item Via simple algebraic manipulation, we can write:
      \begin{multline} 
  \UmatP ^\intercal \UmatA - \VmatP^\intercal \VmatA = [(\UmatP^\intercal \UmatA \DmatA - \DmatP \VmatP^\intercal \VmatA) + (\DmatP \UmatP^\intercal \UmatA - \VmatP^\intercal \VmatA \DmatA)]\DmatA^{-1} \\ - \DmatP(\UmatP^\intercal \UmatA - \VmatP^\intercal \VmatA)\DmatA^{-1}.
    \end{multline}
    Therefore, the following identity holds: 
  \begin{multline} 
  \UmatP ^\intercal \UmatA - \VmatP^\intercal \VmatA + \DmatP(\UmatP^\intercal \UmatA - \VmatP^\intercal \VmatA)\DmatA^{-1} = \\ [(\UmatP^\intercal \UmatA \DmatA - \DmatP \VmatP^\intercal \VmatA) + (\DmatP \UmatP^\intercal \UmatA - \VmatP^\intercal \VmatA \DmatA)]\DmatA^{-1}.
  \label{eq:identity_upua_vpva}
    \end{multline}
  From the definition of $\DmatA$ and $\DmatP$, the absolute value of the $(\ell,h)$-th entry of the left-hand side of the identity above can be written as: 
    \begin{equation}
    \left\vert\left[\UmatP ^\intercal \UmatA - \VmatP^\intercal \VmatA + \DmatP(\UmatP^\intercal \UmatA - \VmatP^\intercal \VmatA)\DmatA^{-1}\right]_{\ell,h}\right\vert = \abs{(\UmatP ^\intercal \UmatA - \VmatP^\intercal \VmatA)_{\ell,h}}\left[1+\frac{\sigma_\ell(\Pmat)}{\sigma_h(\Amat)}\right]
  \end{equation}
Also, the absolute value of the $(\ell,h)$-th entry of the matrix $[(\UmatP^\intercal \UmatA \DmatA - \DmatP \VmatP^\intercal \VmatA) + (\DmatP \UmatP^\intercal \UmatA - \VmatP^\intercal \VmatA \DmatA)]\DmatA^{-1}$ on the right-hand side of \eqref{eq:identity_upua_vpva} can be bounded by the Frobenius norm of the matrix itself, which implies that:
    \begin{multline}
        \abs{(\UmatP ^\intercal \UmatA - \VmatP^\intercal \VmatA)_{\ell,h}}\left[1+\frac{\sigma_\ell(\Pmat)}{\sigma_h(\Amat)}\right] \\ \leq \left(\frob{\UmatP^\intercal \UmatA \DmatA - \DmatP \VmatP^\intercal \VmatA} + \frob{\DmatP \UmatP^\intercal \UmatA - \VmatP^\intercal \VmatA \DmatA}\right) \frob{\DmatA^{-1}}.
    \end{multline}
    Since $[1+ \sigma_\ell(\Pmat)/\sigma_h(\Amat)] > 1$, and using the rate from Part 3 along with Proposition  \ref{prop:ASingValOrder}, we get: 
    \begin{equation}
        \frob{\UmatP^\intercal \UmatA - \VmatP^\intercal \VmatA} = \Op \left\{\frac{\maxKT \log(n)}{\Kn \Tn \rho_n n}\right\},
    \end{equation}
which corresponds to the result.
\end{enumerate}
\end{proof}

\begin{proposition}\label{prop:RatesforUminW}
Let $(\Amat, \Xmat, \Ymat) \sim \mathrm{DMPRDPG}(F_{\rho_n})$ with $\Kn$ layers and $\Tn$ time points, defined as in Definition~\ref{def:DMPRDPG}, where $\Amat$ has a singular value decomposition $\Amat = \UmatA\DmatA\VmatA^\intercal + \mat{U}_{\mat{A} \perp} \mat{D}_{\mat{A} \perp} \mat{V}_{\mat{A} \perp}^\intercal$.
Also, let $\Pmat=\Xmat\Ymat^\intercal$, with singular value decomposition $\Pmat = \UmatP\DmatP\VmatP^\intercal$. 
Let $\UmatP^\intercal \UmatA + \VmatP^\intercal \VmatA$ admit the singular value decomposition
\begin{equation}
\UmatP^\intercal \UmatA + \VmatP^\intercal \VmatA = \mat{W}_1 \mat{D} \mat{W}_2,
\end{equation}
and let $\mat{W} = \mat{W}_1\mat{W}_2^\intercal$. Then
\begin{equation}
\max\left\{ \frob{\UmatP^\intercal \UmatA - \mat{W}}, \frob{\VmatP^\intercal \VmatA - \mat{W}} \right\} = \Op\left\{\frac{\maxKT \log(n)}{\Kn \Tn \rho_n n}\right\}.
\end{equation}
\end{proposition}

\begin{proof}
From \citeSM{schonemann1966generalized} we have that:
\begin{equation}
\mat{W} = \mat{W}_1\mat{W}_2^\intercal = \min_{\mat{Q} \in \Od(d)} \left[ ||\UmatP^\intercal \UmatA - \mat{Q}||^2_F + ||\VmatP^\intercal \VmatA - \mat{Q}||^2_F\right].
\label{eq:proc_min}
\end{equation}
Next, denote the SVD of $\UmatP^\intercal \UmatA$ by $\UmatP^\intercal \UmatA = \mat{W}_{\mat{U},1} \mat{D}_{\mat{U}} \mat{W}_{\mat{U},2}^\intercal$ and define the $d \times d$ orthogonal matrix $\mat{W}_{\mat{U}} = \mat{W}_{\mat{U},1} \mat{W}_{\mat{U},2}^\intercal$. Also, consider $\sigma_1, \dots, \sigma_d$ to be the singular values of $\UmatP^\intercal \UmatA$ as in Proposition \ref{prop:RatesforVariousTerms1}. Then by using the rates provided in Proposition \ref{prop:RatesforVariousTerms1}, we find:
\begin{align}
    ||\UmatP^\intercal \UmatA - &\Wmat_{\mat{U}}||_F = ||\Dmat - \Id||_F =\sqrt{\sum_{i=1}^d 
(1- \sigma_i)^2} \leq \sum_{i=1}^d (1-\sigma_i) \leq  \sum_{i=1}^d (1-\sigma_i^2) \\ &= \sum_{i=1}^d \sin^2(\theta_i) \leq d~\norm{\UmatA \UmatA^\intercal - \UmatP \UmatP^\intercal}^2 = \Op \left\{ \frac{\maxKT \log(n)}{\Kn \Tn \rho_n n} \right\}.
\label{eq:upua_w}
\end{align}
Furthermore, we obtain the rate for $\frob{\VmatP^\intercal \VmatA - \Wmat_{\mat{U}}}$ by applying the triangle inequality, and using the rate in \eqref{eq:upua_w} above as well as Proposition \ref{prop:RatesforVariousTerms1}, to find:
\begin{equation}
\begin{split}
\frob{\VmatP^\intercal \VmatA - \Wmat_{\mat{U}}}  = \frob{\VmatP^\intercal \VmatA - \UmatP^\intercal \UmatA + \UmatP^\intercal \UmatA - \Wmat_{\mat{U}}} \\  \leq \frob{\VmatP^\intercal \VmatA - \UmatP^\intercal \UmatA} + \frob{\UmatP ^\intercal \UmatA - \Wmat_{\mat{U}}} &= \Op\left\{ \frac{ \maxKT \log(n)}{\Kn \Tn \rho_n n} \right\}.
\end{split}
\end{equation}
Therefore, by definition of $\Wmat$ as the minimiser in \eqref{eq:proc_min}, it follows that
\begin{equation}    
\frob{\UmatP^\intercal \UmatA - \Wmat}^2 + \frob{\VmatP^\intercal \VmatA - \Wmat}^2 \leq \frob{\UmatP^\intercal \UmatA - \Wmat_{\mat{U}}}^2 + \frob{\VmatP^\intercal \VmatA - \Wmat_{\mat{U}}}^2.
\end{equation}
Hence, 
\begin{equation}
\max\left\{ ||\UmatP^\intercal \UmatA - \Wmat||_F, ||\VmatP^\intercal \VmatA - \Wmat||_F\right\} = \Op\left\{ \frac{\maxKT \log(n)}{\Kn \Tn \rho_n n} \right\},
\end{equation}
which gives the result. 
\end{proof}

\begin{proposition}\label{prop:RatesforVariousTerms2}
Let $(\Amat, \Xmat, \Ymat) \sim \mathrm{DMPRDPG}(F_{\rho_n})$ with $\Kn$ layers and $\Tn$ time points, defined as in Definition~\ref{def:DMPRDPG}, where $\Amat$ has a singular value decomposition $\Amat = \UmatA\DmatA\VmatA^\intercal + \mat{U}_{\mat{A} \perp} \mat{D}_{\mat{A} \perp} \mat{V}_{\mat{A} \perp}^\intercal$.
Also, let $\Pmat=\Xmat\Ymat^\intercal$, with singular value decomposition $\Pmat = \UmatP\DmatP\VmatP^\intercal$. 
Let $\UmatP^\intercal \UmatA + \VmatP^\intercal \VmatA$ admit the singular value decomposition
\begin{equation}
\UmatP^\intercal \UmatA + \VmatP^\intercal \VmatA = \mat{W}_1 \mat{D} \mat{W}_2,
\end{equation}
and let $\mat{W} = \mat{W}_1\mat{W}_2^\intercal$. Then
\begin{enumerate}
  \item $\frob{\Wmat \DmatA - \DmatP \Wmat} = \Op \left\{\maxKT \Kn^{-1/2} \Tn^{-1/2} \log(n)\right\}$,
  \item $\frob{\Wmat \DmatA^{1/2} - \DmatP^{1/2} \Wmat} = \Op \left\{\rho_n^{-1/2} n^{-1/2} \Tn^{-3/4} \Kn^{-3/4} \maxKT \log(n)\right\}$,
  \item $\frob{\Wmat \DmatA^{-1/2} - \DmatP^{-1/2} \Wmat} = \Op \left\{ \Kn^{-5/4} \Tn^{-5/4} \maxKT \rho_n^{-3/2} n^{-3/2} \log(n) \right\}$.
\end{enumerate}
\end{proposition}

\begin{proof}
The proof is divided in three parts, corresponding to the three rates stated in the proposition. The key arguments are based on \cite{lyzinski2016community}, Lemma 17.
    \begin{enumerate}
  \item 
  Via algebraic manipulation, we get: 
    \begin{align}
      \Wmat \DmatA - \DmatP \Wmat &= (\Wmat - \UmatP^\intercal \UmatA)\DmatA + \UmatP^\intercal \UmatA \DmatA - \DmatP \Wmat \\ &= (\Wmat - \UmatP^\intercal \UmatA)\DmatA + (\UmatP^\intercal \UmatA \DmatA - \DmatP \VmatP^\intercal \VmatA ) + \DmatP (\VmatP^\intercal \VmatA - \Wmat).
  \end{align}
  Applying the Frobenius norm, using the triangle inequality, and applying the rates from Propositions~\ref{prop:Psingvalorder}, \ref{prop:RatesforVariousTerms1} and~\ref{prop:RatesforUminW} on each component of the right-hand side of the summation above, we find that 
    \begin{multline}
    ||(\Wmat - \UmatP^\intercal \UmatA)\DmatA + (\UmatP^\intercal \UmatA \DmatA - \DmatP \VmatP^\intercal \VmatA ) + \DmatP (\VmatP^\intercal \VmatA - \Wmat)||_F \\ = \Op\left\{ \maxKT \Kn^{-1/2} \Tn^{-1/2} \log(n) \right\}.
  \end{multline}
  \item 
  From the definition of $\DmatA$ and $\DmatP$, we can write the $(\ell,h)$-th entry of $\Wmat\DmatA^{1/2} - \DmatP^{1/2} \Wmat$ as: 
    \begin{align}
      \left[\Wmat\DmatA^{1/2} - \DmatP^{1/2} \Wmat\right]_{\ell,h} &= \Wmat_{\ell,h}\left[\sigma_h(\Amat)^{1/2} - \sigma_\ell(\Pmat)^{1/2}\right]  = \frac{\Wmat_{\ell,h}[\sigma_h(\Amat) - \sigma_\ell(\Pmat)]}{\sigma_h(\Amat)^{1/2} + \sigma_\ell(\Pmat)^{1/2}} \\
      &=\frac{[\Wmat \DmatA - \DmatP \Wmat]_{\ell,h}}{\sigma_h(\Amat)^{1/2} + \sigma_\ell(\Pmat)^{1/2}},
  \end{align}
  for $\ell,h\in[d]$. 
  By taking the Frobenius norm of the right-hand side and applying the result from Part 1 as well as Proposition \ref{prop:Psingvalorder} we find that each of the $d^2$ element of $\Wmat \DmatA^{1/2} - \DmatP^{1/2} \Wmat$ scales as $\rho_n^{-1/2} n^{-1/2} \Tn^{-3/4}\Kn^{-3/4} \maxKT \log(n)$ and hence
  \begin{equation}
  \frob{\Wmat \DmatA^{1/2} - \DmatP^{1/2} \Wmat} = \Op \left\{ \frac{\maxKT \log(n)}{\rho_n^{1/2} n^{1/2} \Kn^{3/4} \Tn^{3/4}} \right\}.
  \end{equation}

\item Using a similar approach as the previous part, we get: 
\begin{equation}
    \left[\Wmat\DmatA^{-1/2} - \DmatP^{-1/2} \Wmat\right]_{\ell,h} = \frac{\Wmat_{\ell,h}[\sigma_\ell(\Pmat)^{1/2} - \sigma_h(\Amat)^{1/2}]}{\sigma_\ell(\Pmat)^{1/2} \sigma_h(\Amat)^{1/2}} =\frac{[\Wmat \DmatA^{1/2} - \DmatP^{1/2} \Wmat]_{\ell,h}}{\sigma_\ell(\Pmat)^{1/2} \sigma_h(\Amat)^{1/2}}.
\end{equation}
  Once again, taking the Frobenius norm of the right-hand side and applying the result from Part 2 combined with Propositions \ref{prop:Psingvalorder} and \ref{prop:ASingValOrder}, we get:
  \begin{equation}
  \frob{\Wmat \DmatA^{-1/2} - \DmatP^{-1/2} \Wmat} = \Op \left\{ \frac{\maxKT \log(n)}{\Kn^{5/4} \Tn^{5/4}\rho_n^{3/2} n^{3/2}} \right\},
  \end{equation}
  which is the desired result.
\end{enumerate}
\end{proof}
As discussed in Section \ref{sec:theoretical_results} the true latent position matrices $\Xmat$ and $\Ymat$ are identifiable only up to a linear transformation. In the following proof we show the existence of these transformations and provide some insight into their structure.
\begin{proposition}\label{prop:ExistenceofRotations}
Let $(\Amat, \Xmat, \Ymat) \sim \mathrm{DMPRDPG}(F_{\rho_n})$ with $\Kn$ layers and $\Tn$ time points, defined as in Definition~\ref{def:DMPRDPG}, where $\Amat$ has a singular value decomposition $\Amat = \UmatA\DmatA\VmatA^\intercal + \mat{U}_{\mat{A} \perp} \mat{D}_{\mat{A} \perp} \mat{V}_{\mat{A} \perp}^\intercal$.
Also, let $\Pmat=\Xmat\Ymat^\intercal$, with singular value decomposition $\Pmat = \UmatP\DmatP\VmatP^\intercal$, and define $\XmatP=\UmatP\DmatP^{1/2}$ and $\YmatP=\VmatP\DmatP^{1/2}$.
If both ${\Xmat}$ and ${\Ymat}$ are rank $d$, then there exist matrices $ \Ltilde \in \mathrm{GL}(d)$ and $\Rtilde \in \mathrm{GL}(d)$ such that $\XmatP = {\Xmat} \Ltilde$ and $\YmatP = {\Ymat} \Rtilde$. Furthermore, $\Ltilde \Rtilde^\intercal = \Id_d$.

\end{proposition}

\begin{proof}
Define $\mat{\Pi}_X = ({\Xmat}^\intercal {\Xmat})^{1/2}$ and $\mat{\Pi}_Y = ({\Ymat}^\intercal {\Ymat})^{1/2}$ where we take the unique positive-definite square root for both. For the first result, note that 
\begin{equation}
(\XmatP \DmatP^{1/2}) (\XmatP \DmatP^{1/2})^\intercal = \UmatP \DmatP^2 \UmatP^\intercal  = \Pmat \Pmat^\intercal = \Xmat {\Ymat}^\intercal \Ymat {\Xmat}^\intercal = (\Xmat \mat{\Pi}_Y)(\Xmat \mat{\Pi}_Y)^\intercal.
\end{equation}
This implies the existence of an orthogonal matrix $\Qmat \in \Od(d)$ such that 
\begin{equation}
    \XmatP \DmatP^{1/2} = \Xmat \mat{\Pi}_Y \Qmat.
    \end{equation}
Therefore, the matrix $\mat{\Pi}_Y \Qmat \DmatP^{-1/2} \in \mathrm{GL}(d)$ satisfies the definition of $\Ltilde$. Similarly, for $\Rtilde$ we have that 
\begin{equation}
    (\YmatP \DmatP^{1/2}) (\YmatP \DmatP^{1/2})^\intercal = \VmatP \DmatP^2 \VmatP^\intercal  = \Pmat^\intercal \Pmat = \Ymat {\Xmat}^\intercal \Xmat {\Ymat}^\intercal = (\Ymat \mat{\Pi}_X)(\Ymat \mat{\Pi}_X)^\intercal.
\end{equation}
Therefore, there exists a matrix $\Qmat^\ast \in \Od(d)$ such that 
\begin{equation}
    \YmatP \DmatP^{1/2} = \Ymat \mat{\Pi}_X \Qmat^\ast.
\end{equation}
Hence, $\Rtilde = \mat{\Pi}_X \Qmat^* \DmatP^{-1/2} \in \mathrm{GL}(d)$.
In order to show that $\Ltilde \Rtilde^\intercal = \Id_d$, we write 
\begin{equation}
\Xmat \Ltilde \Rtilde^\intercal {\Ymat}^\intercal = \XmatP \YmatP^\intercal = \Pmat = \Xmat {\Ymat}^\intercal.
\end{equation}
Multiplying both sides of this equality by $({\Xmat}^\intercal \Xmat)^{-1}{\Xmat}^\intercal$ on the left and by $\Ymat({\Ymat}^\intercal \Ymat)^{-1}$ on the right yields the desired result. 
\end{proof}
We now derive asymptotic rates for the transformations defined in the previous proposition. 
\begin{proposition}\label{prop:OrderOfRotations} 
The matrices $\Ltilde \in \mathrm{GL}(d)$ and $\Rtilde \in \mathrm{GL}(d)$ in Proposition~\ref{prop:ExistenceofRotations} satisfy the following: $||\Ltilde|| = \Op (\Tn^{1/4} \Kn^{-1/4})$,\ $||\Ltilde^{-1}|| = \Op(\Kn^{1/4} \Tn^{-1/4})$, $||\Rtilde|| = \Op( \Kn^{1/4} \Tn^{-1/4})$, and $||\Rtilde^{-1}|| = \Op(\Tn^{1/4} \Kn^{-1/4} )$. 

\end{proposition}
\begin{proof}
    From Proposition~\ref{prop:ExistenceofRotations}, recall that $\Ltilde = \mat{\Pi}_Y \Qmat \DmatP^{-1/2}$ and that $\Rtilde = \mat{\Pi}_X \Qmat^\ast \DmatP^{-1/2}$. By Proposition \ref{prop:Psingvalorder} we have $||\DmatP|| = \Op(\Kn^{1/2} \Tn^{1/2} \rho_n n)$ and $||\DmatP^{-1}|| = \Op( \Kn^{-1/2} \Tn^{-1/2} \rho_n^{-1} n^{-1} )$. Following a similar line of reasoning to Proposition \ref{prop:Psingvalorder}, we get $||\mat{\Pi}_X|| = \Op( \Kn^{1/2} \rho_n^{1/2} n^{1/2})$ and $||\mat{\Pi}_Y|| = \Op( \Tn^{1/2} \rho_n^{1/2} n^{1/2})$, by noting the convergence of $\rho_n^{-1} n^{-1} \Kn^{-1}{\Xmat ^\intercal \Xmat} \to \tilde \bDelta_X$ and $\rho_n^{-1} n^{-1} \Tn^{-1}{\Ymat ^\intercal \Ymat} \to \tilde \bDelta_Y$ for $n\to\infty$. Therefore, $\norm{\Ltilde} \leq \norm{\mat{\Pi}_Y}\,\norm{\Qmat}\,\norm{\DmatP^{-1/2}} = \Op( \Tn^{1/4} \Kn^{-1/4} )$, $\norm{\Ltilde^{-1}} \leq \norm{\DmatP^{1/2}}\,\norm{\Qmat^{-1}}\,\norm{\mat{\Pi}_Y^{-1}} = \Op( \Kn^{1/4} \Tn^{-1/4} )$, $\norm{\Rtilde} \leq \norm{\mat{\Pi}_X}\,\norm{\Qmat^\ast}\,\norm{\DmatP^{-1/2}} = \Op( \Kn^{1/4} \Tn^{-1/4} )$, and $\norm{\Rtilde^{-1}} \leq \norm{\DmatP^{1/2}}\,\norm{\Qmat^{\ast -1}}\,\norm{\mat{\Pi}_X^{-1}} = \Op( \Tn^{1/4} \Kn^{-1/4} )$. 
\end{proof}

\begin{proposition}\label{prop:XTX_YTY_ident} 
Let $(\Amat, \Xmat, \Ymat) \sim \mathrm{DMPRDPG}(F_{\rho_n})$ with $\Kn$ layers and $\Tn$ time points, defined as in Definition~\ref{def:DMPRDPG}, , where $\Amat$ has a singular value decomposition $\Amat = \UmatA\DmatA\VmatA^\intercal + \mat{U}_{\mat{A} \perp} \mat{D}_{\mat{A} \perp} \mat{V}_{\mat{A} \perp}^\intercal$.
Also, let $\Pmat=\Xmat\Ymat^\intercal$, with singular value decomposition $\Pmat = \UmatP\DmatP\VmatP^\intercal$, and consider matrices $\Ltilde, \Rtilde\in\GL(d)$ as in Proposition~\ref{prop:ExistenceofRotations}.
If each $\Xmat^k$ and each $\Ymat^t$ is of rank $d$ then
\begin{align}
\Rtilde \DmatP^{-1} \Ltilde^{-1} = ({\Ymat}^\intercal \Ymat)^{-1}, & & \Ltilde \DmatP^{-1} \Rtilde^{-1} = ({\Xmat}^\intercal \Xmat)^{-1}.
\label{eq:prop13_results}
\end{align}
\end{proposition}
\begin{proof}
    For the first result in \eqref{eq:prop13_results}, recall that $\Ltilde \Rtilde^\intercal = \Id_d$ and  
    $$\Xmat \Ltilde \DmatP \Ltilde^\intercal {\Xmat}^\intercal = \XmatP \DmatP \XmatP^\intercal = \Pmat \Pmat^\intercal = \Xmat {\Ymat}^\intercal \Ymat {\Xmat}^\intercal.$$
    Hence, $\Ltilde \DmatP \Ltilde^\intercal = {\Ymat}^\intercal \Ymat$ and 
    $$(\Rtilde \DmatP^{-1} \Ltilde^{-1} )^\intercal = \Ltilde^{-1 \intercal} \DmatP^{-1} \Ltilde^{-1} \Ltilde \Rtilde^\intercal  = ({\Ymat}^\intercal \Ymat)^{-1} \Ltilde \Rtilde^\intercal = ({\Ymat}^\intercal \Ymat)^{-1}.$$
    This proves the first statement in \eqref{eq:prop13_results}. For the second statement, we use the identities $\Ltilde = ({\Xmat}^\intercal \Xmat)^{-1} {\Xmat}^\intercal \XmatP$ and $\XmatP^\intercal \XmatP  = \DmatP$, to get 
    \begin{align}
        (\Ltilde \DmatP^{-1} \Rtilde^{-1})^{\intercal} &= \Rtilde^{-1 \intercal}\DmatP^{-1} \XmatP ^\intercal \Xmat ({\Xmat}^\intercal \Xmat)^{-1} = \Rmat^{-1 \intercal}\DmatP^{-1} \XmatP ^\intercal \XmatP \Ltilde^{-1} ({\Xmat}^\intercal \Xmat)^{-1} \\ &= \Rtilde^{-1 \intercal}\DmatP^{-1} \DmatP \Ltilde^{-1} ({\Xmat}^\intercal \Xmat)^{-1} = \Rtilde^{-1 \intercal} \Ltilde^{-1} ({\Xmat}^\intercal \Xmat)^{-1} = ({\Xmat}^\intercal \Xmat)^{-1}, \notag
    \end{align}
    which gives the result. 
\end{proof}



We now show that our model guarantees desirable incoherence properties for the matrices $\Amat$ and $\Pmat$. In particular, we show a result bounding the two to infinity norm of the right and left singular vectors, 
which is closely related to the incoherence parameters for the matrices $\Amat$ and $\Pmat$ \citep[see, for example,][Definition 3.1]{SpectralMethodsforDS}. 
\begin{proposition}\label{prop:incoherence}
    Let $(\Amat, \Xmat, \Ymat) \sim \mathrm{DMPRDPG}(F_{\rho_n})$ with $\Kn$ layers and $\Tn$ time points, defined as in Definition~\ref{def:DMPRDPG} and define $\Pmat=\Xmat\Ymat^\intercal$, and denote its singular value decomposition by $\Pmat = \UmatP\DmatP\VmatP^\intercal$. Additionally, for $i\in [n\Kn]$ let $\Amat^{(i)}$ denote the partially denoised analogue of $\Amat$ such that:
\begin{equation}
  \Amat^{(i)}_{k,j} =
    \begin{cases}
      \Amat_{k,j} & \text{if $k \neq i$}, \\
      \Pmat_{k,j} & \text{if $k = i$},\\
    \end{cases} \label{eq:aikj}
\end{equation}
for $k \in [n\Kn]$ and $j \in [n\Tn]$. Let $\Amat^{(i)} = \UmatA^{(i)} \DmatA^{(i)} \VmatA^{(i)\intercal} + \mat{U}^{(i)}_{\Amat \perp} \Dmat^{(i)}_{\Amat \perp} \Vmat_{\Amat\perp}^{(i)\intercal}$ denote the singular value decomposition of $\Amat^{(i)}$. Then the following bounds hold. 
    \begin{enumerate}
        \item \begin{enumerate}
            \item $\norm{\UmatP}_{\tti} = \Op\{\Kn^{-1/2} n^{-1/2}\};$
            \item $\norm{\VmatP}_{\tti} = \Op\{\Tn^{-1/2} n^{-1/2}\};$
        \end{enumerate}
        \item \begin{enumerate}
            \item $\norm{\UmatA}_{\tti} = \Op\{\Kn^{-1/2} \Tn^{-1/2} \maxKT^{1/2} \rho_n^{-1/2} n^{-1/2}\log^{1/2}(n) \};$
            \item $\norm{\VmatA}_{\tti} = \Op\{\Kn^{-1/2} \Tn^{-1/2} \maxKT^{1/2} \rho_n^{-1/2} n^{-1/2}\log^{1/2}(n)\};$
        \end{enumerate}
        \item \begin{enumerate}
            \item $\norm{\UmatA^{(i)}}_{\tti} = \Op\{\Kn^{-1/2}\Tn^{-1/2} \maxKT^{1/2} \rho_n^{-1/2} n^{-1/2}\log^{1/2}(n) \};$
            \item $\norm{\VmatA^{(i)}}_{\tti} = \Op\{\Kn^{-1/2} \Tn^{-1/2} \maxKT^{1/2} \rho_n^{-1/2} n^{-1/2}\log^{1/2}(n)\}.$
        \end{enumerate}
    \end{enumerate}
\end{proposition}
\begin{proof}
We prove the bounds above for the left singular vectors of $\Pmat$, $\Amat$ and $\Amat^{(i)}$. The corresponding bounds for the right singular vectors follow analogously. 
\begin{enumerate}
    \item For the bound on $\norm{\UmatP}_{\tti}$. Recall that $\UmatP \DmatP^{1/2} = \Xmat \Ltilde$. Applying the relation $\twoinf{\mat A\mat B} \leq \twoinf{\mat A}\, \norm{\mat B}$ yields $\twoinf{\UmatP} \leq \twoinf{\Xmat}\, \norm{\Ltilde}\, \norm{\DmatP^{-1/2}]}$. By applying Propositions \ref{prop:Psingvalorder} and \ref{prop:OrderOfRotations} and using the fact that the rows of $\Xmat$ are by definition $\Op(\rho_n^{1/2})$, we get $\twoinf{\UmatP} = \Op( \Kn^{-1/2} n^{-1/2})$.
    \item Let $e_i$ denote the $i$-th standard basis vector. Note that for a matrix $\mat{B}$ with orthonormal columns, the quantity $\norm{\mat{B} \mat{B}^\intercal e_i}$ is equal to the spectral norm of the $i$-th row of $\mat{B}$ since
    \begin{equation}
        \norm{\mat{B} \mat{B}^\intercal e_i}^2 =e_i^\intercal \mat{B} \mat{B}^\intercal \mat{B} \mat{B}^\intercal e_i = e_i^\intercal \mat{B} \mat{B}^\intercal e_i = (\mat{B} \mat{B}^\intercal)_{ii} = \norm{\mat{B}_{i}}^2.
    \end{equation}
    Using this fact, proceed to bound the norm of $\norm{(\mat{U}_{\Amat})_i}$, the $i$-th row of $\UmatA$, as
    \begin{align}
        \norm{(\mat{U}_{\Amat})_i} & = \norm{(\UmatA \UmatA^\intercal)e_i} \leq \norm{(\UmatA \UmatA^\intercal - \UmatP \UmatP^\intercal) e_i} + \norm{(\mat{U}_{\mat{P}})_i} \\ &\leq \norm{\UmatA \UmatA^\intercal - \UmatP \UmatP^\intercal} + \twoinf{\UmatP}.
    \end{align}
    Applying the rate from Part 1 as well as Proposition \ref{prop:RatesforVariousTerms1} shows that, for all $i \in [n \Kn]$
    \begin{equation}
        \norm{(\mat{U}_{\Amat})_i} = \Op \{ \Kn^{-1/2} \Tn^{-1/2} \maxKT^{1/2} \rho_n^{-1/2} n^{-1/2}\log^{1/2}(n)\},
    \end{equation}
    and hence 
    \begin{equation}
        \twoinf{\UmatA} = \Op \{ \Kn^{-1/2}\Tn^{-1/2} \maxKT^{1/2}\rho_n^{-1/2} n^{-1/2}\log^{1/2}(n)\}.
    \end{equation}
    \item The bound for $\twoinf{\UmatAi}$ begins with a similar decomposition to that in Part 2.
    \begin{equation}
        \norm{(\mat{U}^{(i)}_{\mat{A}})_i} = \norm{(\UmatAi \UmatA^{(i)\intercal})e_i} \leq \norm{\UmatAi \UmatA^{(i)\intercal} - \UmatP \UmatP^\intercal} + \twoinf{\UmatP}
    \end{equation}
    We can apply the bound from Part 1 to the second term while for the first term we apply Wedin's Theorem \citep[see][Theorem 2.9]{SpectralMethodsforDS} to write 
    \begin{equation}\label{eq:UAi-min-Up-bound}
        \norm{\UmatAi \UmatA^{(i)\intercal} - \UmatP \UmatP^\intercal} \leq \frac{\sqrt{2} \max\{ \norm{(\Amat^{(i)} - \Pmat)^\intercal\UmatP},\norm{(\Amat^{(i)} - \Pmat)\VmatP}\}}{\sigma_d(\Pmat) + \sigma_{d+1}(\Pmat) + \norm{\Amat^{(i)} - \Pmat}}
    \end{equation}
    Because $\norm{\UmatP} = \norm{\VmatP} = 1$, both terms in the numerator can be bounded by $\norm{\Amat^{(i)} - \Pmat}$. To bound this  quantity, we write
    \begin{equation*}
        \norm{\Amat^{(i)} - \Pmat} \leq \norm{\Amat^{(i)} - \Amat} + \norm{\Amat - \Pmat} \leq \frob{\Amat^{(i)} - \Amat} + \norm{\Amat - \Pmat}.
    \end{equation*}
    The first term is $\Op(n^{1/2} \Tn^{1/2})$ as it contains 
    $n \Tn$ non-zero elements which are bounded and the second term can be bounded by applying Proposition~\ref{prop:AminPorder}. Hence,
    \if1\arxiv
    $\norm{\Amat^{(i)} - \Pmat} = \Op \{ \rho_n^{1/2} \maxKT^{1/2} \\ n^{1/2} \log^{1/2}(n)\}$. 
    \else
    $\norm{\Amat^{(i)} - \Pmat} = \Op \{ \rho_n^{1/2} \maxKT^{1/2} n^{1/2} \log^{1/2}(n)\}$.
    \fi
    Plugging this rate into \eqref{eq:UAi-min-Up-bound} and applying Proposition \ref{prop:Psingvalorder} yields 
    \begin{equation*}
        \norm{\UmatAi \UmatA^{(i)\intercal} - \UmatP \UmatP^\intercal} = \Op \left\{ \frac{\maxKT^{1/2} \log^{1/2}(n)}{\Kn^{1/2} \Tn^{1/2} \rho_n^{1/2} n^{1/2}} \right\}.
    \end{equation*}
    Therefore, for all $i \in [n \Kn]$
    \begin{equation*}
        \norm{(\mat{U}^{(i)}_{\mat{A}})_i} = \Op \left\{ \frac{\maxKT^{1/2}\log^{1/2}(n)}{\Kn^{1/2}\Tn^{1/2}\rho_n^{1/2} n^{1/2}} \right\}.
    \end{equation*}
    Hence, 
    \begin{equation*}
        \twoinf{\UmatA^{(i)}} = \Op \left\{ \frac{\maxKT^{1/2} \log^{1/2}(n)}{ \Kn^{1/2}\Tn^{1/2} \rho_n^{1/2} n^{1/2}} \right\}.
    \end{equation*}
\end{enumerate}
The bounds for the right singular vectors follow the same arguments. 
\end{proof}

Using the asymptotic rates derived above we are now prepared to bound a number of residual terms which will appear in the proofs of Theorems \ref{result:TwotoInfNorm} and \ref{result:CLT}.
\begin{proposition}\label{prop:RatesforVariousTerms3}
Let $(\Amat, \Xmat, \Ymat) \sim \mathrm{DMPRDPG}(F_{\rho_n})$ with $\Kn$ layers and $\Tn$ time points, defined as in Definition~\ref{def:DMPRDPG}.
Also, let $\Pmat=\Xmat\Ymat^\intercal$, with singular value decomposition $\Pmat = \UmatP\DmatP\VmatP^\intercal$, and define the matrix $\Wmat$ as in Proposition~\ref{prop:RatesforUminW}. 
Additionally, let:
\begin{enumerate}
    \item \begin{enumerate}
        \item $\Rmat_{1,1} = \UmatP(\UmatP^\intercal \UmatA \DmatA^{1/2} - \DmatP^{1/2}\Wmat)$,
        \item $\Rmat_{2,1} = \VmatP(\VmatP^\intercal \VmatA \DmatA^{1/2} - \DmatP^{1/2}\Wmat)$,
    \end{enumerate} 
    \item \begin{enumerate}
        \item $\Rmat_{1,2} = (\Id - \UmatP \UmatP^\intercal)(\Amat - \Pmat)(\VmatA - \VmatP \Wmat)\DmatA^{-1/2}$,
        \item $\Rmat_{2,2} = (\Id - \VmatP \VmatP^\intercal)(\Amat - \Pmat)(\UmatA - \UmatP \Wmat)\DmatA^{-1/2}$,
    \end{enumerate}
    \item \begin{enumerate}
        \item $\Rmat_{1,3} = -\UmatP \UmatP^\intercal (\Amat - \Pmat)\VmatP \Wmat \DmatA^{-1/2}$,
        \item $\Rmat_{2,3} = -\VmatP \VmatP^\intercal (\Amat - \Pmat)\UmatP \Wmat \DmatA^{-1/2}$,
    \end{enumerate} 
    \item \begin{enumerate}
        \item $\Rmat_{1,4} = (\Amat - \Pmat)\VmatP(\Wmat \DmatA^{-1/2} - \DmatP^{-1/2}\Wmat)$,
        \item $\Rmat_{2,4} = (\Amat - \Pmat)\UmatP(\Wmat \DmatA^{-1/2} - \DmatP^{-1/2}\Wmat)$.
    \end{enumerate} 
\end{enumerate}
Then the following bounds hold:
\begin{align*}
    & \twoinf{\Rmat_{1,1}} = \Op \left\{ \frac{\maxKT \log(n)}{\rho_n^{1/2} n \Kn^{5/4} \Tn^{3/4}}\right\}, & &  
    \twoinf{\Rmat_{2,1}} = \Op \left\{ \frac{\maxKT \log(n)}{\rho_n^{1/2} n \Kn^{3/4} \Tn^{5/4}}\right\}, \\
    & \twoinf{\Rmat_{1,2}} = \Op\left\{ \frac{\maxKT^{3/2}\log^{3/2}(n)}{\Kn^{5/4}\Tn^{5/4} \rho_n^{2} n} \right\}, & & \twoinf{\Rmat_{2,2}}  = \Op\left\{ \frac{\maxKT^{3/2}\log^{3/2}(n)}{\Kn^{5/4}\Tn^{5/4} \rho_n^{2} n} \right\}, \\
    & \twoinf{\Rmat_{1,3}} = \Op\left\{ \frac{\log^{1/2}(n)}{\rho_n^{1/2} n \Kn^{3/4} \Tn^{1/4}} \right\}, & & \twoinf{\Rmat_{2,3}} = \Op\left\{ \frac{\log^{1/2}(n)}{\rho_n^{1/2} n \Kn^{1/4} \Tn^{3/4}} \right\}, \\
    & \twoinf{\Rmat_{1,4}} = \Op\left\{ \frac{ \maxKT^{3/2} \log^{3/2}(n)}{\Kn^{5/4} \Tn^{5/4}\rho_n n } \right\}, & & \twoinf{\Rmat_{2,4}}  = \Op\left\{ \frac{ \maxKT^{3/2} \log^{3/2}(n)}{\Kn^{5/4} \Tn^{5/4}\rho_n n } \right\}.
\end{align*}
\end{proposition}
\begin{proof}
 The following are the proofs for the terms $\Rmat_{1,\ell},\ \ell=1,2,3,4$. Unless otherwise noted, the proofs for $\Rmat_{2,\ell},\ \ell=1,2,3,4$, follow analogously. 
\begin{enumerate}
    \item We begin by applying the relation $\twoinf{\mat A\mat B} \leq \twoinf{\mat A}\, \norm{\mat B}$ to write  
    \begin{align}
    \twoinf{\Rmat_{1,1}} & \leq \twoinf{\UmatP}\, \norm{\UmatP^\intercal \UmatA \DmatA^{1/2} - \DmatP^{1/2}\Wmat} \\ & \leq \twoinf{\UmatP} \left[\frob{(\UmatP^\intercal \UmatA - \Wmat) \DmatA^{1/2}} + \frob{\Wmat\DmatA^{1/2} - \DmatP^{1/2}\Wmat}\right].
    \end{align}
    By Propositions~\ref{prop:ASingValOrder} and~\ref{prop:RatesforUminW}, the first term is $\Op\{ {\rho_n^{-1/2} n^{-1/2}\Tn^{-3/4}}\Kn^{-3/4}\maxKT \log(n)\}$, and by Proposition~\ref{prop:RatesforVariousTerms2}, the second term is  $\Op\{{\rho_n^{-1/2} n^{-1/2}\Tn^{-3/4}}{\Kn^{-3/4} \maxKT \log(n)}\}$. Applying the rate from Proposition \ref{prop:incoherence} to bound $\twoinf{\UmatP}$ yields: 
    \begin{equation}
        \Rmat_{1,1} = \Op \left\{ \frac{\maxKT \log(n)}{\rho_n^{1/2} n \Kn^{5/4} \Tn^{3/4}}\right\}.
    \end{equation}
    
    \item Define $\mat{M}_1 = (\UmatP \UmatP^\intercal)(\Amat - \Pmat)(\VmatA - \VmatP \Wmat)\DmatA^{-1/2}$ and $\mat{M}_2 = (\Amat - \Pmat )(\VmatA - \VmatP \Wmat ) \DmatA^{-1/2}$. Hence, $\Rmat_{1,2} = \mat{M}_2 - \mat{M}_1$, which implies that $\twoinf{\Rmat_{1,2}} \leq \twoinf{\mat{M}_2} + \twoinf{\mat{M}_1}$. We bound these terms individually. For $\mat{M}_1$:
    \begin{equation}
    \twoinf{\mat{M}_1} \leq \twoinf{\UmatP}\, \norm{\Amat - \Pmat}\, \norm{\VmatA - \VmatP \Wmat}\, \norm{\DmatA^{-1/2}}.
    \end{equation}
    Propositions~\ref{prop:AminPorder}, \ref{prop:ASingValOrder} and~\ref{prop:incoherence} give bounds for $\norm{\Amat - \Pmat}$, $\norm{\DmatA^{-1/2}}$, and $\twoinf{\UmatP}$ respectively. To bound 
    $\norm{\VmatA - \VmatP \Wmat}$, we make use of Propositions~\ref{prop:RatesforVariousTerms1} and~\ref{prop:RatesforUminW} as follows: 
    \begin{align}
        \norm{\VmatA - \VmatP \Wmat} &\leq \norm{\VmatA - \VmatP \VmatP^\intercal \VmatA} + \norm{\VmatP (\VmatP^\intercal \VmatA - \Wmat)} \\ &= \Op\left\{ \frac{\maxKT^{1/2} \log^{1/2}(n)}{\rho_n^{1/2}n^{1/2} \Kn^{1/2} \Tn^{1/2}}\right\} + \Op\left\{ \frac{\maxKT \log(n)}{\rho_n n \Kn \Tn}\right\}.
    \end{align}
    Under Assumptions \ref{ass:rho-growth} and \ref{ass:KT-growth} which dictate the asymptotic growth of $\rho_n$, $\Kn$ and $\Tn$, both terms converge to zero and the left summand dominates. Hence, 
    \begin{equation}
    \label{eq:M1-rate}
    \twoinf{\mat{M}_1} = \Op\left\{ \frac{\maxKT \log(n)}{\rho_n^{1/2} n \Kn^{5/4} \Tn^{3/4}} \right\}.
    \end{equation}
    For $\mat{M}_2$, we have:
    \begin{equation}
        \mat{M}_2 = (\Amat - \Pmat)(\Id - \VmatP \VmatP^\intercal )\VmatA \DmatA^{-1/2} + (\Amat - \Pmat) \VmatP (\VmatP^\intercal \VmatA - \Wmat)\DmatA^{-1/2}. 
        \label{eq:m2_dec}
    \end{equation}
    Using the fact that $\twoinf{\cdot} \leq \norm{\cdot}$, and applying 
    Propositions~\ref{prop:AminPorder}, \ref{prop:ASingValOrder}, and~\ref{prop:RatesforUminW}, shows that the two-to-infinity norm of the right summand is $\Op\{\Kn^{-5/4} \Tn^{-5/4} \maxKT^{3/2} \rho_n^{-1} n^{-1} \log^{3/2}(n)\}$. For the left term in \eqref{eq:m2_dec}, define $\mat{M} = (\Amat - \Pmat ) (\Id - \VmatP \VmatP^\intercal)\VmatA \VmatA^\intercal$, and observe that we can then rewrite this term as $\mat{M} \VmatA \DmatA^{-1/2}$. Therefore: 
    \begin{equation}
    \twoinf{(\Amat - \Pmat)(\Id - \VmatP \VmatP^\intercal )\VmatA \DmatA^{-1/2}}  \leq \twoinf{\mat{M}}\, \norm{\VmatA \DmatA^{-1/2}}.\end{equation}
    By Proposition \ref{prop:ASingValOrder}, $\norm{\VmatA \DmatA^{-1/2}} = \Op \{\rho_n^{-1/2}n^{-1/2} \Kn^{-1/4} \Tn^{-1/4} \}$. Hence, it remains only to bound $\twoinf{\mat{M}}$. To do this, we follow \cite{corneck2026spectral} by making use of leave-one-out analysis in order to disentangle the 
dependence between $\Amat - \Pmat$ and $\VmatA \VmatA^\intercal$. For $i\in [n\Kn]$ let $\Amat^{(i)}$ be defined 
as in Equation~\eqref{eq:aikj}
for $k \in [n\Kn]$ and $j \in [n\Tn]$, 
corresponding to the matrix $\Amat$, except that the $i$-th row has been stripped of the 
noise corresponding to $(\Amat - \Pmat)_i$. Let $\VmatA^{(i)}$ denote the right singular vectors of the matrix $\Amat^{(i)}$. We can then decompose $\norm{(\Amat - \Pmat)(\mat I - \VmatP \VmatP^\intercal)\VmatA \VmatA^\intercal}_{\tti}$ as:
\begin{multline}
\label{eq:fixed-bound-decomp}
    \big\| (\Amat - \Pmat)(\mat I - \VmatP \VmatP^\intercal)\VmatA \VmatA^\intercal\big\|_{\tti} \leq 
    \norm{(\Amat - \Pmat)(\mat I - \VmatP \VmatP^\intercal)\VmatA^{(i)} \VmatA^{(i)\intercal}}_{\tti} \\ 
    + \norm{(\Amat - \Pmat)(\mat I - \VmatP \VmatP^\intercal)(\VmatA \VmatA^\intercal - \VmatA^{(i)} \VmatA^{(i)\intercal})}_{\tti}
    \end{multline}
and bound these two terms separately. We begin with several intermediate results. As a first step we bound $\|\VmatA^{(i)} \VmatA^{(i)\intercal}-\VmatP \VmatP^\intercal\|$. By applying the Davis-Kahan theorem in a manner similar to the proof of Proposition  \ref{prop:RatesforVariousTerms1}, we find that:
\begin{equation*}
    \norm{\VmatA^{(i)} \VmatA^{(i)\intercal}-\VmatP \VmatP^\intercal} \leq \frac{2\sqrt{d}\ \|\Amat^{(i)}-\Pmat\|\{2\sigma_1(\Pmat) + \|\Amat^{(i)} - \Pmat\|\}}{\sigma_d(\Pmat)^2}.
\end{equation*}
Using the fact that by definition $\|\Amat^{(i)}-\Pmat\| \leq \norm{\Amat-\Pmat} + \Op\{n^{1/2} \Tn^{1/2} \}$ we can apply the rates from Propositions~\ref{prop:Psingvalorder}~and~\ref{prop:AminPorder} to find that: 
\begin{equation}
\label{eq:fixed-bound-inter-1}
    \norm{\VmatA^{(i)} \VmatA^{(i)\intercal}-\VmatP \VmatP^\intercal} = \Op \{ 
    \Kn^{-1/2}\Tn^{-1/2}\maxKT^{1/2}\rho_n^{-1/2}n^{-1/2}\log^{1/2}{(n)}
    \}.
\end{equation}
We then proceed by bounding $\|(\mat I - \VmatP \VmatP^\intercal)\VmatA^{(i)}\|_F$ as
\begin{multline}
\label{eq:fixed-bound-inter-2}
 \norm{(\mat I - \VmatP \VmatP^\intercal)\VmatA^{(i)}}_F = \norm{\VmatA^{(i)} - \VmatP \VmatP^\intercal\VmatA^{(i)}}_F
 \\ = \norm{(\VmatA^{(i)} \VmatA^{(i)\intercal}-\VmatP \VmatP^\intercal)\VmatA^{(i)}}_F
 \leq \norm{\VmatA^{(i)} \VmatA^{(i)\intercal}-\VmatP \VmatP^\intercal}_F \cdot d^{1/2} \\
 \leq \norm{\VmatA^{(i)} \VmatA^{(i)\intercal}-\VmatP \VmatP^\intercal} \cdot 2^{1/2} d =
 \Op\{\Kn^{-1/2}\Tn^{-1/2}\maxKT^{1/2}\rho_n^{-1/2}n^{-1/2}\log^{1/2}{(n)}\}),
\end{multline}
where in the last line we have used the rate from Equation~\eqref{eq:fixed-bound-inter-1} and the relation $\norm{\mat M}_F \leq \norm{\mat M} \cdot \mathrm{rank}^{1/2}(\mat M)$. We can now proceed with the bound of the first term of right-hand side of Equation~\eqref{eq:fixed-bound-decomp}. We use the fact that $\norm{\cdot}_{\tti}$ corresponds to be maximum Frobenius norm of the individual rows and proceed to bound the Frobenius norm of the $i$-th row of $(\Amat - \Pmat)(\mat I - \VmatP \VmatP^\intercal)\VmatA^{(i)} \VmatA^{(i)\intercal}$ using the relation
\begin{equation}
    \norm{(\Amat - \Pmat)_i(\mat I - \VmatP \VmatP^\intercal)\VmatA^{(i)} \VmatA^{(i)\intercal}}_F \leq \norm{(\Amat - \Pmat)_i(\mat I - \VmatP \VmatP^\intercal)\VmatA^{(i)}}_F \cdot d^{1/2}.
\end{equation}
If we examine the term on the right-hand side of this expression, we see that each of its $d$ entries is a weighted sum of $n\Tn$ bounded independent random variables with weights given by the columns of $(\mat I - \VmatP \VmatP^\intercal)\VmatA^{(i)}$. Because this term contains only $\VmatA^{(i)}$ rather than $\VmatA$, the weights are statistically independent of the $i$-th row of $(\Amat - \Pmat)$. If we denote the $j$-th element of $(\Amat - \Pmat)_i(\mat I - \VmatP \VmatP^\intercal)\VmatA^{(i)}$ as $Z_{i,j}$ we can apply Hoeffding's inequality to bound this term.  
\begin{equation}
    \mathbb{P}\left(\abs{Z_{i,j}} > t \right) \leq 2\exp\left\{\frac{-t^2}{
    \sum_{l=1}^{n\Tn} ((\mat I - \VmatP \VmatP^\intercal)\VmatA^{(i)})^2_{l,j}
    } \right\}.
\end{equation}
By substituting
$t = C_{\alpha}^{1/2}\Kn^{-1/2}\Tn^{-1/2}\maxKT^{1/2}\rho_n^{-1/2} n^{-1/2} \log{(n)}$ and applying 
Equation~\eqref{eq:fixed-bound-inter-2}
we see that 
\begin{equation}
    \mathbb{P}\left(\abs{Z_{i,j}} >
    C_{\alpha}^{1/2}\Kn^{-1/2} \Tn^{-1/2} \maxKT^{1/2} \rho_n^{-1/2} n^{-1/2} \log{(n)}
    \right) \leq 2\exp\left\{- C_{\alpha}\log(n)\right\},
\end{equation}
and therefore that 
\begin{equation}
    \norm{(\Amat - \Pmat)_i(\mat I - \VmatP \VmatP^\intercal)\VmatA^{(i)}\VmatA^{(i)\intercal}}_F =
    \Op\{\Kn^{-1/2} \Tn^{-1/2} \maxKT^{1/2}\rho_n^{-1/2} n^{-1/2} \log{(n)}\}.
\end{equation}
By taking the union bound over all rows we conclude that 
\begin{equation}
\label{eq:fixed-bound-T1-rate}
     \norm{(\Amat - \Pmat)(\mat I - \VmatP \VmatP^\intercal)\VmatA^{(i)}\VmatA^{(i)\intercal}}_{\tti} = \Op\{\Kn^{-1/2} \Tn^{-1/2}\maxKT^{1/2}\rho_n^{-1/2} n^{-1/2} \log{(n)}\}.
\end{equation}
It now remains to bound the second term of the right-hand side of Equation~\eqref{eq:fixed-bound-decomp}. To analyze the $i$-th row of the matrix we use the relation $\norm{\mat B \mat C}_F \leq \norm{\mat B} \norm{\mat C}_F$ to write
\begin{multline}
    \norm{(\Amat - \Pmat)_i(\mat I - \VmatP \VmatP^\intercal)(\VmatA \VmatA^\intercal - \VmatA^{(i)} \VmatA^{(i)\intercal})}_{F} \\ \leq \norm{(\Amat - \Pmat)_i(\mat I - \VmatP \VmatP^\intercal)} \norm{\VmatA \VmatA^\intercal - \VmatA^{(i)}  \VmatA^{(i)\intercal}}_F.
\end{multline}
We have $\norm{(\Amat - \Pmat)_i(\mat I - \VmatP \VmatP^\intercal)} =
\Op\{n^{1/2}\Tn^{1/2}\}$ as $(\Amat - \Pmat)_i$ is a vector of $n \Tn$ elements with bounded entries and $(\mat I - \VmatP \VmatP^\intercal)$ has constant order spectral norm. To bound $\|\VmatA \VmatA^\intercal - \VmatA^{(i)}  \VmatA^{(i)\intercal}\|_F$ we make use of Wedin's theorem \citep[see, for example, Theorem 2.9 in][]{SpectralMethodsforDS} to write 
\begin{equation}
    \label{eq:wedin}
    \frob{\VmatA \VmatA^\intercal - \VmatA^{(i)}  \VmatA^{(i)\intercal}} \leq \frac{\sqrt{2}\mathrm{max}(\frob{{{(\UmatA^{(i)}})_i}^\intercal(\Amat - \Pmat)_i },\|(\Amat - \Pmat)_i \VmatA^{(i)}\|_F)}{\sigma_d(\Amat^{(i)}) - \sigma_{d+1}(\Amat^{(i)}) -\norm{(\Amat - \Pmat)_i} }.
\end{equation}
Applying Hoeffding's inequality to the second term of the numerator shows that this term is
\if1\arxiv
\\$\Op\{\log^{1/2}(n)\}$,
\else
$\Op\{\log^{1/2}(n)\}$,
\fi
while for the first term we use the fact that this is the outer product of the $i$-th row of $\UmatA^{(i)}$ and the $i$-th row of $(\Amat - \Pmat)$. For two vectors $\mat{a}$ and $\mat{b}$ it holds that $\frob{\mat{a}\mat{b}^\intercal} \leq \frob{\mat{a}} \frob{\mat{b}}$. Using this relation, we can bound the first term of the numerator as $\frob{{{(\UmatA^{(i)}})_i}^\intercal(\Amat - \Pmat)_i } \leq \frob{{{(\UmatA^{(i)}})_i}} \frob{(\Amat - \Pmat)_i }$. We have already established that $\frob{(\Amat - \Pmat)_i } = \Op\{n^{1/2} \Tn^{1/2}\}$ because it is a vector with $n\Tn$ bounded elements. For $\frob{{{(\UmatA^{(i)}})_i}}$ we note that $\frob{{{(\UmatA^{(i)}})_i}} \leq \norm{{\UmatA^{(i)}}}_{\tti}$ and make use of the rate from Proposition \ref{prop:incoherence} to find that  
\if1\arxiv
$\frob{{(\UmatA^{(i)})_i}} = \Op\{ \Kn^{-1/2}\Tn^{-1/2}\maxKT^{1/2}\rho_n^{-1/2} n^{-1/2} \\ \log^{1/2}(n)\}$. 
\else
$\frob{{(\UmatA^{(i)})_i}} = \Op\{ \Kn^{-1/2}\Tn^{-1/2}\maxKT^{1/2}\rho_n^{-1/2} n^{-1/2}\log^{1/2}(n)\}$.
\fi
We therefore conclude that the numerator of Equation \eqref{eq:wedin} is 
\if1\arxiv 
$\Op\{\Kn^{-1/2} \maxKT^{1/2} \\ \rho_n^{-1/2} \log^{1/2}(n)\}$. 
\else
$\Op\{\Kn^{-1/2} \maxKT^{1/2} \rho_n^{-1/2}\log^{1/2}(n)\}$.
\fi
For the denominator of Equation \eqref{eq:wedin} we apply Corollary~7.3.5 from \cite{horn2012matrix} along with Proposition~\ref{prop:ASingValOrder} to show that this term is $\OmegaP\{\rho_n n \Kn^{1/2} \Tn^{1/2}\}$. 
We therefore find that for all $i \in [n\Kn]$:
\begin{multline*}
    \norm{(\Amat - \Pmat)_i(\mat I - \VmatP \VmatP^\intercal)(\VmatA \VmatA^\intercal - \VmatA^{(i)} \VmatA^{(i)\intercal})}_{F} \\ = \Op \left\{\Kn^{-1}\maxKT^{1/2}\rho_n^{-3/2} n^{-1/2}\log^{1/2}(n)\right\}.
\end{multline*}    
Taking the union bound over all rows yields :
\begin{multline*}
    \norm{(\Amat - \Pmat)(\mat I - \VmatP \VmatP^\intercal)(\VmatA \VmatA^\intercal - \VmatA^{(i)} \VmatA^{(i)\intercal})}_{\tti} \\ = \Op \left\{\Kn^{-1}\maxKT{1/2}\rho_n^{-3/2} n^{-1/2}\log^{1/2}(n)\right\}.
\end{multline*}   

We have now bounded both terms on the right-hand side of Equation \eqref{eq:fixed-bound-decomp}. We therefore conclude that
\begin{equation}
    \norm{\mat{M}}_{\tti} = \Op\left\{\Kn^{-1}\maxKT^{1/2}\rho_n^{-3/2} n^{-1/2}\log^{1/2}(n)\right\},
\end{equation}
and hence
\begin{equation}
    \twoinf{(\Amat - \Pmat)(\Id - \VmatP \VmatP^\intercal )\VmatA \DmatA^{-1/2}} = \Op\left\{ \frac{\maxKT^{1/2}\log^{1/2}(n)}{\Kn^{5/4}\Tn^{1/4} \rho_n^{2} n} \right\}.
\end{equation}
Applying this rate to Equation \eqref{eq:m2_dec} we find that 
\begin{equation}
    \label{m2-rate}
    \twoinf{\mat{M}_2} = \Op\left\{ \frac{\maxKT^{3/2}\log^{3/2}(n)}{\Kn^{5/4}\Tn^{5/4} \rho_n^{2} n} \right\}.
    \end{equation}
Combining the rates from Equations \eqref{eq:M1-rate} and \eqref{m2-rate} we conclude that 
\begin{equation}
    \twoinf{\Rmat_{1,2}} = \Op\left\{ \frac{\maxKT^{3/2}\log^{3/2}(n)}{\Kn^{5/4}\Tn^{5/4} \rho_n^{2} n} \right\}.
\end{equation}

\item Using  the rate for $\twoinf{\UmatP}$ from Proposition \ref{prop:incoherence}, as well as Propositions  \ref{prop:ASingValOrder} and \ref{prop:UpAminPVporder}, we see that 
\begin{align}
    \twoinf{\Rmat_{1,3}} &\leq \twoinf{\UmatP}\, \norm{-\UmatP \UmatP^\intercal (\Amat - \Pmat)\VmatP \Wmat \DmatA^{-1/2}} \\ &\leq \twoinf{\UmatP}\, \frob{\UmatP^\intercal (\Amat - \Pmat)\VmatP}\, \frob{\Wmat \DmatA^{-1/2}} = \Op \left\{\frac{\log^{1/2}(n)}{\rho_n^{1/2} n \Kn^{3/4} \Tn^{1/4}}\right\}.
\end{align}
\item By Propositions \ref{prop:AminPorder} and \ref{prop:RatesforVariousTerms2}, we get: 
\begin{align}
    \twoinf{\Rmat_{1,4}} \leq \frob{\Rmat_{1,4}} \leq & \norm{\Amat - \Pmat}\, \frob{\Vmat_P}\, \frob{\Wmat \DmatA^{-1/2} - \DmatP^{-1/2} \Wmat} \\& = \Op\left\{ \frac{ \maxKT^{3/2} \log^{3/2}(n)}{\Kn^{5/4} \Tn^{5/4}\rho_n n }\right\}.
\end{align}
\end{enumerate}
    
\end{proof}

\section{Proof of Theorem~\ref{result:TwotoInfNorm}} \label{sec:proof_tti}

\begin{proof}
    
For the left embedding, we write 
\begin{align}
    \hat \Xmat - \XmatP\Wmat &= \UmatA \DmatA^{1/2} - \UmatP \DmatP^{1/2}\Wmat \\ 
    &= \UmatA \DmatA^{1/2} - \UmatP \UmatP^\intercal \UmatA \DmatA^{1/2} + \UmatP (\UmatP^\intercal \UmatA \DmatA^{1/2} - \DmatP^{1/2} \Wmat) \\ 
    &= \UmatA \DmatA^{1/2} - \UmatP \UmatP^{\intercal} \UmatA \DmatA^{1/2} + \Rmat_{1,1},
\end{align}
where $\Rmat_{1,1}$ is defined in Proposition~\ref{prop:RatesforVariousTerms3}.
Using the fact that $\UmatA \DmatA^{1/2} = \Amat \VmatA \DmatA^{-1/2}$ (which can be seen by expanding $\Amat$ via its singular value decomposition), and $\UmatP \UmatP^\intercal \Pmat = \Pmat$, we have:
\begin{align}
    \hat \Xmat - \XmatP \Wmat &= \Amat \VmatA \DmatA^{-1/2} - \UmatP \UmatP^\intercal \Amat \VmatA \DmatA^{-1/2} + \Rmat_{1,1} 
    \\ &= (\Amat - \Pmat) \VmatA \DmatA^{-1/2} - (\UmatP \UmatP^\intercal \Amat - \Pmat)\VmatA \DmatA^{-1/2} + \Rmat_{1,1} 
    \\ &= (\Amat - \Pmat) \VmatA \DmatA^{-1/2} - \UmatP \UmatP^\intercal(\Amat - \Pmat)\VmatA \DmatA^{-1/2} + \Rmat_{1,1} 
    \\ &= (\Id - \UmatP \UmatP^\intercal )(\Amat - \Pmat)\VmatA \DmatA^{-1/2} + \Rmat_{1,1}
    \\ &= (\Id - \UmatP \UmatP^\intercal )(\Amat - \Pmat)[\VmatP \Wmat + (\VmatA - \VmatP \Wmat)]\DmatA^{-1/2} + \Rmat_{1,1}
    \\ &= (\Amat - \Pmat)\VmatP\Wmat\DmatA^{-1/2} + \Rmat_{1,3} + \Rmat_{1,2} + \Rmat_{1,1}
    \\ &= (\Amat - \Pmat)\VmatP\DmatP^{-1/2} \Wmat + \Rmat_{1,4} + \Rmat_{1,3} + \Rmat_{1,2} + \Rmat_{1,1},
\label{eq:XminXPdecomp}
\end{align}
where the $\Rmat_{\cdot,\cdot}$ are defined in Proposition~\ref{prop:RatesforVariousTerms3}.
By grouping the residual terms into the matrix $\Rmat_\Xmat = \Rmat_{1,4} + \Rmat_{1,3} + \Rmat_{1,2} + \Rmat_{1,1}$, we see that 
\begin{align}
    \hat \Xmat - \XmatP \Wmat &= (\Amat- \Pmat) \VmatP \DmatP^{-1/2} \Wmat + \Rmat_\Xmat
\end{align}
where $\twoinf{\Rmat_\Xmat} = \Op \{\Kn^{-5/4}\Tn^{-5/4} \maxKT^{3/2} \rho_n^{-2} n^{-1}\log^{3/2}(n) \}$ by Proposition \ref{prop:RatesforVariousTerms3}. We now define $\Amat^k$ and $\Pmat^k$ to be the $(n \times n \Tn)$ matrices consisting of rows $n(k-1)+1$ through $nk$ of $\Amat$ and $\Pmat$ respectively, and define $\Rmat_\Xmat^k$ to be the corresponding rows of $\Rmat_\Xmat$. We can then write 
\begin{equation}
    \hat \Xmat^k - \XmatP^k \Wmat = (\Amat^k - \Pmat^k) \VmatP \DmatP^{-1/2}\Wmat + \Rmat_\Xmat^k.
\end{equation}
Because the two to infinity norm corresponds to the maximum Euclidean row norm, it follows that for each $k\in[\Kn]$, $\twoinf{\Rmat_\Xmat^k} \leq \twoinf{\Rmat_\Xmat}$, and hence
\begin{equation}
    \twoinf{\hat \Xmat^k - \XmatP^k \Wmat} \leq \sigma_d(\Pmat)^{-1/2} \twoinf{(\Amat^k - \Pmat^k) \VmatP} + \Op \left\{ \frac{\maxKT^{3/2} \log^{3/2}(n)}{\Kn^{5/4}\Tn^{5/4} \rho_n^{2} n} \right\}. \label{eq:global_to_local_Xk}
\end{equation}
By Proposition \ref{prop:Psingvalorder}, we have $\sigma_d(\Pmat)^{-1/2} = \Op(\rho_n^{-1/2}n^{-1/2} \Kn^{-1/4} \Tn^{-1/4})$. On the other hand, for $\twoinf{(\Amat^k- \Pmat^k) \VmatP}$ we proceed in a similar manner to the proof of Proposition \ref{prop:AminPorder} and condition on fixed latent positions. For $i \in [n]$ and $j \in [d]$
\begin{equation}
 [(\Amat^k - \Pmat^k)\VmatP]_{i,j} = \sum_{r=1}^{\Tn} \sum_{l=1,\ l \neq i}^n (\Amat^{k,r}_{i,l} - \Pmat^{k,r}_{i,l}) v_{j,l}^r - \sum_{r=1}^{\Tn} (\Pmat^{k,r}_{i,i}) v_{j,i}^r,
 \label{eq:apvpij}
\end{equation}
where $v^r_{i,j}$ is used to denote the $i$-th entry of the $j$-th column of $\VmatP^r$. For the second summation on the right-hand side of \eqref{eq:apvpij}, we can apply the Cauchy-Schwarz inequality to show that the term is at most $\Op (\rho_n \Tn^{1/2})$, since each $\vec v^r_i$ comes from the same column of $\VmatP$, which has unit Frobenius norm by construction. On the other hand, the first  summation in \eqref{eq:apvpij} is a sum of independent random variables with mean zero with absolute value bounded by $\abs{ v_{j,l}^r}$. Therefore, Hoeffding's inequality can be applied, yielding:
\begin{equation}
    \mathbb{P} \left( \left|\ \sum_{r=1}^{\Tn} \sum_{l=1, l \neq i}^n (\Amat^{k,r}_{i,l} - \Pmat^{k,r}_{i,l})\, v^r_{j,l})\  \right|  > \tau \right) \leq 2 \exp \left( - \frac{\tau^2}{2 \sum_{r=1}^{\Tn} \sum_{l=1}^n (v^r_{j,l})^2}  \right) = 2 \exp \left( -\frac{\tau^2}{2}  \right).
\end{equation}
Setting $\tau = \sqrt{2 \alpha \log(n)}$ shows that $\abs{[(\Amat^k - \Pmat^k)\VmatP]_{i,j}} = \Op\{\log^{1/2}(n)\}$. 
Taking the union bound over the $d$ entries of each row $[(\Amat^k - \Pmat^k) \VmatP]_i$ shows that $\norm{((\Amat^k - \Pmat^k) \VmatP)_i} = \Op\{\log^{1/2}(n)\}$ for all $k\in[\Kn]$. Taking the union bound over all of the $n$ rows of $(\Amat^k - \Pmat^k) \VmatP$ shows that the rows  scale uniformly as $\log^{1/2}(n)$ with overwhelming probability for any choice of latent positions. Because the two to infinity norm corresponds to the maximum Euclidean
row norm it follows that $\twoinf{(\Amat^k - \Pmat^k) \VmatP} = \Op\{\log^{1/2}(n)\}$ and we conclude from \eqref{eq:global_to_local_Xk} that:
\begin{equation}
    \twoinf{\hat{\Xmat}^k - \XmatP^k \Wmat} = \Op \left\{ \frac{\log^{1/2}(n)}{\rho_n^{1/2} n^{1/2} \Kn^{1/4} \Tn^{1/4}}\right\}.
\end{equation}
Furthermore, defining $\WX = \Ltilde \Wmat$ where $\Ltilde$ is as defined in Proposition \ref{prop:ExistenceofRotations}, and applying the rate from Proposition \ref{prop:OrderOfRotations} yields:
\begin{equation}
\twoinf{\hat \Xmat^k \WX^{-1} - \Xmat^k} = \Op \left\{ \frac{\log^{1/2}(n)}{\rho_n^{1/2} n^{1/2} \Tn^{1/2}}\right\}.
\label{eq:WX_first}
\end{equation}
An analogous argument is used to prove the result in \eqref{eq:tti_bound} for the right embedding:  
\begin{align}
    \hat \Ymat - \YmatP\Wmat &= \VmatA \DmatA^{1/2} - \VmatP \DmatP^{1/2}\Wmat \\ 
    &= \VmatA \DmatA^{1/2} - \VmatP \VmatP^\intercal \VmatA \DmatA^{1/2} + \VmatP (\VmatP^\intercal \VmatA \DmatA^{1/2} - \DmatP^{1/2} \Wmat) \\ 
    &= \VmatA \DmatA^{1/2} - \VmatP \VmatP^{\intercal} \VmatA \DmatA^{1/2} + \Rmat_{2,1}.
\end{align}
Using the fact that $\Amat^\intercal \UmatA \DmatA^{-1/2} = \VmatA \DmatA^{1/2}$ and $\VmatP \VmatP^\intercal \Pmat^\intercal = \Pmat^\intercal$, and following the same algebraic steps as \eqref{eq:XminXPdecomp}, we have: 
\begin{align}
    \hat \Ymat - \YmatP \Wmat &= \Amat^\intercal \UmatA \DmatA^{-1/2} - \VmatP \VmatP^\intercal \Amat^\intercal \UmatA \DmatA^{-1/2} + \Rmat_{2,1} 
    \\ &= (\Amat - \Pmat)^\intercal \UmatA \DmatA^{-1/2} - (\VmatP \VmatP^\intercal \Amat^\intercal - \Pmat^\intercal)\UmatA \DmatA^{-1/2} + \Rmat_{2,1} 
    \\ &= (\Amat - \Pmat)^\intercal \UmatA \DmatA^{-1/2} - \VmatP \VmatP^\intercal(\Amat - \Pmat)^\intercal \UmatA \DmatA^{-1/2} + \Rmat_{2,1} 
    \\ &= (\Id - \VmatP \VmatP^\intercal )(\Amat - \Pmat)^\intercal \UmatA \DmatA^{-1/2} + \Rmat_{2,1}
    \\ &= (\Id - \VmatP \VmatP^\intercal )(\Amat - \Pmat)^\intercal [\UmatP \Wmat + (\UmatA - \UmatP \Wmat)]\DmatA^{-1/2} + \Rmat_{2,1}
    \\ &= (\Amat - \Pmat)^\intercal \UmatP\Wmat\DmatA^{-1/2} + \Rmat_{2,3} + \Rmat_{2,2} + \Rmat_{2,1}
    \\ &= (\Amat - \Pmat)^\intercal \UmatP\DmatP^{-1/2} \Wmat + \Rmat_{2,4} + \Rmat_{2,3} + \Rmat_{2,2} + \Rmat_{2,1}.
\label{eq:YminYPdecomp}
\end{align}
Hence, by Proposition \ref{prop:RatesforVariousTerms3}:
\begin{align}
    \twoinf{\hat \Ymat - \YmatP \Wmat} &= \twoinf{(\Amat- \Pmat)^\intercal \UmatP \DmatP^{-1/2}} + \Op \left\{ \frac{\maxKT^{3/2} \log^{3/2}(n)}{\Kn^{5/4}\Tn^{5/4} \rho_n^{2} n} \right\} \\ &\leq \sigma_d(\Pmat)^{-1/2} \twoinf{(\Amat- \Pmat)^\intercal \UmatP} + \Op \left\{ \frac{\maxKT^{3/2} \log^{3/2}(n)}{\Kn^{5/4}\Tn^{5/4} \rho_n^{2} n} \right\}.
\end{align}
Using an identical argument to \eqref{eq:global_to_local_Xk}, we obtain: 
\begin{equation}
    \twoinf{\hat \Ymat^t - \YmatP^t \Wmat} \leq \sigma_d(\Pmat)^{-1/2} \twoinf{(\Amat^t - \Pmat^t)^\intercal \UmatP} + \Op \left\{ \frac{\maxKT^{3/2} \log^{3/2}(n)}{\Kn^{5/4}\Tn^{5/4} \rho_n^{2} n} \right\}.
\end{equation}
The term $\twoinf{(\Amat^t- \Pmat^t)^\intercal \UmatP}$ can be bounded using an identical approach to the proof for the left DUASE, based on conditioning on a set of latent positions and making use of Hoeffding's inequality. This gives $\twoinf{(\Amat^t- \Pmat^t)^\intercal \UmatP} = \Op\{\log^{1/2}(n)\}$. In addition, $\sigma_d(\Pmat)^{-1/2} = \Op( \rho_n^{-1/2}n^{-1/2}\Kn^{-1/4} \Tn^{-1/4})$, hence for $\WY = \Rtilde \Wmat$:
\begin{equation}
\twoinf{\hat \Ymat^t \WY^{-1} - \Ymat^t } = \Op \left\{ \frac{\log^{1/2}(n)}{\rho_n^{1/2} n^{1/2} \Kn^{1/2}} \right\},
\end{equation}
which completes the proof. 
\end{proof}

\section{Proof of Theorem~\ref{result:CLT}}
\label{sec:proof_clt}
\begin{proof}
     Once again, beginning with the left embedding, we make use of the identities $\Xmat=\XmatP\Ltilde^{-1}$, $\WX = \Ltilde \Wmat$ and the decomposition in \eqref{eq:XminXPdecomp} to write
    \begin{multline}    
    n^{1/2} \Tn^{1/2} (\hat \Xmat \WX^{-1} - \Xmat) =
    n^{1/2} \Tn^{1/2} (\hat \Xmat - \XmatP\Ltilde^{-1}\WX) \WX^{-1} =
    \\n^{1/2} \Tn^{1/2} (\Amat - \Pmat )\VmatP \DmatP^{-1/2} \Ltilde^{-1} + n^{1/2} \Tn^{1/2} (\Rmat_{1,4} + \Rmat_{1,3} + \Rmat_{1,2} + \Rmat_{1,1})\WX^{-1}.
    \end{multline}
    By Proposition \ref{prop:RatesforVariousTerms3}, the term $||n^{1/2} \Tn^{1/2} (\Rmat_{1,4} + \Rmat_{1,3} + \Rmat_{1,2} + \Rmat_{1,1})\WX^{-1}||_{\tti} \to 0$ with overwhelming probability for $n\to\infty$. To analyse the term $n^{1/2} \Tn^{1/2} (\Amat - \Pmat )\VmatP \DmatP^{-1/2} \Ltilde^{-1}$, we express it in terms of its block-wise components. By unstacking both sides of the equation above (and grouping the residual terms for simplicity), we find an analogous expression for the left embedding of each layer $k$:
    \begin{equation}
    n^{1/2} \Tn^{1/2} (\hat \Xmat^k \WX^{-1} - \Xmat^k) = n^{1/2} \Tn^{1/2} (\Amat^k - \Pmat^k )\VmatP \DmatP^{-1/2} \Ltilde^{-1} + \Op\left\{ \log^{-1/2}(n)\right\}.
    \end{equation}
    We further decompose the first term on the right-hand side of this expression by writing it as a sum of terms related to individual time points $t$. Using the identity $\VmatP^t \DmatP^{-1/2} = \YmatP^t \DmatP^{-1} = \Ymat^t \Rtilde \DmatP^{-1}$ (where $\VmatP^t$ is defined as in Proposition~\ref{prop:UpAminPVporder}), we obtain:
    \begin{equation}
    n^{1/2} \Tn^{1/2} (\Amat^k - \Pmat^k )\VmatP \DmatP^{-1/2} \Ltilde^{-1} = n^{1/2} \Tn^{1/2} \sum_{t=1}^{\Tn} (\Amat^{k,t} - \Pmat^{k,t})\Ymat^t \Rtilde \DmatP^{-1} \Ltilde^{-1} .
    \end{equation}
    Breaking this expression down further to the individual rows (corresponding the left embedding of the individual nodes) and transposing both sides, we see that
    \begin{equation}
    n^{1/2} \Tn^{1/2} (\hat \Xmat^k \WX^{-1} - \Xmat^k)_i^\intercal = n^{1/2} \Tn^{1/2} \sum_{t=1}^{\Tn} (\Rtilde \DmatP^{-1} \Ltilde^{-1})^\intercal [(\Amat^{k,t} - \Pmat^{k,t})\Ymat^t]_i^\intercal + \Op\left\{ \log^{-1/2}(n) \right\}.
    \end{equation}
    Therefore, using the identity derived in Proposition \ref{prop:XTX_YTY_ident} and the result above, we obtain: 
    \begin{equation}
    n^{1/2} \Tn^{1/2} (\hat \Xmat^k \WX^{-1} - \Xmat^k)_i^\intercal = n^{1/2} \Tn^{1/2} ({\Ymat}^\intercal \Ymat)^{-1} \sum_{t=1}^{\Tn} [(\Amat^{k,t} - \Pmat^{k,t})\Ymat^t]_i^\intercal + \Op\left\{ \log^{-1/2}(n) \right\}.
    \end{equation}
    We now define $\boldsymbol{\nu}^{\ast t}$ to be a random $n \times d$ matrix where each row is sampled independently from $F_{Y,t}$. By the definition of $\mat{Y}^t$ in Section~\ref{sec:sparsity_considerations} we  substitute $\rho_n^{1/2} \boldsymbol{\nu}^{\ast t}$ for $\mat{Y}^t$. Furthermore, we break down $[(\Amat^{k,t} - \Pmat^{k,t})\Ymat^t]_i^\intercal$ into the sum of individual entries. We also absorb the diagonal term into the residual, since these are $\Op(n^{-1/2}\Tn^{-1/2})$. This yields:
    \begin{multline}
    n^{1/2} \Tn^{1/2} (\hat \Xmat^k \WX^{-1} - \Xmat^k)_i^\intercal = \\ \rho_n n \Tn ({\Ymat}^\intercal \Ymat)^{-1} \sum_{t=1}^{\Tn} \left\{ \frac{1}{\rho_n^{1/2} n^{1/2} \Tn^{1/2}} \sum_{j=1; j\neq i}^n [(\Amat^{k,t}_{i,j} - \Pmat^{k,t}_{i,j})\, \boldsymbol{\nu}^{\ast t}_j] \right\} + \Op\left\{ \log^{-1/2}(n) \right\}.
    \end{multline}
    By the law of large numbers, the quantity $\rho_n n \Tn (\Ymat^{\intercal} \Ymat)^{-1}$ converges in probability to the constant matrix $ \tilde \bDelta_{Y}^{-1}$. To conclude the proof via Slutsky's theorem, it remains to show that the double summation above converges in distribution. By reordering this double sum and conditioning on $\boldsymbol{\xi}^k_i = \vec{x}$ we can express it as a the sum of $n-1$ independent and identically distributed random variables $Z_{j} = \rho_n^{-1/2} \Tn^{-1/2} \sum_{t=1}^{\Tn} [(\Amat^{k,t}_{i,j} - \Pmat^{k,t}_{i,j})\, \boldsymbol{\nu}^{\ast t}_j]$. Hence,
    $$\frac{1}{n^{1/2}}   \sum_{j=1; j\neq i}^n \frac{1}{\rho_n^{1/2} \Tn^{1/2}} \sum_{t=1}^{\Tn} [(\Amat^{k,t}_{i,j} - \Pmat^{k,t}_{i,j})\, \boldsymbol{\nu}^{\ast t}_j]  = \frac{1}{n^{1/2}}   \sum_{j=1; j\neq i}^n Z_j.$$
    By definition, each $Z_j$ is a random variable with mean $0$ and Assumption \ref{ass:clt-matrices} ensures that the covariance matrix $\mat{V}_{Y, n}(\vec{x}) = \mathrm{Var} (Z_j)$ converges to a constant matrix $\mat{V}_Y ( \vec{x}) \in \mathbb{R}^{d\times d}$ for $n\to\infty$, such that: 
    \begin{equation}
    \mat{V}_{Y}( \vec{x}) = \lim_{n \to \infty} \mathbb{E} \left[ \frac{1}{\Tn} \sum_{t=1}^{\Tn}  \vec{x}^\intercal \boldsymbol{\nu}^t (1 - \rho_n  \vec{x}^\intercal \boldsymbol{\nu}^t) \cdot \boldsymbol{\nu}^t {\boldsymbol{\nu}^t}^\intercal \right],
    \end{equation}
    where $\boldsymbol\nu^t\sim F_{Y,t}$. 
    Because the average of a convergent sequence converges to the limit of the sequence, the average variance of the $Z_j$ converges to $\mat{V}_Y ( \vec{x})$. We therefore apply the multivariate Lindeberg-Feller central limit theorem to find that
    \begin{equation}
    \frac{1}{n^{1/2}}   \sum_{j=1; j\neq i}^n \frac{1}{\rho_n^{1/2} \Tn^{1/2}} \sum_{t=1}^{\Tn} [(\Amat^{k,t}_{i,j} - \Pmat^{k,t}_{i,j})\, \boldsymbol{\nu}^{\ast t}_j] \to \mathcal{N} \left\{\mat{0}, \mat{V}_Y ( \vec{x}) \right\}
    \end{equation}
    in distribution for $n\to\infty$. 
    Hence, by Slutsky's Theorem we conclude that: 
    \begin{equation}
    n^{1/2}\Tn^{1/2} (\hat \Xmat^k \WX^{-1} - \Xmat^k)_i^\intercal \to \mathcal{N} \left\{\mat{0}, \tilde \bDelta_{Y}^{-1} \mat{V}_Y( \vec{x}) \tilde \bDelta_{Y}^{-1} \right\}\end{equation}
    in distribution for $n\to\infty$. This proves the first of the two limit theorems stated in Theorem~\ref{result:CLT}.  
    
    The proof for the result on the right embedding follows a similar argument. 
    From \eqref{eq:YminYPdecomp}, Proposition \ref{prop:RatesforVariousTerms3}, and the identities $\Ymat=\YmatP\Rtilde^{-1}$, $\WY = \Rtilde \Wmat$, we have
    \begin{equation}
        n^{1/2}\Kn^{1/2}(\hat \Ymat \WY^{-1} - \Ymat ) = n^{1/2}\Kn^{1/2}(\Amat - \Pmat)^\intercal \UmatP\DmatP^{-1/2} \Rtilde^{-1} + \Op\left\{ \log^{-1/2}(n)\right\}.
    \end{equation}
    Let $\Amat^t$ and $\Pmat^t$ denote the $[n(t-1)+1]$-th through $nt$-th columns of $\Amat$ and $\Pmat$ respectively. Then:
    \begin{equation}
        n^{1/2}\Kn^{1/2}(\hat \Ymat^t \WY^{-1} - \Ymat^t) = n^{1/2}\Kn^{1/2}(\Amat^t - \Pmat^t)^\intercal \UmatP\DmatP^{-1/2}  \Rtilde^{-1} + \Op\left\{ \log^{-1/2}(n)\right\}.
    \end{equation}
    By writing the first term on the right-hand side as the sum of terms related to each layer $k$ and plugging in the identity $\UmatP^k \DmatP^{-1/2} = \XmatP^k \DmatP^{-1} = \Xmat^k \Ltilde \DmatP^{-1}$, we find:  
    \begin{equation}
        n^{1/2}\Kn^{1/2}(\Amat^t - \Pmat^t)^\intercal \UmatP\DmatP^{-1/2}  \Rtilde^{-1} = n^{1/2}\Kn^{1/2} \sum_{k=1}^{\Kn} (\Amat^{k,t} - \Pmat^{k,t})^\intercal\Xmat^k \Ltilde \DmatP^{-1} \Rtilde^{-1}.
    \end{equation}
    Hence, for the $i$-th row transposed, we get:
    \begin{equation}
        n^{1/2}\Kn^{1/2}(\hat \Ymat^t \WY^{-1} - \Ymat^t)_i^\intercal = n^{1/2}\Kn^{1/2} (\Ltilde \DmatP^{-1} \Rtilde^{-1})^\intercal \sum_{k=1}^{\Kn} [(\Amat^{k,t} - \Pmat^{k,t})^\intercal\Xmat^k]_i^\intercal + \Op\left\{ \log^{-1/2}(n)\right\}.
    \end{equation}
    Plugging in the identity from Proposition \ref{prop:XTX_YTY_ident}, as well as the identity $\Xmat^k = \rho_n^{1/2} \boldsymbol{\xi}^{\ast k}$ gives
    \begin{multline}
        n^{1/2}\Kn^{1/2}(\hat \Ymat^t \WY^{-1} - \Ymat^t)_i^\intercal = \\ \rho^{-1}n^{-1}\Kn^{-1} (\Xmat^\intercal \Xmat)^{-1} \frac{1}{\rho^{1/2}n^{1/2}\Kn^{1/2}}\sum_{k=1}^{\Kn} [(\Amat^{k,t} - \Pmat^{k,t})^\intercal  \boldsymbol{\xi}^{\ast k}]_i^\intercal + \Op\left\{ \log^{-1/2}(n)\right\}.
    \label{eq:rightCLTexpansion}
    \end{multline}
    By conditioning on a fixed right latent position $\vec \nu^t_i = \vec y$, a process identical to that for the left embedding shows that: 
    \begin{equation}
        \frac{1}{\rho^{1/2}n^{1/2}\Kn^{1/2}}\sum_{k=1}^{\Kn} [(\Amat^{k,t} - \Pmat^{k,t})^\intercal  \boldsymbol{\xi}^{\ast k}]_i^\intercal \to \mathcal{N}\left\{0, \Vmat_X (\vec y) \right\}.
    \end{equation}
    in distribution for $n\to\infty$.
    Furthermore, by the law of large numbers, the quantity $\rho^{-1}n^{-1}\Kn^{-1} (\Xmat^\intercal \Xmat)^{-1}$ converges in probability to $\tilde \bDelta_X^{-1}$. Therefore, it follows from Slutsky's Theorem and \eqref{eq:rightCLTexpansion} that conditional on $\vec \nu^t_i = \vec y$; 
    \begin{equation}
        n^{1/2}\Kn^{1/2}(\hat \Ymat^t \WY^{-1} - \Ymat^t)_i^\intercal
        \to \mathcal{N}\left\{0, \tilde \bDelta_X^{-1} \Vmat_X (\vec y) \tilde \bDelta_X^{-1} \right\}
    \end{equation}
    in distribution for $n\to\infty$. This proves the second of the two limit theorems stated in Theorem~\ref{result:CLT}.  
\end{proof}

\bibliographystyleSM{rss}
\bibliographySM{references}



\end{document}